\documentclass[fleqn,usenatbib]{mnras}

\usepackage[T1]{fontenc}    
\usepackage{ae,aecompl}

\usepackage{soul}
\usepackage{cancel}
\usepackage{graphicx}	
\usepackage{verbatim}
\usepackage[dvipsnames]{xcolor}

\usepackage{amsmath}	
\usepackage{amssymb}	
\usepackage{newtxtext,newtxmath}

\usepackage{ragged2e}
\usepackage{subfiles}

\title[SKA Science Data Challenge 2]{SKA Science Data Challenge 2: analysis and results}

\author[Hartley et al.]{\Large P. Hartley$^{1}$\thanks{E-mail: philippa.hartley@skao.int}, A. Bonaldi$^{1,2}$, R. Braun$^{1}$, 
J. N. H. S. Aditya$^{3}$, 
S. Aicardi$^{4}$,
L. Alegre$^{1,5}$, 
A. Chakraborty$^{6}$,\newauthor \Large
X. Chen$^{7}$,
S. Choudhuri$^{8,9}$,
A. O. Clarke$^{1}$
% ...
\\
Affiliations can be found after the references}

\date{Accepted XXX. Received YYY; in original form ZZZ}

\pubyear{2022}

\begin{document}
\label{firstpage}
\pagerange{\pageref{firstpage}--\pageref{lastpage}}
\maketitle

\begin{abstract}
The Square Kilometre Array Observatory (SKAO) will explore the radio sky to new depths in order to conduct transformational science. SKAO data products made available to astronomers will be correspondingly large and complex, requiring the application of advanced analysis techniques to extract key science findings. To this end, SKAO is conducting a series of Science Data Challenges, each designed to familiarise the scientific community with SKAO data and to drive the development of new analysis techniques. We present the results from Science Data Challenge 2 (SDC2), which invited participants to find and characterise 233245 neutral hydrogen (H{\sc i}) sources in a simulated data product representing a 2000\,h SKA MID spectral line observation from redshifts 0.25 to 0.5. Through the generous support of eight international supercomputing facilities, participants were able to undertake the Challenge using dedicated computational resources. Alongside the main challenge, `reproducibility awards' were made in recognition of those pipelines which demonstrated Open Science best practice. The Challenge saw over 100 participants develop a range of new and existing techniques, with results that highlight the strengths of multidisciplinary and collaborative effort. The winning strategy -- which combined predictions from two independent machine learning techniques to yield a 20 percent improvement in overall performance -- underscores one of the main Challenge outcomes: that of method complementarity. It is likely that the combination of methods in a so-called ensemble approach will be key to exploiting very large astronomical datasets.

\end{abstract}

\begin{keywords}
methods: data analysis -- radio lines: galaxies -- techniques: imaging spectroscopy -- galaxies: statistics -- surveys -- software: simulations
\end{keywords}



\section{Introduction}
\label{intro}

The Square Kilometre Array (SKA) project was born from an ambition to create a telescope sensitive enough to trace the formation of the earliest galaxies. Observing this era via the very weak emission from neutral hydrogen atoms will be possible only by using a collecting area of unprecedented size: large enough not only to provide a window onto {\it Cosmic Dawn} but -- thanks to its increase in sensitivity over current instruments -- also to explore new frontiers in  galaxy evolution and cosmology, cosmic magnetism, the laws of gravity, extraterrestrial life and -- in the strong tradition of radio astronomy \citep{wilkinson2004exploration} -- the unknown (see the SKA Science Book, \citealt{braun2015advancing} for a comprehensive description of the full SKA science case). 

First light at the SKA Observatory (SKAO) will mark a paradigm shift not only in the way we see the Universe but also in how we undertake scientific investigation. In order to perform such sensitive observations and extract scientific findings, huge amounts of data will need to be captured, transported, processed, stored, shared and analysed. Innovations developed in order to enable the SKAO data journey will drive forward data technologies across software, hardware and logistics. In a truly global collaborative effort, preparations are underway under the guidance of the SKA Regional Centre Steering Committee to build the required data infrastructure and prepare the community for access to it \citep{2020SPIE11449E..0XC}. Alongside operational planning,  scientific planning -- undertaken by the SKAO Science Working Groups -- is underway in order to maximise the exploitation of future SKAO datasets. The SKA model of data delivery will provide science users with data in the form of science-ready image and non-image SKAO products, with calibration and pre-processing having been performed by the Observatory within the Science Data Processor (SDP) and at the SKA Regional Centres (SRCs). While this model reduces by many orders of magnitude the burden of data volume on science teams, the size and complexity of the final data products remains unprecedented \citep{2020RSPTA.37890060S}.

The primary goal of the SKAO Science Data Challenge (SDC) series is defined thus: 
\begin{enumerate}
   \item To support future observers to prepare for SKAO data.
\end{enumerate}

\noindent This goal is achieved via the following objectives:

\begin{description}
       \item \textbullet \;To familiarise the astronomy community with the nature of SKAO data products.
    \item \textbullet \;To drive forward the development of data analysis techniques.
\end{description}

\noindent The first objective allows participants not only to gain familiarity with the size and complexity of SKAO data, but also with the provision of data products in science-ready form. It is achieved through the distribution of publicly available\footnote{ \url{https://astronomers.skatelescope.org/ska-data-challenges/}} real or simulated datasets designed to represent as closely as possible future SKAO data. A successful Challenge will see engagement and participation representing a broad range of geography and expertise, and a step forward by participants in the understanding and skills involved in analysing SKA-like data. The second objective is achieved through the application of new or existing methods in order to extract findings from the data. Standardised cross-comparisons of methods, which would require a strict set of running conditions and constraints on participants, are not performed. Instead, the focus is on inclusion, training, and the generation of ideas. A successful Challenge will see the application of diverse ideas and methods to the problem, and an understanding of the ability of respective methods to produce useful findings.

The SKAO is committed to Open Science values and the FAIR data principles \citep{Wilkinson2016,katz2021fresh} of Findability, Accessibility, Interoperability and Reproducibility. Accordingly, we aim to ensure equal accessibility to the Challenges for all participants. In the latest Challenge, teams were able to access the $\sim$1 TB Challenge dataset and computational resources at one of eight partner supercomputing facilities, at which each could deploy their own analysis pipelines (Section~\ref{hpcs}). This model also served as a test bed for a number of future SRC technologies.  Throughout the Challenge, a strong emphasis was placed on the reproducibility and reusability of software solutions. All teams taking part in the Challenge were eligible to receive a reproducibility prize, awarded against a set of pre-defined criteria. We thus identified two secondary goals for this Challenge:

\begin{enumerate}
        \item To test SKA Regional Centre prototyping.
    \item To encourage Open Science best practice.
\end{enumerate}

Science Data Challenge 1 (SDC1, \citealt{Bonaldi_2020}) saw participating teams find and characterise sources in simulated SKA-MID continuum images, with results that demonstrate the complementarity of methods, the challenge of finding sources in crowded fields, and the importance of careful image partitioning. Domain knowledge proved important not only in the design of pipelines but in the application of correct unit conversions specific to radio astronomy. SDCs benefit from additional domain reference material to support participants who do not have a radio astronomy background. 

Science Data Challenge 2\footnote{\url{https://sdc2.astronomers.skatelescope.org/}} (SDC2) involved a simulated spectral line observation designed to represent the SKAO view of neutral hydrogen (H{\sc i}) emission up to $z=0.5$, again inviting participants to attempt source finding and characterisation within a very large data product. Resulting from the `spin-flip' of an electron in a neutral hydrogen atom, 21cm spectral line emission and absorption traces the distribution of H{\sc i} across the history of the Universe. This cold gas exists in and around galaxies, fueling star-formation via ongoing infall from the cosmic web. Observations of individual H{\sc i} sources can reveal the interactions between galaxies and the surrounding intergalactic medium (IGM; \citealt{2015aska.confE.132P}), can probe stellar feedback processes within the interstellar medium (ISM, \citealt{2015arXiv150101211D}), and can allow us to study the impact of AGN on the large-scale gas distribution in galaxies \citep{2015aska.confE.134M}. H{\sc i} dynamics also provide a measurement of the dark matter content of galaxies \citep{2015aska.confE.133P}. Deep H{\sc i} surveys are therefore crucial for our understanding of galaxy formation and evolution over cosmic time \citep{2015aska.confE.128B,2010MNRAS.406...43P,2017PASA...34...52M,2021arXiv211206488D}. 

The faintness of H{\sc i} emission has until recently limited survey depths to up to $z\sim0.25$  (see  \citealt{2008A&ARv..15..189S}, \citealt{2013pss6.book..183V} and \citealt{2020Ap&SS.365..118K} for  reviews of the results so far).  H{\sc i} emission has now been imaged in a starburst galaxy at $z\sim0.376$ \citep{2016ApJ...824L...1F} using the Very Large Array within the COSMOS H I Large Extragalactic Survey (CHILES), and signals observed using the Giant Metrewave Radio Telescope (GMRT) have been stacked in order to make a successful measurement of the cosmic H{\sc i} mass density at $0.2<z<0.4$ \citep{Bera_2019} and to detect the H{\sc i} 21 cm signal from 2841 galaxies at average redshift $z\sim1.3$ \citep{2021ApJ...913L..24C}.  The MeerKAT telescope -- a precursor to the SKAO -- has now launched the Looking At the Distant Universe with the MeerKAT Array (LADUMA) survey \citep{blyth2016laduma}, which will image H{\sc i} emission in the Chandra Deep Field-South out to $z\sim1$. The SKAO MID telescope will survey to depths of $z\sim1$ in emission and $z\sim3$ in absorption across a wider field. Comparing both surveys over 2000 hours of observation, an SKAO MID survey is likely to increase by 0.8 dex the number of detected galaxies, probing a cosmic volume $V_{\rm c} \approx 185$ Mpc versus $V_{\rm c} \approx 74$ Mpc and significantly reducing the sensitivity of the results to cosmic variance. The size of resulting datasets necessitates the use of automated source finding methods; several software tools are currently available for H{\sc i} source detection and characterisation \citep{2012PASA...29..244F,Jurek_2012,2012ascl.soft01011W,westerlund_harris_2014,2015MNRAS.448.1922S,2021MNRAS.506.3962W,teeninga2015improved} and a comparative study based on WSRT data has recently been performed \citep{2022arXiv221112809B}.

In this paper we report on the outcome of SDC2. The structure of the paper is as follows: in Section 2 we define the Challenge; in Section 3 we describe the simulation of the SDC2 datasets; in Section 4 we present the methods used by participating teams to complete the Challenge; in Section 5 we describe the scoring procedure;  in Sections 6 and 7 we present the Challenge results and analysis, before setting out our conclusions in Section 8.

\section{The Challenge}

In this Section we present an overview of the Challenge delivery and the data product supplied to Challenge teams, followed by the definition of the Challenge undertaken.

\subsection{Challenge overview}
\label{thechallenge}

Participating teams were invited to access a 913\,GB dataset hosted on dedicated facilities provided by the SDC2 computational resource partners (Section~\ref{hpcs}). The dataset, 5851 $\times$ 5851 $\times$ 6668 pixels in size, simulates an H{\sc i} imaging datacube representative of future deep SKA MID spectral line observations, with the following specifications:

\begin{enumerate}
\item 20 square degrees field of view.
\item 7 arcsec beam size, sampled with $2.8\times2.8$ arcsec pixels.
\item 950–1150 MHz bandwidth, sampled with a 30 kHz resolution. This corresponds to rest frame velocity widths 7.8 and 9.5 km s$^{-1}$ at the upper and lower limits, respectively, of the redshift interval $z = $ 0.235--0.495.
\item Noise consistent with a 2000 hour total observation, in the range 26--31 \textmu Jy per channel.
\item Systematics including imperfect continuum subtraction, simulated RFI flagging and excess noise due to RFI. 
\end{enumerate}

\noindent The H{\sc i} datacube was accompanied by a radio continuum datacube covering the same field of view at the same spatial resolution, with a 950-1400 MHz frequency range at a 50 MHz frequency resolution.

Challenge teams were invited to use analysis methods that were any combination of purpose-built and bespoke to existing and publicly available. Together with the full-size Challenge dataset, two smaller datasets were made available for development purposes. Generated using the same procedure as the full-size dataset but with a different statistical realization,  the `development' and `large development' datasets were provided along with truth catalogues listing H{\sc i} source property values. A further, `evaluation', dataset was provided without a truth catalogue, in order to allow teams to validate their methods in a blind way prior to application to the full dataset. The evaluation dataset was also used by teams to gain access to the full-size datacube hosted at an SDC2 partner facility. Access was granted upon submission of a source catalogue based on the evaluation dataset and matching a required format. The development and evaluation datasets were made available for download prior to and during the Challenge. 

The Challenge description, its rules, its scoring method and a description of the data simulations were provided on the Challenge website before and during the Challenge. A dedicated online discussion forum was used throughout the Challenge to provide information to participants, to answer questions about the Challenge and to facilitate participant interaction. Definitions of conventions and units applicable to the challenge were circulated to participants before and during the Challenge.

\subsection{Supercomputing partner facilities}
\label{hpcs}

The following eight supercomputing centres formed an international platform on which the full Challenge dataset was hosted and processed:\newline

\noindent\textit{AusSRC and Pawsey – Perth, Australia, \url{aussrc.org}}\newline

\noindent\textit{China SRC-proto – Shanghai, China, \cite{2022arXiv220613022A}}\newline

\noindent\textit{CSCS – Lugano, Switzerland, \url{www.cscs.ch}}\newline

\noindent\textit{ENGAGE SKA-UCLCA – Aveiro and Coimbra, Portugal, \url{www.engageska-portugal.pt}; \url{www.uc.pt/lca}} \newline

\noindent\textit{GENCI-IDRIS – Orsay, France, \url{www.genci.fr}}\newline

\noindent\textit{IAA-CSIC – Granada, Spain, \cite{10.1117/1.JATIS.8.1.011004}}\newline

\noindent\textit{INAF – Rome, Italy, \url{www.inaf.it}}\newline

\noindent\textit{IRIS (STFC) – UK, \url{www.iris.ac.uk}}\newline

\noindent Collectively, the Challenge facilities provided 15 million CPU hours of processing and 15\,TB of RAM to participating teams.

\subsection{The challenge definition}

The Challenge results were scored on the full-size dataset, on which teams undertook:

\begin{description}
  \item Source finding, defined as the determination of the location in RA (degrees), Dec (degrees) and central frequency (Hz) of the dynamical centre of each source.
  
  \item Source characterisation, defined as the recovery of the following properties:
    \begin{enumerate}
     \item Integrated line flux (Jy Hz): the total flux density $S$ integrated over the signal $\int S {\rm d}_{\nu}$.
     \item H{\sc i} size (arcsec): the H{\sc i} major axis diameter at 1 M$_{\odot}$ pc$^{-2}$.
     \item Line width (km s$^{-1}$): the observed line width at 20 percent of its peak.
     \item Position angle (degrees): the angle of the major axis of the receding side of the galaxy, measured anticlockwise from North. 
     \item Inclination angle (degrees): the angle between line-of-sight and a line normal to the plane of the galaxy.    
   
   \end{enumerate}
 
\end{description}

\noindent Catalogues listing measured properties were submitted via a dedicated scoring service (see Section~\ref{scoring}), which compared each submission with the catalogue of truth values and returned a score. For the duration of the Challenge, scores could be updated at any time; the outcome of the Challenge was  based on the highest scores submitted by each team.  The Challenge opened on 1st February 2021 and closed on 31st July 2021.

\subsection{Reproducibility awards}

Alongside the main challenge, teams were eligible for `reproducibility awards’, which were granted to all teams whose processing pipelines demonstrated best practice in the provision of reproducible methods and Open Science. An essential part of the scientific method, reproducibility leads to better, more efficient science. Open Science generalises the principle of reproducibility, allowing previous work to be built upon for the future. Reproducibility awards ran in parallel and independently from the SDC2 score, and there was no cap on the number of teams to whom the awards were given.

\section{The simulations}
\label{sims}


Simulation of the H{\sc i} datacubes involved three steps: source catalogue generation, sky model creation, and telescope simulation. 


\subsection{Source catalogues}

To produce a catalogue of sources with both continuum and H{\sc i} properties we used the Tiered Radio Continuum Simulation (TRECS; \citealt{bonaldi2019tiered}) as updated by Bonaldi et al. 2023 in prep. Initial catalogues of H{\sc i} emission sources were generated by sampling from an H{\sc i} mass function derived from the ALFALFA survey results \citep{2018MNRAS.477....2J}:

\begin{equation}
\phi(M_{\rm HI},z)=\ln(10)\;\phi_{*}\left(\frac{M_{\rm HI}}{M_*(z)}\right)^{\alpha+1}e^{-\frac{M_{\rm HI}}{M_*(z)}} , 
\label{HIMF_eq}
\end{equation}

\noindent where the knee mass, $M_*= 8.71\times 10^{9} $ M$_{\odot}$, marks the exponential decline from a shallow power law parameterised by $\alpha=-1.25$, and $\phi_{*}=4.5\times 10^{-3} \:{\rm Mpc}^{-3}\; {\rm dex}^{-1} $ is a normalisation constant. A mild redshift dependence was applied by using $\log(M_*(z))= \log(M_*) + 0.075z$.

Conversion from H{\sc i} mass in units solar mass to integrated line flux $F$ followed the relation from \cite{2012MNRAS.426.3385D}:

\begin{equation}
    M_{\rm HI} =  49.8\: F \:{D_{\rm L}}^2 ,
    \label{massderiv}
\end{equation}

\noindent where luminosity distance, ${D_{\rm L}}$,  is measured in Mpc and is obtained via the source redshift. A lower integrated flux limit of 1 Jy Hz was made, such that a fully face-on and unresolved source at this limit  would produce a peak flux density approximately equal to the noise r.m.s.  The catalogue also included a position angle $\theta$ drawn from a uniform distribution between 0--360 degrees, and an inclination angle $i$ from the probability distribution function $f(i)=\sin(i)$.

Catalogues of radio-continuum sources -- star-forming galaxies (SFGs) and Active Galactic Nuclei (AGN)  -- were then generated using the Tiered Continuum Radio Extragalactic Continuum Simulation (T-RECS,  \citealt{bonaldi2019tiered}) for the frequency interval 950-1400 MHz. A flux density limit of $2\times10^{-7}$ Jy at 1150 MHz was applied, corresponding to k-corrected radio luminosities $L_{\rm 1150 \,MHz}= 1.58\times 10^{19}$ W\,Hz$^{-1}$ and $L_{\rm 1150 \,MHz} = 8.59 \times 10^{19}$ W\,Hz$^{-1}$ at the lower and upper redshift limits, respectively, for a source with spectral index $\alpha = -0.7$. Continuum T-RECS catalogue properties included dark matter mass, star-formation rate and redshift.

The H{\sc i} catalogue and the portion of the radio continuum catalogue covering the same redshift interval were then further processed to identify those that would constitute a counterpart, i.e. be hosted by the same galaxy (see Bonaldi et al. 2023 for more details). 

In order to generate source positions in RA ($x$), Dec ($y$) and redshift ($z$) and to provide a realistic clustering signal,  the galaxies were associated with dark matter (DM) haloes from the P-Millennium simulation  \citep{baugh2019galaxy}. Both the mass and environment of host DM haloes were considered; galaxies were associated with available DM haloes having the closest mass in the same redshift interval, and preferential selection of DM haloes with local density lower than 50 objects per cubic Mpc was made for H{\sc i}-containing sources. The redshift of each source was converted to obtain the observed frequency ($\nu$) at its dynamical centre.

\subsection{Sky model}

The sky model was generated using the {\sc python} scripting language, making use of the {\sc astropy}, {\sc scipy} and {\sc scikit-image} libraries for image and cube generation, and using {\sc fitsio} for writing to file.

\subsubsection{H{\sc i} emission datacube}

H{\sc i} sources were injected into the field using an atlas of high quality H{\sc i} source observations. The atlas, containing 55 sources in total,  was collated using samples available from the WSRT Hydrogen Accretion in LOcal GAlaxieS (HALOGAS) survey \citep{2002AJ....123.3124F,2007AJ....134.1019O,2011A&A...526A.118H} -- available online -- and the THINGS survey \citep{2008AJ....136.2563W}, made available after the application of multi-scale beam deconvolution.  The preparation of atlas sources involved the following steps:

\begin{enumerate}

\item Measurement of H{\sc i} major axis diameter at a surface density of 1 M$_{\odot}$ pc$^{-2}$, made after converting source flux to mass per pixel.
\item Masking of all pixels with surface density less than 1 M$_{\odot}$ pc$^{-2}$, in order to produce a positive definite noiseless model.
\item Rotation, using published source position angles, to a common position angle of 0 degrees.
\item Preliminary spatial resampling, such that the physical pixel size of the resampled data would be no lower than required for the lowest redshift simulated sources. A smoothing filter was applied prior to resampling, in order to prevent aliasing.
\item Preliminary velocity resampling after application of a smoothing filter.
\end{enumerate}

Though modestly sized, the atlas sample of real H{\sc i} galaxies represented considerable morphological diversity, containing examples of Hubble stages 2 to 10. The parameter space representing catalogue sources was not completely covered. Physical properties of the atlas sample covered the SFR range 0.004 to 6.05 M$_{\odot}$ y$^{-1}$, the H{\sc i} mass range $1.20\times 10^7$ to $1.41\times 10^{10}$ M$_{\odot}$ and the H{\sc i} major axis diameter 2.29 to 102.23 kpc. Catalogue sources covered the SFR range 0.0039 to 251 M$_{\odot}$ y$^{-1}$ (median 0.97),  H{\sc i} masses $M_{\rm HI}=  6.99\times 10^{7} $  M$_{\odot}$ and  $4.08\times 10^{8} $ M$_{\odot}$ at the lower and upper limits of the simulated redshift range, respectively (with median $1.14 \times 10^{9}$ and maximum  $ 1.08\times 10^{11}$), and H{\sc i} diameters  $S=$ 4.78 to 270 kpc (median 24.7).

For each source from the simulation catalogue, a source from the prepared atlas of sources was chosen from those nearby in normalised H{\sc i} mass-inclination angle parameter space. Once matched with a catalogue source, atlas sources underwent transformations in size in the spatial cube dimensions $x$ and $y$ and in velocity dimension $V$ in order to obtain the H{\sc i} size $S$, minor axis size $b$ and line width $w_{20}$ . An appropriate smoothing filter was applied prior to all scalings, in order to avoid aliasing effects. Transformation scalings were determined using the catalogue source properties of H{\sc i} mass, inclination angle, and redshift, and making use of the following relations:

\begin{figure}
	\centering
	\includegraphics[trim={0.cm 0.cm 0.cm 0.cm},clip,width=1\columnwidth]{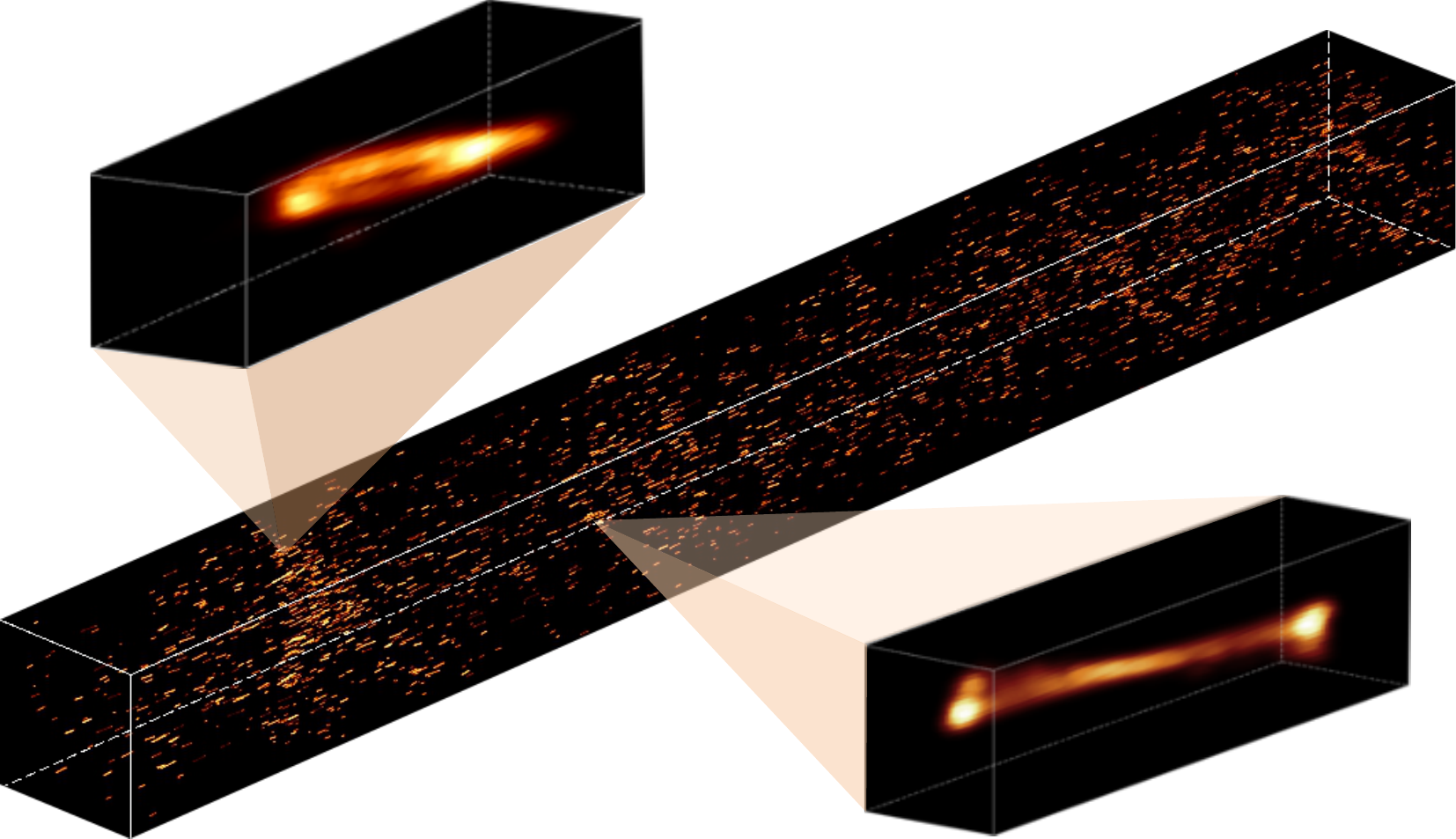}
	\caption{3D view of the ‘development’ H{\sc i} emission datacube, containing 2683 H{\sc i} sources. The cube uses 1286 $\times$ 1286 $\times$ 6668 pixels to represent a 1 square degree field of view across the full Challenge frequency range 0.95–1.15 GHz (redshift 0.235–0.495). A log scaling has been applied to image pixel values. The two shorter axes represent the spatial dimensions and the longer axis the frequency dimenion. }
	\label{HIcube}
\end{figure}

\begin{equation}
 S= 0.51 \log M_{\rm HI} -3.32,   
\end{equation}

\noindent from \cite{1997A&A...324..877B}, in order to determine spatial scalings for mass ;

\begin{equation}
V^2_{\rm rot} = \frac{{\rm G}M_{\rm dyn}}{r},     
\end{equation}

\noindent where $V_{\rm rot}$ is the rest frame rotational velocity at radius $r$ and $M_{\rm dyn}$  is the dynamical mass and is set using $M_{\rm dyn}/M_{\rm HI}=10$, in order to determine frequency scalings for H{\sc i} mass;

\begin{equation}
\cos^2(i) =  \frac{(b/ S)^2 - \alpha^2 }{( 1 - \alpha^2)}, 
\end{equation}

\noindent where $\alpha = 0.2$, in order to determine spatial scalings for inclination;

\begin{equation}
V_{\rm rad} = V_{\rm rot}\, \sin(i), 
\label{vrot}
\end{equation}

\noindent where $V_{\rm rad}$ is the rest frame radial velocity, and 

\begin{equation}
w_{20} = \sqrt(V_T^2 + 2V_{\rm rad}^2),
\end{equation}

where $V_T$ is the contribution to line width from turbulence, in order to determine velocity scalings for inclination. While a best fit to ALFALFA data finds a value $V_T=90$ km s$^{-1}$, a lower value,  $V_T=40$ km s$^{-1}$, is chosen in order to avoid excessive  scaling between peaks in velocity. Spatial scalings for redshift were determined by calculating the angular diameter distance $D_{\rm A}$,  assuming a standard flat cosmology with $\Omega_{\rm m}=0.31$ and H$_0=67.8$ km s$^{-1}$ Mpc$^{-1}$ (Planck Collaboration XIII \citeyear{2016A&A...594A..13P}).



Finally, each transformed object was rotated to its catalogued position angle, convolved with a circular Gaussian of 7 arcsec FWHM and scaled according to total integrated H{\sc i} flux, before being placed in the full H{\sc i} emission field at its designated position in RA, Dec and central frequency (Fig.~\ref{HIcube}).

\subsubsection{Continuum emission datacube}
\label{continuum}
The treatment of continuum counterparts of H{\sc i} objects was dependent on the full width at half maximum (FWHM) continuum size. An empty datacube with spatial resolution matching the H{\sc i} datacube and an initial frequency sampling of 50 MHz was first generated. Each counterpart was then injected into the simulated field as either:

\begin{enumerate}
    \item an extended source, for those objects with a continuum size greater than 3 pixels;
    \item a compact source, for those objects with a continuum size smaller than 3 pixels.
\end{enumerate}

\noindent All compact sources were modelled as unresolved, and added as Gaussians of the same size as the synthesised beam. Images of all extended sources were generated according to their morphological parameters and then added as ``postage stamps'' to an image of the full field, after applying a Gaussian convolving kernel corresponding to the beam. 

The morphological model for the extended SFGs is an exponential Sersic profile, projected into an ellipsoid with a given axis ratio and position angle. The AGN population comprises steep-spectrum AGN, exhibiting the typical double-lobes of FRI and FRII sources, and flat-spectrum AGN,  exhibiting a compact core component together with a single lobe viewed end-on. Within both classes of AGN all sources are treated as the same object type viewed from a different angle. For the steep-spectrum AGN we used the Double Radio-sources Associated with Galactic Nucleus (DRAGNs) library of real, high-resolution AGN images \citep{2013atdr..cat.....L}, scaled in total intensity and size, and randomly rotated and reflected, to generate the postage stamps. All flat-spectrum AGN were added as a pair of Gaussian components: one unresolved and with a given ``core fraction'' of the total flux density, and one with a specified larger size. 

The continuum catalogues accompanying the Challenge datasets report the continuum size of objects as the Largest Angular Size (LAS) and the exponential scale length of the disk for AGN and SFG populations, respectively.

\subsubsection{Net emission and absorption cube}
\label{netcube}

The HI emission cube described in Section~\ref{continuum} was further processed to introduce absorption features and the effect of imperfect continuum subtraction. HI absorption occurs if a radio continuum source is at a higher redshift along the same line of sight as an HI source. The intensity of the effect depends on both the brightness temperature of the continuum source and the HI opacity $\tau\Delta V$ of the HI source.  Absorption features were introduced on the pixels of the HI model cube only if a background continuum source was present having at least a brightness temperature  $T_{\rm min} = 100$\,K. This corresponds to a flux density of $   S_{\rm min} = 7.35 \times 10^{-4}  T_{\rm min} \Delta\phi^2 / \lambda^2 $, with $\Delta\phi$ the beam size in arcsec and $\lambda$ the observing wavelength in cm, yielding $S_{\rm min}$ in Jy beam$^{-1}$.

The absorption signature, $S_{\rm HIA}(\nu)$, was calculated as:
\begin{equation}
S_{\rm HIA}(\nu) = S_{\rm C} [1 - e^{(- \tau \Delta V/{\rm d} V)}],
\end{equation}

\noindent where $S_{\rm C}$ is the continuum model flux density at this frequency and d$V$ is the actual channel sampling in units of km s$^{-1}$. When observed with 100 pc or better physical resolution, the apparent H{\sc i} column density $N_{\rm HI}$, can be related to an associated H{\sc i} opacity \citep{2012ApJ...749...87B}, as

\begin{equation}
    N_{\rm HI} = N_0 e^{-\tau \Delta V} + N_{\infty} ( 1 - e^{-\tau \Delta V}) ,
\end{equation}
\noindent where $N_0 = 1.25 x 10^{20} $ cm$^{-2}$, $N_{\infty} = 7.5 x 10^{21}$ cm$^{-2}$ and a nominal $\Delta V = 15$ km$^{-1}$ provide a good description of the best observational data in hand. In turn, the hydrogen column density, $N_{\rm HI}$, associated with every pixel in the H{\sc i} model cube can be obtained with 
\begin{equation}
N_{\rm HI} = 49.8 \,S_{\rm L}(\nu) \,\Delta \nu\, {\rm M}_{\odot} {(1+z)^4}/ ({ N_{\rm p}\, m_{\rm H}\, \Delta\theta^2\, C_{\rm M}^2}),    
\end{equation}

\noindent where $S_{\rm L}$ is the H{\sc i} brightness in the pixel in Jy beam$^{-1}$, $\Delta \nu\ $ the channel spacing in Hz, ${\rm M}_{\odot}$ a solar mass, $z$ the redshift of the H{\sc i} 21cm line that applies to this pixel, $N_{\rm p}$ the number of pixels per spatial beam, $m_{\rm H}$ the hydrogen atom mass, $\Delta\theta$ the spatial pixel size in radians and $C_{\rm M}$ a Mpc expressed in cm. The preceding constant in the equation follows the flux density to H{\sc i} mass conversion of \cite{2012MNRAS.426.3385D}.

In the current case, the physical resolution is too coarse -- some 10 kpc per pixel -- to resolve the individual cold atomic clouds that give rise to significant H{\sc i} absorption opacity. The apparent column densities per pixel have therefore been subjected to an arbitrary power law rescaling designed to render a plausible amount of observable absorption signatures. We used 

\begin{equation}
N_{\rm HI}'= 10^{19 + [\log10(N_{\rm HI}) - 19] \beta } ,
\end{equation}

\noindent if $N_{\rm HI} > 10^{19}$, with power law index $\beta
= 1.9$. This is followed by a hyperbolic tangent asymptotic filtering:

\begin{equation}
    N_{\rm HI}''= N_{\infty}[{\rm e}^{2 N_{\rm HI}'/N_{\infty} } - 1] / [{\rm e}^{2 N_{\rm HI}'/N_{\infty}} + 1],
\end{equation}

\noindent in order to avoid numerical problems when solving for the opacity.

In order to simulate imperfect continuum emission subtraction within  the final H{\sc i} datacube, a noise cube representing gain calibration errors was produced. We first interpolated the simulated continuum sky model, $S_{\rm C}$($\nu$), to a frequency sampling of 10~MHz, before producing for each channel a two dimensional image of uncorrelated noise to represent a r.m.s. gain calibration error of  $\sigma = 1\times 10^{-3}$ and with spatial sampling 515 $\times$ 515 arcsec.   The spatial and frequency samplings were chosen in order to represent the residual bandpass calibration errors that might result from the typical spectral standing wave pattern of an SKA dish at these frequencies, together with the angular scale over which direction dependent gain differences might be apparent. 

The coarsely sampled noise field was then interpolated up to the 2.8 $\times$ 2.8 arcsec sampling of the sky model and a deliberately imperfect version of the continuum sky model, $S_{\rm NC}$($\nu$), was constructed by multiplying each pixel in the perfect model by $(1+N)$, where $N$ is the value of the corresponding pixel in the noise cube. Finally, both the perfect and imperfect continuum models were downsampled to the final simulation frequency interval of 30 kHz. The net continuum-subtracted H{\sc i} emission and absorption cube, $S(\nu))$ is finally calculated from the sum

\begin{equation}
S(\nu) = S_{\rm L}(\nu) + S_{\rm C}(\nu) - S_{\rm NC}(\nu) - S_{\rm HIA}(\nu).
\end{equation}

\subsection{Telescope simulation}
    
The simulation of telescope sampling effects has been implemented by using {\sc python} to script tasks from the {\sc miriad} package \citep{1995ASPC...77..433S}. Multi-processing parallelisation is exploited by applying the procedure over multiple frequency channels simultaneously. 

\subsubsection{Calculation of effective PSF and noise level}

The synthesized telescope beam was based on a nominal 8 hour duration tracking observation of the complete SKA MID configuration. A one minute time sampling interval was used in order to make beam calculations sufficiently realistic while limiting computational costs. The thermal noise level was based on nominal system performance \citep{2019arXiv191212699B} for an effective on-sky integration time of 2000 hours distributed uniformly over the 20 deg$^2$ survey field. The effective integration time per unit area of the survey field increases towards lower frequencies in proportion to wavelength squared. This is due to the variation in the primary beam size in conjunction with an assumed survey sampling pattern that is fine enough to provide a uniform noise level even at the highest frequency channel. The nominal r.m.s. noise level, $\sigma_{N}$, therefore declines linearly with frequency between 950 and 1150 MHz. 

Observations of the South Celestial Pole (Experiment ID 20190424-0024) using MeerKAT, which is located on the future SKA MID site and will constitute part of the SKA MID array, have been used to obtain a real world total power spectrum. With this power spectrum we can estimate the system noise temperature floor of the MeerKAT receiver system as a function of frequency, in addition to an estimate of any excess average power due to Radio Frequency Interference (RFI). The ratio of excess RFI to system noise temperature, $\gamma_{\rm RFI}$, was used to scale the nominal noise in each frequency channel and to determine the degree of simulated RFI flagging to apply to the nominal visibility sampling. Flagging was applied to all baselines from a minimum B$_{\rm min} = 0$ up to a maximum $B_{\rm max}$ according, in units of wavelength, to

\begin{equation}
B_{\rm max} = 71 \times 10^{(\gamma_{\rm RFI}- 1)^{1/3}},
\end{equation}

\noindent which produced maximum baseline lengths ranging from under 15~m to around 10~km across the relevant range of observing frequencies. The duration of RFI flagging, $\Delta$HA, was determined, in hours, from

\[
    \Delta {\rm HA}= 
\begin{cases}
    0,& \text{if } \gamma_{\rm RFI}< \gamma_{\rm min} \\
    8 \,(\gamma_{\rm RFI} - \gamma_{\rm min})/(\gamma_{\rm max} - \gamma_{\rm min} ),   & \text{if } \gamma_{\rm min}> \gamma_{\rm RFI} > \gamma_{\rm max}\\
    8, &   \text{if }\gamma_{\rm RFI} > \gamma_{\rm max} 
\end{cases}
\]

\noindent where $\gamma_{\rm min} = 1.1$ and $\gamma_{\rm max} = 2$, are used to define the ranges of RFI ratios over which flagging is absent, intermittent or continuous. Intermittent flagging intervals were placed randomly within the nominal HA = $-4$h to +4h tracking window.

After application of flagging to the nominal visibility sampling, the synthesized beam and corresponding ``dirty'' noise image were generated for each frequency channel. During imaging, a super-uniform visibility weighting algorithm was employed that makes use of a 64$\times$64 pixel FWHM Gaussian convolution of the gridded natural visibilities in order to estimate the local density of visibility sampling. The super-uniform re-weighting was followed by a Gaussian tapering of the visibilities to achieve the final target dirty PSF properties, namely the most Gaussian possible dirty beam central lobe with 7$\times$7 arcsec FWHM. The effective PSF is then modified to account for the fact that the survey area will be built up via the linear combination of multiple, finely spaced, telescope pointings on the sky. The effective PSF in this case was formed from the product of the calculated dirty PSF with a model of the telescope primary beam at this frequency, as documented in \cite{2019arXiv191212699B}. The dirty noise image for each channel was then rescaled to have an r.m.s. fluctuation level, $\sigma_i$, corresponding to the nominal sensitivity level of the channel degraded by its RFI noise ratio:

\begin{equation}
 \sigma_i  = \sigma_N \gamma_{\rm RFI}  .
\end{equation}

\subsubsection{Simulated sampling and deconvolution}

The H{\sc i} net absorption and emission datacube (Section~\ref{netcube}) was subjected to simulated deconvolution and residual degradation by the relevant synthesized dirty beam. Any signal, both positive and negative, in excess of three times the local noise level, 3$\sigma_i$, was extracted as a ``clean'' image with the threshold signal retained to form a residual sky image. The residual sky image was subjected to a linear deconvolution (via FFT division) with a 7$\times$7 arcsec Gaussian, truncated at 10 percent of the peak and then convolved with the dirty beam. The final data product cube was formed by summing for each channel the dirty residuals, the previously extracted clean feature image and the dirty noise image.

\subsection{Limitations of the simulated data products}
 \label{limitations}
While significant effort has been expended to make a realistic data product for the Challenge analysis,
there are many limitations to the degree of realism that could be achieved. Some of the most
apparent are outlined below.

\begin{enumerate}
    \item Telescope sampling limitations, arising from the adoption of image plane sky model convolution to approximate the actual imaging process. This forms the most significant limitation to the simulations, but is necessitated by the fact that working instead in the visibility plane would require processing of datasets 7.4~PB in size, far exceeding current capabilities.
    \item Realism of the noise properties: systematic effects such as residual RFI, bandpass ripples, residual continuum sidelobes and deconvolution artifacts were not included in the simulation. Additionally, the properties of the errors that have been included feature mostly Gaussian, uncorrelated noise, which may not represent the complexity of those those found in real interferometric data. 
    \item H{\sc i} emission model limitations, arising from the limited number of real H{\sc i} observations used to generate simulated H{\sc i} sub-cubes. 
    \item Catalogue limitations, arising from the independent generation of H{\sc i} and continuum catalogues.
    \item H{\sc i} absorption model limitations, due to very coarse sampling used to assess physical properties along the line of sight in order to introduce H{\sc i} absorption signatures. Further, the relatively low resolution of the simulated observation results in a low apparent brightness temperature of continuum sources (< 100~K), such that the occurrence of absorption signatures has been restricted only to those continuum sources that exceed this brightness limit.
    \item Continuum emission model limitations, arising from the use of simple models to describe SFGs and flat-spectrum AGN sources, and from the limited number of real images used to generate steep spectrum sources.
   \item An assumption of negligible H{\sc i} self-opacity which, although widely adopted in the current literature, is unlikely to be the case in reality (see e.g. \citealt{2012ApJ...749...87B}).
  \item The overall translation of truth catalogue inputs to simulated source morphologies: the Challenge scoring definition measures the recovery of truth catalogue inputs, while teams themselves measure properties from a simulated realisation of those inputs. This could introduce a degeneracy in the evaluation of method performance.

\end{enumerate}

The limitations listed above would in turn place limits on how well teams' performances on this dataset would transfer to real data.

\section{Methods}
\label{methods}

Participating teams made use of a range of methods to tackle the problem, first making use of the smaller development dataset and truth catalogue in order to investigate techniques. Twelve teams made a successful submission entry using the full Challenge dataset. The methods employed by each of those finalist teams are presented below.

\subsection{Coin}
\textit{C. Heneka, M. Delli Veneri, A. Soroka, F. Gubanov,  A. Meshcheryakov, B. Fraga,  C.R. Bom,  M. Brüggen }\newline

\begin{figure}
	\centering
	\includegraphics[width=0.9\columnwidth]{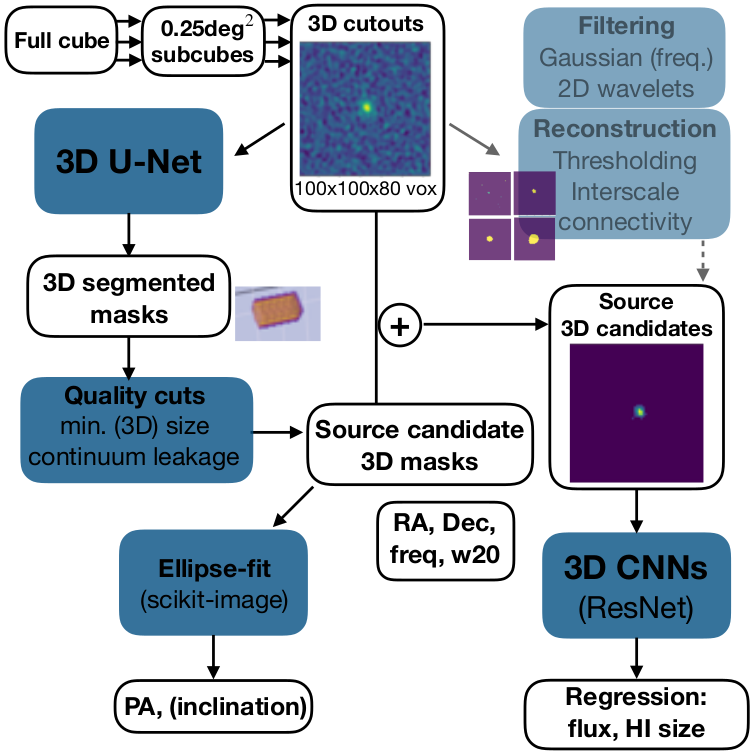}
	\caption{Data processing pipeline used by the Coin team.}
	\label{coin_im}
\end{figure}

\noindent During the Challenge the Coin team tested several modern ML algorithms from scratch alongside the development our own wavelet-based `classical' baseline detection algorithm. For all approaches we first flagged the first 324 channels in order to remove residual RFI, as measured by the per-channel signal mean and variance. We considered the following ML architectures for object detection: 2D/3D U-Nets, R-CNN and an inception-style network that mimics filtering with wavelets. The to-date best-performing architecture was a comparably shallow segmentation U-Net that translated the 2D U-Net in~\citet{ronneberger2015u} to 3D. It was trained on 3D cubic patches taken from the development cube, each containing a source and with no preprocessing applied. We mitigated High ($> 90\%$) rates of false positives to moderate levels ($\sim 50\%$; see Fig.~\ref{coin_im}) by imposing interconnectivity and size cuts on the potential sources and discarding continuum-bright areas. We obtained a roughly constant $\sim$50:50 ratio between true and false positives for 0.25 deg$^2$ cutouts across the development cube and the full Challenge cube. Our `classical' baseline performed an alternative detection procedure, first using Gaussian filtering in the frequency dimension followed by wavelet filtering and thresholding.  Interscale connectivity~\citep{scherzer2010handbook} and reconstruction were performed on the denoised and segmented output. This pipeline detected $< 10\%$ true positives for the Challenge data release: an order of magnitude higher false positive rate than the ML-based pipeline.

Source positions (RA, Dec, central frequency, line width) were directly inferred from the obtained segmentation maps via the \texttt{regionprops} function of the {\sc scikit-image python} package~\citep{scikit-image}.  Source properties (flux, size) were derived through a series of ResNet CNNs~\citep{He2016DeepRecognition} applied to the source candidate 3D cutouts. The position angle was measured using the {\sc scikit-image} package to fit ellipses to sources masks; inclination could not be fitted for most objects.

We conclude that further cleaning and denoising and the application of techniques from the `classical' baseline, such as wavelet filtering, is needed to improve on our machine learning pipeline method. Alternatively, further steps that include classification and a more curated training set could be desirable. Lessons learned in these `from-scratch' developments can give valuable insights into the performance and application of said algorithms, such as the suitability of 3D U-Nets for segmentation of tomographic H{\sc i} data and the need for additional cleaning algorithms jointly with networks or multi-step procedures, such as a classification step, when faced with low S/N data.

\subsection{EPFL}

\textit{ E. Tolley, D. Korber, A. Peel,  A. Galan, M. Sargent, G. Fourestey, C. Gheller, J.-P. Kneib, F. Courbin, J.-L.Starck}\newline

\begin{figure}
	\centering
	\includegraphics[width=0.95\linewidth]{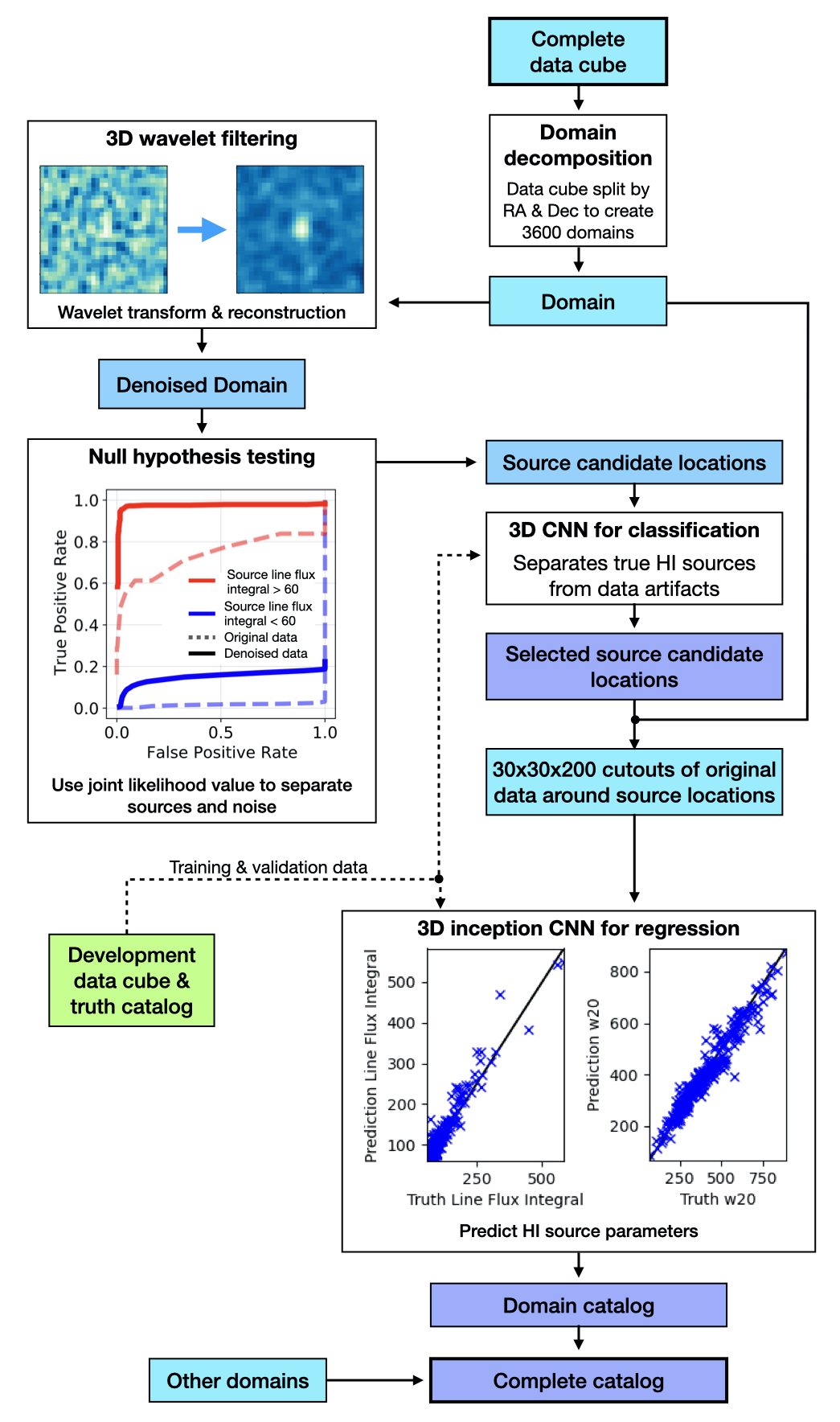}
	\caption{Data processing pipeline used by the EPFL team.}
	\label{fig_epfl}
\end{figure}

\noindent The EPFL team used a variety of techniques developed specifically for the Challenge and which have been collected into the \texttt{LiSA} library \citep{TOLLEY2022100631} publically available on github\footnote{\url{https://github.com/epfl-radio-astro/LiSA}}. The pipeline (Fig.~\ref{fig_epfl}) first decomposed the Challenge data cube into overlapping domains by dividing along RA and Dec. Each domain was then analysed by a separate node on the computing system. A pre-processing step used 3D wavelet filtering to denoise each domain: decomposition in the 2D spatial dimensions used the Isotropic Undecimated Wavelet Transform \citep{4060954}, while the decimated 9/7 wavelet transform \citep{vonesch2007generalized} was applied to the 1D frequency axis. A joint likelihood model was then calculated from the residual noise and used to identify H{\sc i} source candidates through null hypothesis testing in a sliding window along the frequency axis. Pixels with a likelihood score below a certain threshold (i.e. unlikely to be noise) were grouped into islands. The size and arrangement of these islands were used to reject data artefacts. Ultimately the location of the pixel with the highest significance was kept as an H{\sc i} source candidate location.

A classifier CNN was used to further distinguish true H{\sc i} sources from the set of candidates. The final H{\sc i} source locations were then used to extract data from the original, non-denoised domain to be passed to an Inception CNN which calculated the source parameters. The Inception CNN used multiple modules to examine data features at different scales. Finally, the H{\sc i} source locations and features for each domain were concatenated to create the full catalogue. Both CNNs were trained on the development dataset using extensive data augmentation.

\subsection{FORSKA-Sweden}
\textit{H. Håkansson, A. Sjöberg, M. C. Toribio, M. Önnheim, M. Olberg, E. Gustavsson, M. Lindqvist, M. Jirstrand, J. Conway}\newline

\noindent The FORSKA-Sweden team performed source detection using a U-Net \citep{Ronneberger2015U-Net:Segmentation} CNN with a ResNet  \citep{He2016DeepRecognition} encoder. Our methods are presented more in detail in \cite{hakansson_utilization_2023}, and all related code is published on GitHub \footnote{\url{https://github.com/FraunhoferChalmersCentre/ska-sdc-2/tree/cb3d34ebd944f3332de661cfb8fd7d3403cf9a45}}.A training set was generated from the lower 80\% of the development cube, split along the x-axis, by applying a binary mask to all pixels within range of a source defined by a cylinder using source properties (major axis, minor axis, line width) from the truth catalogue. Batches of 128 cubes of size $32\times32\times32$ pixels were sampled from the training area. Half of these cubes contained pixels assigned to a source in the target mask, which caused galaxy pixels to be over-represented in a training batch compared to the full development cube. This over-representation made training more efficient. The remaining 20\% of the development cube was used for frequent validation and tuning of model hyperparameters. 

\begin{figure}
    \centering
    \includegraphics[width=1.05\linewidth]{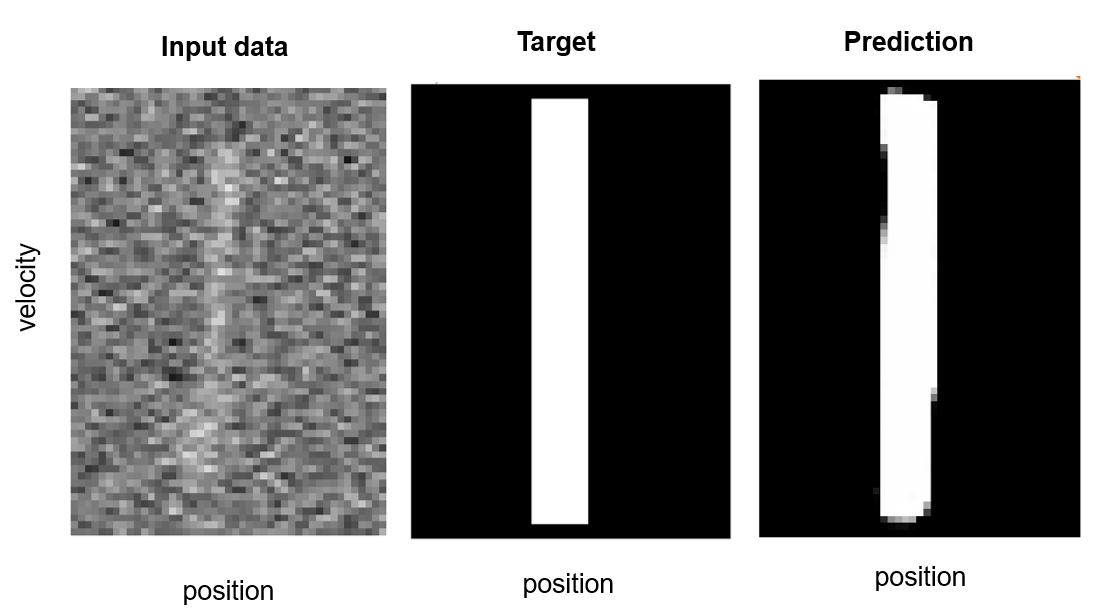}
    \caption{Cross-section images of input data, target and prediction with velocity and one positional dimension for one of the sources in the cube by team FORSKA-Sweden. The position axis is aligned with the major axis of the source.}
    \label{fig:forska}
\end{figure}

We used the soft Dice loss as the objective function \citep{Milletari2016V-Net:Segmentation,Khvedchenya_Eugene_2019_PyTorch_Toolbelt}. The initial weights of the model, pretrained from ImageNet, were provided by the {\sc PyTorch}-based {\sc Segmentation Models} package \citep{Yakubovskiy:2019}. Each 2D $k \times k$-filter of the pretrained model was converted to a 3D filter with a procedure similar to \cite{Yang2021ReinventingImages}. We aligned two dimensions to the spatial plane, and repeated the same 2D filter for $k$ frequencies, which resulted in a $k \times k \times k$ filter. The Adam optimizer \citep{Kingma2014Adam:Optimization} with an initial learning rate of $10^{-3}$ was used for training the model. The trained CNN was applied to the raw Challenge data cube to produce a binary segmentation mask assigning each pixel either to a galaxy or not (Fig.~\ref{fig:forska}). 

The {\it merging} and {\it mask dilation} modules from {\sc SoFiA 1.3.2} \citep{Serra2015SoFiA:Data} were employed to post-process the mask and extract coherent segments into a list of separated sources. The last step of the pipeline was to compute the characterisation properties for each extracted source. Some source properties were estimated in the aforementioned {\sc SoFiA} modules, while others had to be computed outside in our code. The most recent weights obtained from CNN training and a fixed set of hyperparameters from the post-processing step were used to compute a score intended to mimic the scoring of the Challenge. The best model from training was then used as a basis for hyperparameter tuning, again using the mimicked scoring.

\subsection{HI FRIENDS}
\label{hifriends}
\textit{J. Moldón, L. Darriba, L. Verdes-Montenegro, D. Kleiner, S. Sánchez, M. Parra, J. Garrido, A. Alberdi, J. M. Cannon, Michael G. Jones, G. Józsa, P. Kamphuis, I. Márquez, M. Pandey-Pommier, J. Sabater, A. Sorgho}\newline

\noindent The HI-FRIENDS team implemented a workflow \citep{moldon_javier_2021_5172930} based on a combination of {\sc SoFiA-2} \citep{2021MNRAS.506.3962W} and {\sc python} scripts to process the data cube. The workflow, which is publicly available in GitHub\footnote{\url{https://github.com/HI-FRIENDS-SDC2/hi-friends}}, is managed by the workflow engine {\sc snakemake} \citep{10.12688/f1000research.29032.2}, which orchestrates the execution of a series of steps (called rules) and parallelizes the data analysis jobs.  {\sc snakemake} also manages the installation of the software dependencies of each rule in isolated environments using conda  \cite{anaconda}. Each rule executes a single program, script, shell command or {\sc jupyter} notebook. With this methodology, each step can be developed, tested and executed independently from the others, facilitating modularisation and reproducibility of the workflow.

First, the cube is divided into smaller subcubes using the {\sc spectral-cube} library. Adjacent subcubes include an overlap of 40 pixels (112 arcsec) in order to avoid splitting large galaxies. In the second rule, source detection and characterisation is performed on each subcube using {\sc Sofia-2} \citep{2021MNRAS.506.3962W}. We optimised the {\sc Sofia-2} input parameters based on visual inspection of plots of the statistical quality of the fit and of some individual sources. In particular, we found that the parameters \texttt{scfind.threshold}, \texttt{reliability.fmin}, and \texttt{reliability.threshold} were key to optimising our solution. We found that using the spectral noise scaling in {\sc SoFiA-2} dealt well with the effects of RFI-contaminated channels and we did not include any flagging step.

The third rule converts the {\sc Sofia-2} output catalogues to new catalogues containing the relevant SDC2 source parameters in the correct physical units. We computed the inclination of the sources based on the ratio of minor to major axis of the ellipse fitted to each galaxy, including a correction factor dependent on the intrinsic axial ratio distribution from a sample of galaxies, as described in \cite{1992MNRAS.258..334S}. The next two rules produce a concatenated catalogue for the whole cube: we concatenate the individual catalogues into a main, unfiltered catalogue containing all the measured sources, and then we remove the duplicates coming from the overlapping regions between subcubes using the r.m.s. as a quality parameter to discern the best fit. Because the cube was simulated based on real sources from catalogues in the literature we further filtered the detected sources to eliminate outliers using a known correlation between derived physical properties of each galaxy. In particular, we used the correlation in Fig.~1 in \cite{2016MNRAS.460.2143W} that relates the H{\sc i} size and H{\sc i} mass of nearby galaxies. Several plots are produced during the workflow execution, and a final visualization rule generates a {\sc jupyter} notebook with a summary of the most relevant plots. 

Our workflow aims to follow FAIR principles \citep{Wilkinson2016,katz2021fresh} to be as open and reproducible as possible. To make it findable, we uploaded the code for the general workflow to Zenodo  \citep{hifriendswf} and WorkflowHub \citep{hifriendswf2}, which includes metadata and globally unique and persistent identifiers. To make the code accessible, we made derived products and containers available on Github and Zenodo as open source. To make it interoperable, our workflow can be easily deployed on different platforms with dependencies either automatically installed (e.g., in a virtual machine instance in myBinder \citep{project_jupyter-proc-scipy-2018} or executed through singularity, podman or docker containers. Finally, to make it reusable we used an open license, we included workflow documentation\footnote{\url{https://hi-friends-sdc2.readthedocs.io/en/latest/}} that contains information for developers, the workflow is modularized as {\sc snakemake} rules, we included detailed provenance of all dependencies and we followed The Linux Foundation Core Infrastructure Initiative (CII) Best Practices\footnote{\url{https://bestpractices.coreinfrastructure.org/en/projects/5138}}. Therefore, the workflow can be used to process other data cubes and should be easy to adapt to include new methodologies or adjust the parameters as needed.

\subsection{HIRAXers}
\textit{A. Vafaei Sadr, N. Oozeer  }\newline

\noindent The HIRAXers team used a multi-level deep learning approach to address the Challenge. The approach extends to 3D a method applied to a similar, 2D, challenge  \citep{vafaei2019deepsource} and uses multiple levels of supervision. Prior to source finding, a pre-processing step is used to detect regions of interest. Motivated by the recent progress in image-to-image translation techniques, one can utilize prior knowledge about source shapes to magnify signals, effectively suppressing background noise in a manner similar to image cleaning. We investigated two pre-processing approaches to reconstruct a `clean' image. For both approaches we used a training set generated by using 2D spatial slices of the development dataset to produce a source map containing masks and probability values. The output of the trained model can then be interpreted as a probability map.  

Our first preprocessing approach used 2D slices in frequency as grayscale images. The model learns to retrieve information employing only transverse information. For the second approach, we extended the inputs into 3D to benefit from longitudinal patterns by adding different frequencies as convolutional channels, thus forming a multichannel image. We used a $128\times128$ sliding window to manage memory consumption, a mean squared error loss function, and a decaying learning rate. We used the standard image processor in {\sc TensorFlow} \citep{tensorflow2015-whitepaper} for minimal data augmentation, with ranges of one degree for rotation and one percent for zoom range, in addition to horizontal and vertical flips. 

We developed our pipeline to examine the following architectures: 
V-Net \citep{milletari2016v};
Attention U-Net \citep{oktay2018attention};
R2U-Net \citep{alom2018recurrent};
U$^2$net \citep{qin2020u2};
UNet$3+$ \citep{huang2020unet};
TransUNet \citep{chen2021transunet} and 
and ResUNet-a \citep{DIAKOGIANNIS202094}. One can find most of the implementations in the {\sc keras-unet-collection} \citep{keras-unet-collection} package. The learning rate was initiated at $1\times 10^{-3}$ with a 0.95 decay per 10 epochs using the Adam optimizer. Our results using the development dataset found that the U$^2$net architecture achieved the best performance. U$^2$net employs residual U-blocks in a `U-shaped' architecture. It applies the deep-supervision technique to supervise training at all scales by downgrading the output. 

In the second step of our method we trained a model to find and characterise the objects. To find the objects, we applied a peak finder algorithm to the 3D output of U$^2$net. A peak is simply the pixel that is larger than all its 27 neighbours. The `found' catalogue was then passed into a modified 8-layer HighRes3DNet \citep{li2017compactness} as a regressor for characterisation before generating the final catalogue.

\subsection{JLRAT}
\textit{L. Yu, B. Liu , H. Xi, R. Chen, B. Peng}\newline

\noindent The JLRAT team first divided the whole dataset into small cubes of size  $320\times320\times160$ (RA, Dec, frequency) before applying to each cube a CNN containing a fully convolutional layer and a softmax layer. The CNN used 1D spectra from the cube as inputs and produced a masked output of candidate spectral signals. Using the inner product, we computed the correlation in the space domain between each candidate spectrum and known spectra from the SDC2 development cube. The result provided us with a set of 3D cubes, each containing a predicted galaxy with approximate position and size, and accurate line width. A two-dimensional Gaussian function was used to fit the moment zero map with an intensity cutoff at 1 M$_{\odot}$ pc$^{-2}$. The fit produced an ellipse with central position (RA, Dec), major axis and position angle, and the inclination of the galaxy.  The flux integral was obtained by integrating the spectra within the ellipse in both space and frequency.

\subsection{MINERVA}

\textit{D. Cornu, B. Semelin, X. Lu, S. Aicardi, P. Salomé, A. Marchal, J. Freundlich, F. Combes, C. Tasse\newline}

\noindent The MINERVA team developed two pipelines in parallel. The final catalogue merges the results from the two pipelines.

\begin{figure}
	\centering
	\includegraphics[width=1.0\linewidth]{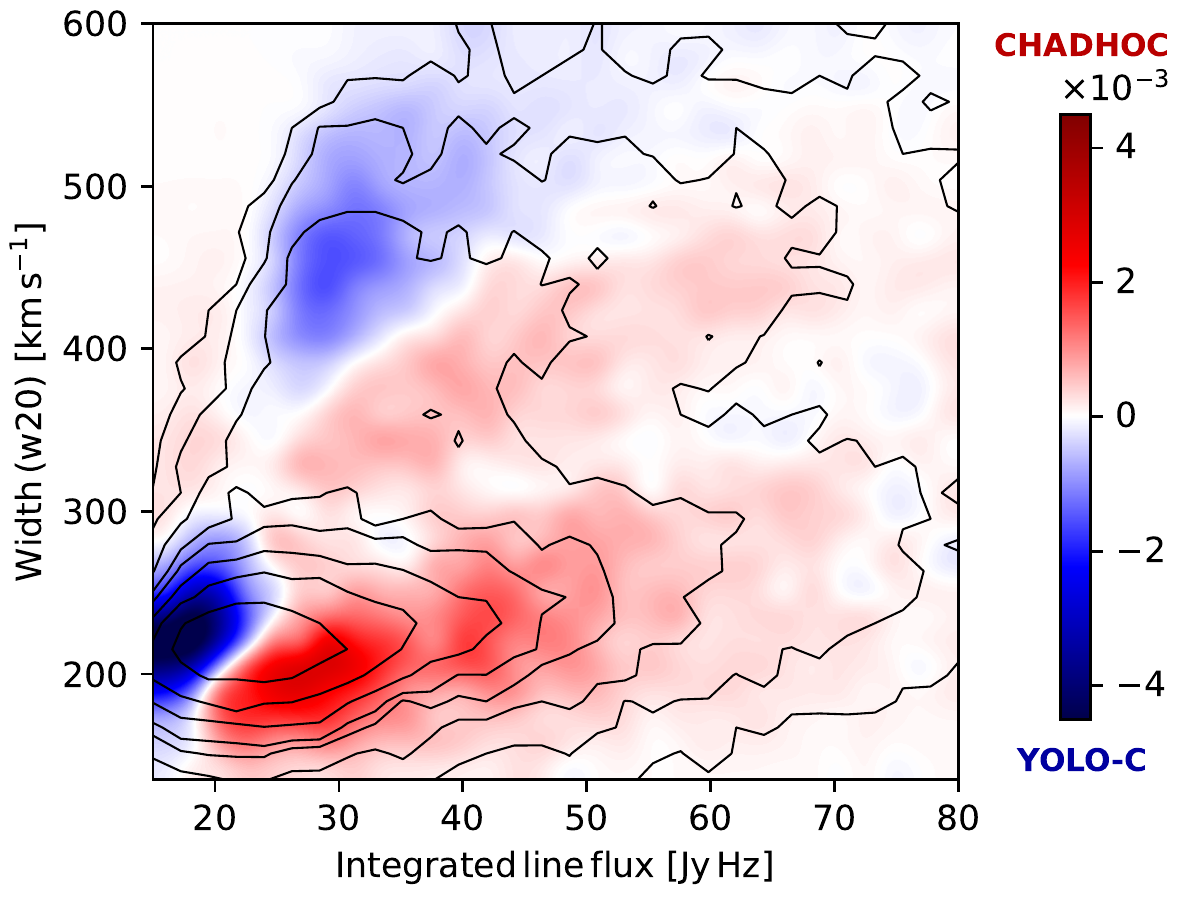}
	\caption{Team Minerva: Difference in the number of sources found between CHADHOC and YOLO-C catalogues in a flux against line width parameter space. The color encodes the difference in the local number of sources as a proportion of the total merged catalogue size (32652 predicted sources). The contours are the local number of sources averaged between the two catalogues with values: 6, 14, 30, 50, 64, 92, 128, 192. The density heatmap is computed on a 30$\times$30 grid and plotted with interpolation.}
	\label{fig_minerva}
\end{figure}

\subsubsection{YOLO-CIANNA}
\textit{}\newline

\noindent The YOLO-CIANNA pipeline implemented a highly customised version of a YOLO \citep[You Only Look Once,][]{redmon_yolo_v1, redmon_yolo_v2, redmon_yolo_v3} network, which is a regression-based object detector and classifier with a CNN architecture. Our YOLO implementation is part of our general-purpose CNN framework, CIANNA\footnote{\url{https://github.com/Deyht/CIANNA}}  (Convolutional Interactive Artificial Neural Networks by/for Astrophysicists).

The definition of the training sample was of major importance to get good results. Most of the sources in the large development dataset are impossible for the network to detect, and tagging them as positive detections would lead to a poorly trained model. For YOLO we used a combination of criteria to define a training set: i) the CHADHOC classical detection algorithm (see Section~\ref{chadhoc}); ii) a volume brightness threshold; iii) a local signal-to-noise ratio estimation. Our refined training set contains around $\sim$1500 `true' objects, with 10\% set aside for validation. All inputs were augmented using position and frequency offsets and flips. Our retained network architecture for this challenge operates on sub-volumes of $48 \times 48 \times 192$ (RA, Dec, Frequency) pixels. The network was trained by selecting either a sub-volume that contains at least one true source or a random empty field, in order to learn to exclude all types of noise aggregation and artefacts. 

The network maps each sub-volume to a $6\times6\times12$ grid, where each element corresponds to a region of $8\times8\times16$ pixels inside the input sub-volume. We chose to have the network predict a single possible detection box per grid element, producing the following parameters: $x$, $y$, $z$ the bounding-box central position in the grid element; $w$, $h$, $d$ the bounding-box dimension. We modified the YOLO loss function to allow us to predict the required H{\sc i} flux, size, line width, position angle and inclination in a single network forward for each possible box. The retained network architecture is made of 21 3D-convolutional layers, which alternate several `large' filters (usually $3\times3\times5$) that extract morphological properties and fewer `smaller' filters (usually $1\times1\times3$) that force a higher degree feature space while preserving a manageable number of weights to optimise. Some of the layers include a higher stride value in order to progressively reduce the dimensions down to the $6\times6\times12$ grid. The last few layers include dropout for regularisation and error estimation. In total the network has of the order of $2.3\times10^6$ parameters. When applying on the full datacube, predicted boxes are filtered using an ``objectness'' score threshold to maximize the SDC2 metric.  

Despite the fact that YOLO networks are known for their computational performance, our retained architecture still requires up to 36 hours of training on a single RTX 3090 GPU using FP16/FP32 Tensor Core mixed precision training. The trained network has an inference speed of 76 sub-volumes per second using a V100 GPU on Jean-Zay/IDRIS, but due to necessary partial overlap and RAM limitations, it still requires up to 20 GPU hours to process the full $\sim$1\,TB data cube.

\subsubsection{CHADHOC}\label{chadhoc}
\textit{}\newline
The Convolutional Hybrid Ad-Hoc pipeline (CHADHOC) has been developed specifically for SDC2. It is composed of three steps: a traditional detection algorithm, a CNN for identifying true sources among the detections, and a set of CNNs for source parameter estimation. 

For detection, we first smooth the signal cube by a 600 kHz width along the frequency dimension and convert to a signal-to-noise ratio on a per channel basis. Pixels below a fixed SNR of $\sim$2.2 are filtered out, and the remaining pixels are aggregated into detected sources using a simple friend-of-friend linking process with a linking length of 2 pixels. The position of each detection is computed by averaging the positions of the aggregated pixels. A catalogue of detections is then produced, ordered according to the summed source SNR values. When applied to the full Challenge dataset, we divide the cube into 25 chunks and produce one catalogue for each chunk.

The selection step is performed with a CNN. A training sample is built by cross-matching with the truth catalogue the $10^5$ brightest detections in the development cube, thus assigning a True/False label to each detection. Unsmoothed signal-to-noise cutouts of $38 \times 38 \times 100$ pixels around the position of each detection are the inputs for the network. The learning set is augmented by flipping in all three dimensions, and one third of the detections are set aside as a test set. The comparatively light network is made of 5 3D convolutional layers, containing 8, 16, 32, 32 and 8 filters, and 3 dense layers, containing 96, 32 and 2 neurons. Batch normalisation, dropouts and pooling layers are inserted between almost every convolutional and dense layer. In total the network has of the order of $10^5$ parameters. The training is performed on a single Tesla V100 GPU in at most a few hours, reaching best performances after a few tens of epochs. The model produces a number between 0 (False) and 1 (True) for each detection. The threshold where the source is labelled as True is a parameter that must be tuned to maximise the metric defined by the SDC2. This optimisation is performed independently of the training.

A distinct CNN has been developed to predict each of the source properties and includes a correction to the source position computed during the detection step. The architecture is similar to the one of the selection CNN, with small variations: for example, no dropout is used between convolutional layers for predicting the line flux. Cutouts around the $\sim$1300 brightest sources in the truth catalogue of the development cube are augmented by flipping and used to build the learning and tests sets. The networks are trained for at most a few hundred epochs in a few to 20 minutes each on a Tesla V100 GPU. Training for longer results in overfitting and a drop in accuracy. 
    
Many details impact the final performance of the pipeline. Among them, the centering of the sources in the cutouts. Translational invariance is not trained into the networks. This is dictated by the nature of the detection process and is possibly the main limitation of the pipeline: the selection CNN will never be asked about sources that have not been detected by the traditional algorithm.

\subsubsection{ Merging the catalogues}
If we visualize the catalogues produced by YOLO and CHADHOC in the sources parameter space (Fig.~\ref{fig_minerva}), we find that they occupy slightly different regions. For example, CHADHOC tends to find a (slightly) larger number of typical sources compared to YOLO, but misses more low-brightness sources because of the hard SNR threshold applied during the detection step. Thus, merging the catalogues yields a better catalogue.

Since both pipelines provide a confidence level for each source to be true, we can adjust the thresholds after cross-matching the two catalogues. In case of a cross-match we lower the required confidence level while when no cross-match is found we increase the required threshold. The different thresholds must be tuned to maximise purity and completeness. Finally, the errors on the source properties are at least partially uncorrelated between the two pipelines. Thus averaging the predicted values also improves the resulting catalogue properties.

\subsection{NAOC-Tianlai}
\textit{K. Yu, Q. Guo, W. Pei, Y. Liu, Y. Wang, X. Chen, X. Zhang, S. Ni, J. Zhang, L. Gao, M. Zhao, L. Zhang, H. Zhang, X. Wang, J. Ding, S. Zuo, Y. Mao}\newline

\noindent After testing several methods, the NAOC-Tianlai team used the {\sc SoFiA-2} software to process of the SDC2 datasets. We optimised the  {\sc SoFiA-2} input parameters by first performing a grid search in parameter space before refining the result using an MCMC simulation. We are currently developing a dedicated cosmological simulation on which to test our methods. However, during the Challenge time frame we mainly used the development and large development datasets to perform the optimisation. The optimised parameters were then used for the processing of the full Challenge dataset.

Due to the memory constraints and the consideration of avoiding excessive division along the spectral axis, the dasets were split into subcubes of size  $\sim$$330\times330\times3340$ pixels for processing. Adjacent subcubes had an overlap of 10 or 20 pixels along each axis to ensure that H{\sc i} galaxies on the border region were not missed. The full Challenge dataset was therefore divided into $18\times18\times2$ subcubes when processing.

Our main parameter selection procedure is as follows:
\begin{enumerate}
\item\label{step1} We set a list of values to be searched for each parameter of interest, such as: \texttt{replacement}, \texttt{threshold} in the {\it scfind} module; \texttt{minSizeZ}, \texttt{radiusZ} in the {\it linker} module; and \texttt{minSNR}, \texttt{threshold}, \texttt{scaleKernel} in the {\it reliability} module. We then processed in parallel the development dataset with the different combinations of parameters values.

\item Next, we selected the optimal parameter combination by comparing the output catalogues from the previous step with the  development dataset truth catalogue. To choose the optimal parameters, thresholds were applied to  the {\it total detection number}, to the {\it match rate} (true detection/total detection), and to the final {\it score}.

\item To make the found optimal parameter combination more robust, different subcubes were processed following the procedure given above, and the combination that performed well on all subcubes was selected. 

\end{enumerate}

For reference, our trial produced the following optimised parameter settings: {\tt scaleNoise.windowXY/Z = 55} for normalising the noise across the whole datacube; {\tt kernelsXY = [0, 3, 7]}, {\tt kernelsZ = [0, 3, 7, 15, 21, 45]}, {\tt threshold = 4.0}, {\tt replacement = 1.0} in the {\it scfind} module for the S+C finder in {\sc SoFiA-2}; {\tt radiusXY/Z = 2}, {\tt minSizeXY = 5}, {\tt minSizeZ = 20} in the {\it linker} module for merging the masked pixels detected by the finder; and {\tt threshold = 0.5}, {\tt scaleKernel = 0.3}, {\tt minSNR = 2.0} in the {\it reliability} module for reliability calculation and filtering. In our processing, each parameter combination  instance took $\sim$5 minutes with one CPU thread to process one subcube.

Finally, we applied the optimal parameter combination to the processing of all subcubes from the Challenge dataset, and merged the results.

\subsection{SHAO}
\textit{S. Jaiswal, B. Lao, J. N. H. S. Aditya, Y. Zhang, A. Wang,  X. Yang}\newline

\noindent The SHAO team developed a fully-automated pipeline in {\sc python} to work on the Challenge dataset. Our method involved the following steps: 1) We first sliced the datacube into individual frequency channel images and used SExtractor \citep{1996A&AS..117..393B} to  perform source finding on each image. We used a 2.5 sigma detection threshold (for $\sim$99\% detection confidence) and minimum detection area of 2 pixels. 2) We cross-matched the sources found in consecutive channel images using the software TOPCAT \citep{2005ASPC..347...29T} with a search radius of 1 pixel $=2.8$ arcsec. 3) For each source detected in at least two consecutive channel images we estimated the range of channels for each source, adding 1 extra channel on both sides. 4) We extracted a subcube across the channel range obtained in the previous step, using a spatial size of 12 pixels around each identified source. 5) We made a moment-0 map for each extracted source using its subcube, after first masking negative flux densities. 6) We used SExtractor on the moment-0 map of each extracted H{\sc i} source to estimate the source RA and Dec coordinates, major axis, minor axis, position angle and integrated flux. Inclination angle was estimated using the relations given by \cite{1926ApJ....64..321H} and \cite{1946MeLuS.117....3H}. 7) We constructed a global H{\sc i} profile for each source by estimating the flux densities within a box of size 6 pixels around the source position in every channel of its subcube. 8) We finally fit a single Gaussian model to estimate the central frequency of H{\sc i} emission and line width at 20\% of the peak.

The score obtained by this method is not very satisfactory. However, our investigations gave us confidence in dealing with a large H{\sc i} cube and making the pipeline for the analysis. We will try to improve our pipeline by optimising the input parameters and implementing different algorithms in the future. The use of machine learning techniques could be a good choice for such datasets.

\subsection{Spardha}

\textit{A. K. Shaw, N. N. Patra, A. Chakraborty, R. Mondal, S. Choudhuri, A. Mazumder, M. Jagannath}\newline
\begin{figure}
    \centering
    \includegraphics[trim={0cm 0.5cm 0.cm 0cm},clip,width=0.8\columnwidth]{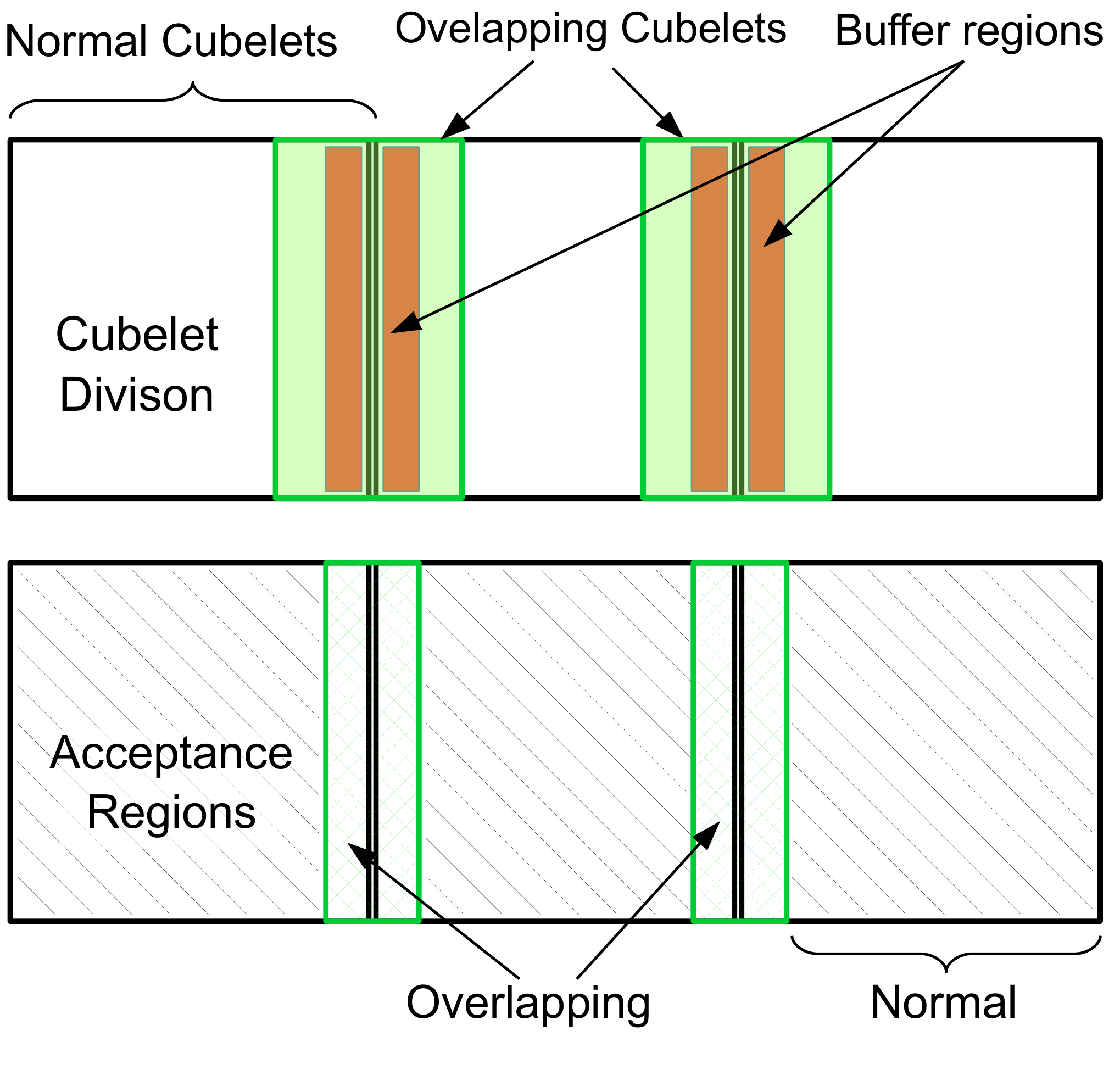}
    \caption{Team Spardha: The 2D projection along one axis of the schematic division of the data into {\it Normal} and {\it Overlapping} cubelets (top row), and the corresponding Acceptance regions (black hashing; bottom row). Normal cubelets are illustrated by black outlined boxes. Overlapping cubelets are centred at the common boundaries of Normal cubelets and are illustrated by green boxes. Orange regions (top row) represent buffer zones. }
    \label{fig:spardha}
\end{figure}

\noindent The SPARDHA team developed a {\sc python}-based pipeline which starts by dividing the 1\,TB Challenge dataset into several small cubelets. We performed source finding using an MPI-based implementation to run parallel instances of {\sc SoFiA}-$2$ on each cubelet. We tuned the parameters of {\sc SoFiA}-$2$ to maximize the number of detected sources. A total of $118$ cubelets were analysed, which were categorised into two groups, namely: 1) Normal cubelets and 2) Overlapping cubelets. The whole datacube was first divided into consecutive blocks of equal dimensions to create Normal cubelets (Fig.~\ref{fig:spardha}). Overlapping cubelets were then centred at the common boundaries of Normal cubelets in order to detect sources that fall at their common boundaries.

In order to avoid source duplication, buffer regions were defined around the faces of each cubelet (see Fig.~\ref{fig:spardha}, top row).  We always accepted any source whose centre was detected within the cubelet but not in the buffer zone (see Fig.~\ref{fig:spardha}, bottom row).  We conservatively set the width of buffer zones based on the physically motivated values of the spatial and frequency extent of typical galaxies scaled at the desired redshifts. We chose the maximum extent of the galaxy on the sky plane to be $\sim$80~kpc \citep{2016MNRAS.460.2143W}, corresponding to $\sim$10 pixels in the nearest frequency channel. The buffer region was set to be twice this extent, \textit{i.e.} $20$ pixels. Overlapping regions were therefore $4\times 20=80$ pixels wide. Along the frequency direction, galaxies can have a line-width extent of $\sim$500~km/s, which corresponds to  $\sim$72 channels. The widths of the buffer regions and Overlapping regions along the frequency axis were therefore $144$ and 288 channels, respectively. The acceptance regions of the cubelets (normal and overlapping) were such that they spanned the whole data cube contiguously when arranged accordingly. Although this approach increased the computation slightly due to analysing some regions of the data more than once, it ensured that there was no common source present in the list. Analysing cubelets was the most time consuming part in our pipeline. We analysed $118$ cubelets on $472$ cores in parallel in around $15$ minutes.

We used physical equations to convert the {\sc SoFiA}-$2$ catalogue into the SDC-prescribed units and to discard bad detections such as those sources having \textit{NaN} values in the columns or those with negative flux values. In the final stage we put limits on the line width, discarding detections with unusual values. Motivated by physical models and observations of galaxies, we conservatively accepted the sources having $w_{20}\in [60,\, 500]~{\rm km/s}$ \citep{2000ApJ...533L..99M}. We finally arranged the catalogue in descending order of the flux values. Based on tests using the development datacube, for which the exact source properties are known, we chose the top $35\%$ of total sources to generate the final catalogue for submission.

\subsection{Starmech}
\textit{M. J. Hardcastle, J. Forbrich, L. Smith, V. Stolyarov, M. Ashdown, J. Coles}\newline

\noindent The Starmech tackled the Challenge from the point of view of dealing with the Challenge dataset within the constraints of the resources provided to us (a single node with 30 cores and 124~GB RAM, 800~GB root volume and 1\,TB additional data volume). Some computational constraints will be a feature of future working in the field when computing resources are provided as part of shared SKA Regional Centres. 

We considered existing source finding tools: {\sc PyBDSF} \citep{2015ascl.soft02007M}, a continuum source finder, and {\sc SoFiA} and {\sc SoFiA-2}, two generations of a 3D source finder already optimised for H{\sc i} \citep{2021MNRAS.506.3962W}. While {\sc PyBDSF} readily generated a catalogue of the continuum sources and could be run on many slices in frequency, slicing and averaging with fixed frequency steps does not give good results since emission lines have a variety of possible widths in frequency space. Instead we focused on the two publicly available 3D source finders. Our tests showed that {\sc SoFiA-2}’s memory footprint is much lower than that of {\sc SoFiA} for a given data cube and its speed significantly higher, so it became our algorithm of choice. 

In order to work with the available RAM, we needed to slice the full Challenge datacube either in frequency or spatially. We chose to slice spatially because this allows {\sc SoFiA-2} to operate as expected in frequency space; essentially the approach is to break the sky down into smaller angular regions, run {\sc SoFiA-2} on each one in series, and then join and de-duplicate the resulting catalogue. Whether done in parallel (as in the MPI implementation {\sc SoFiA-X}; \citealt{2021MNRAS.506.3962W}), or in series as we describe here, some approach like this will always be necessary for large enough H{\sc i} series in the SKA era since the full dataset sizes will exceed any feasible RAM in a single node for the foreseeable future. 

Our implementation was a simple {\sc python} wrapper around {\sc SoFiA-2}. The code calculates the number of regions into which the input data cube needs to be divided such that each individual sub-cube can fit into the available RAM. Assuming a tiling of $n \times n$, it then tiles the cube with $n^2$ overlapping rectangular spatial regions. We define a guard region width $g$ in pixels: each region passed to {\sc SoFiA} overlaps the adjacent one, unless on an edge, by $2g$ pixels. Looping over the sub-cubes, {\sc SoFiA-2} is run on each one to produce $n^2$ overlapping catalogues in total. For our final submission we used {\sc SoFiA-2} default parameters with an \texttt{scfind.threshold} of 4.5 sigma, $g=20$ pixels, a spatial offset threshold for de-duplication of 1 pixel, and a frequency threshold of 1 MHz. $g$ was chosen to be larger than the typical size in pixels of any real source. We verified that there were no significant differences, using these parameters, between the reassembled catalogue for a smaller test cube and the catalogue directly generated by running {\sc SoFiA-2} on the same cube, using {\sc TOPCAT} for simple catalogue visualization and cross-matching. Due to time constraints, we did not move on to the next obvious step of optimising the parameters used for {\sc SoFiA-2} based on further runs on the test and development datasets.

We removed source duplication arising from overlapping regions by considering catalogues from adjacent sub-cubes pairwise.  We firstly discarded all catalogue entries with pixel position more than $g$ pixels from the edge of a sub-cube; these should already be present in another catalogue. The remaining overlap region, $2g$ pixels in width, height or both, was cross-matched in position and sources whose position and frequency differ by less than user-defined threshold values were considered duplicates and discarded from one of the two catalogues. Finally the resulting $n^2$ de-duplicated catalogues were merged and catalogue values converted according to units specified by the submission format.

We would like to have explored the utility of dimensional compression of the data as part of the source finding, for example by using moment maps in an attempt to eliminate noise and better pinpoint source detection algorithms. A priori, this would have been of rather technical interest since any resulting bias on source detection would need to be considered. However, in this way, it may have been possible to identify candidate sources to then characterise based on observable parameters such as size and linewidth, in a first step as point sources vs resolved sources, and including flags for potential overlap in projection or velocity.

\subsection{Team SoFiA}
\textit{K. M. Hess,	R. J. Jurek, S. Kitaeff, P. Serra, A. X. Shen,	J. M. van der Hulst, T. Westmeier}\newline
\begin{figure}
	\centering
	\includegraphics[trim={0cm 1cm .cm 0.cm},clip,width=0.95\columnwidth]{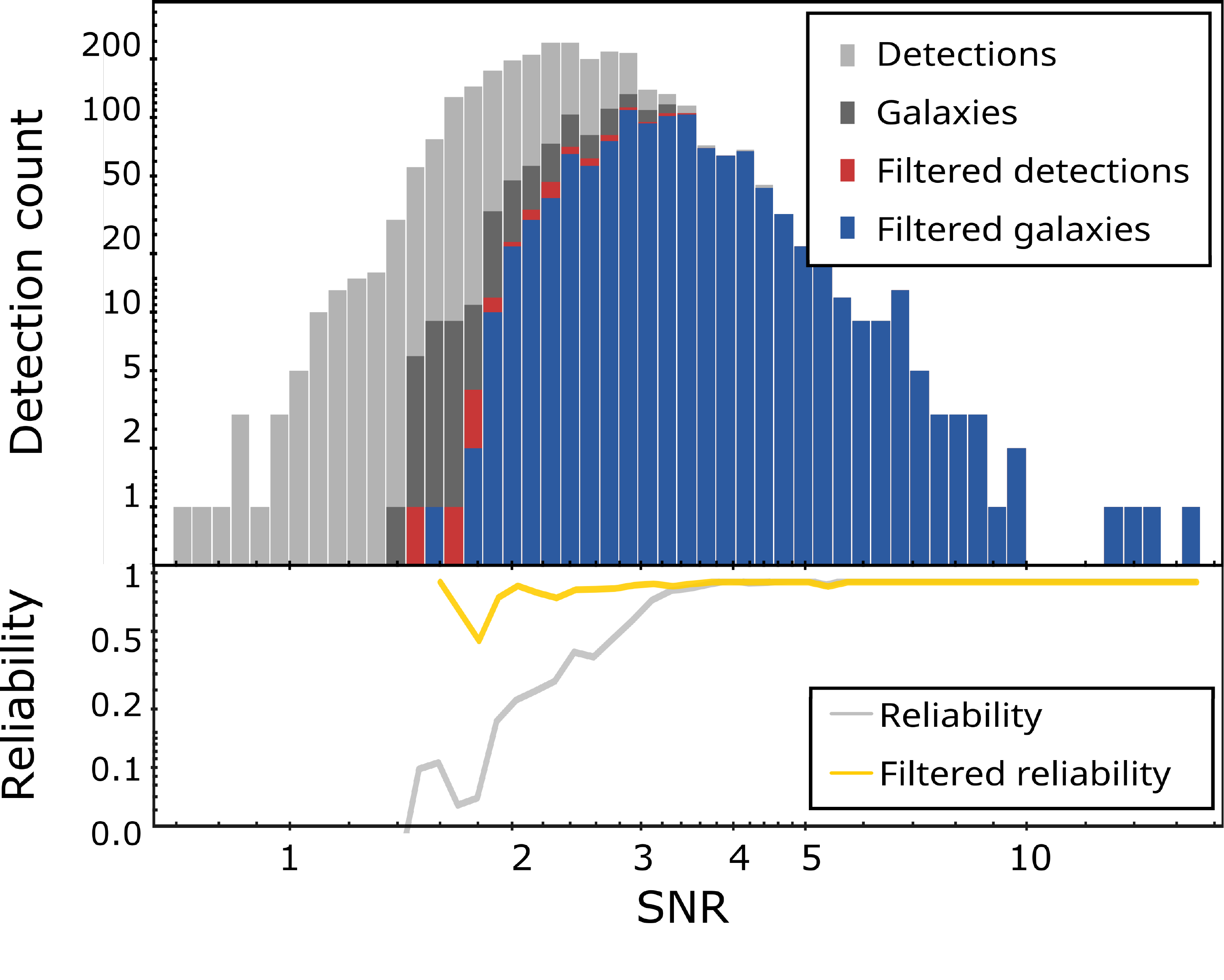}
	\caption{Team SoFiA: Histogram of total detections (light-grey), real galaxies (dark-grey), detections after filtering (red) and real galaxies after filtering (blue) as a function of integrated signal-to-noise ratio from a SoFiA run on the development cube (top panel). The reliability of the original and filtered catalogue is shown as the grey and orange curve, respectively (bottom panel). Parameter space filtering significantly boosts SoFiA's reliability at low SNR. Note that we measure SNR within the actual SoFiA source mask, and the resulting values can not be directly compared with the optimised SNR defined in Section~\ref{snr}.}
	\label{fig_sofia_reliability}
\end{figure}

\noindent Team SoFiA made use of the Source Finding Application (SoFiA; \citealt{2015MNRAS.448.1922S,2021MNRAS.506.3962W}) to tackle the  Challenge. Development version 2.3.1 of the software, dated 22 July 2021,\footnote{\url{https://github.com/SoFiA-Admin/SoFiA-2/tree/11ff5fb01a8e3061a79d47b1ec3d353c429adf33}} was used in the final run submitted to the scoring service. To minimise processing time, 80~instances of SoFiA were run in parallel, each operating on a smaller region ($\approx 11.8~\mathrm{GB}$) of the full cube. The processing time for an individual instance was just under 25~minutes, increasing to slightly more than 2~hours when all 80~instances were launched at once due to overhead from simultaneous file access. The resulting output catalogues were merged and any duplicate detections in areas of overlap between adjacent regions discarded.

We ran SoFiA with with the following options: after flagging of bright continuum sources $> 7~\mathrm{mJy}$ followed by noise normalisation in each spectral channel, the S+C finder was run with a detection threshold of $3.8$ times the noise level, spatial filter sizes of 0, 3 and 6~pixels and spectral filter sizes of 0, 3, 7, 15 and 31~channels. We adopted a linking radius of 2 and a minimum size requirement of 3~pixels/channels. Lastly, reliability filtering was enabled with a reliability threshold of 0.1, an SNR threshold of 1.5 and a kernel scale factor of 0.3.

Based on tests using the development cube, we improved the reliability of the resulting source catalogue from SoFiA by removing all detections with $n_{\rm pix} < 700$, $s < -0.00135 \times (n_{\rm pix} - 942)$ or $f > 0.18 \times \mathrm{SNR} + 0.17$, where $n_{\rm pix}$ is the number of pixels within the 3D source mask, $s$ is the skewness of the flux density values within the mask, $f$ is the filling factor of the source mask within its rectangular 3D bounding box, and $\mathrm{SNR}$ is the integrated signal-to-noise ratio of the detection. Detection counts for the original and filtered catalogue from the development cube are shown in Fig.~\ref{fig_sofia_reliability} as a function of SNR. Our final detection rate peaks at $\mathrm{SNR} \approx 3$, with a reliability of close to $1$ down to $\mathrm{SNR} \approx 2$. The filtered catalogue from the full cube contains almost $25,000$ detections, about $23,500$ of which are real, implying a global reliability of 94.2\%.

It should be emphasised that our strategy of first creating a low-reliability catalogue with SoFiA and then removing false positives through additional cuts in parameter space is based on development cube tests and was adopted to maximise our score. This strategy may not work well for real astronomical surveys which are likely to have different requirements for the balance between completeness and reliability than the one mandated by the scoring algorithm.

Lastly, the source parameters measured by SoFiA were converted to the requested physical parameters. As the calculation of disc size and inclination required spatial deconvolution of the source, we adopted a constant disc size of $8.5$ arcsec and an inclination of $57.3$ degrees for all spatially unresolved detections. In addition, statistical noise bias corrections were derived from the development cube and applied to SoFiA's raw measurement of integrated flux, line width and H{\sc i} disc size.

\section{Scoring}
\label{scoring_section}

\begin{table*}  
	\centering
	\begin{tabular}{llllll} 
        	\hline\noalign{\smallskip}
    	  Team name & Pre-processing & Detection & False-positive rejection & Characterisation & Additional notes \\
    	 	 \hline\noalign{\smallskip}        
        	Coin & RFI flagging & 3D U-Net CNN &  Size cuts &ResNet CNNs & Several CNNs tested \\
         &&Interscale connectivity &Continuum rejection&Ellipse-fitting&\\  
         	EPFL & Wavelet filtering &Joint likelihood & Size cut& Inception CNN & Data augmentation\\
          && &Classifier CNN&\\         
          	FORSKA-Sweden & - & 3D U-Net CNN &SoFiA& SoFiA& -  \\
           &&&&Modelling: check&\\            
           	HI-FRIENDS & SoFiA:& SoFiA & SoFiA &SoFiA &-  \\
                     &\;\;Continuum flagging&&Additional parameter cuts&Ellipse fitting&\\    
            &\;\;Noise normalisation&&  &&\\
           	HIRAXers  & U$^2$~net &  Peak-finding&-& HighRes3DNet&Data augmentation \\            
           	JLRAT & - & CNN & -  &Gaussian-fitting&Spectral inputs to CNN\\
            &&Cross-correlation &&&\\       
			  MINERVA$^*$  &  - | Smoothing  &  YOLO CNN | Friend-of-friend & - | CNN & YOLO CNN | CNNs & Training data refinement\\
     &| SNR mask&&&&Data augmentation\\     
			  NAOC-Tianlai & SoFiA: & SoFiA & Parameter tuning &SoFiA& Gridsearch, MCMC \\	  
                 &\;\;Continuum flagging&&  &&\\
            &\;\;Noise normalisation&&  &&\\
     		SHAO  &  -& SExtractor & -&SExtractor&-\\       
            &&TOPCAT&&Gaussian fitting&\\           
			  Spardha &  SoFiA:& SoFiA &  SoFiA & SoFiA & Partioning buffer zones \\ 
                 &\;\;Continuum flagging&&  Additional parameter cuts &&\\
            &\;\;Noise normalisation&&  &&\\
			  Starmech &  SoFiA: & SoFiA & SoFiA& SoFiA&TOPCAT for verification\\    
                 &\;\;Continuum flagging&&  &&\\
            &\;\;Noise normalisation&&  &&\\
     		Team SoFiA &SoFiA: & SoFiA & SoFiA & SoFiA & Noise bias corrections \\ 
            &\;\;Continuum flagging&& Additional parameter cuts &&\\
            &\;\;Noise normalisation&&  &&\\

		  	\hline \noalign{\smallskip}
	\end{tabular}
        
    	\caption{The main features of the methods applied by each team to SDC2 are summarised for ease of reference. The methodology is divided into pre-processing, source finding, false-positive rejection and source characterisation steps. The asterisk denotes the step taken by team MINERVA to combine the results of two independent methods, demarcated here by the pipe symbol, to form a final catalogue. }
        \label{SDC2methods}
\end{table*}

A live scoring service was provided for the duration of the Challenge. The service allowed teams to self-score catalogue submissions while keeping the truth catalogue hidden, and automatically updated a live leaderboard each time a team achieved an improved score. All participating teams were provided with credentials with which the scoring service could be accessed over the internet using a simple, pip-installable command line client. Participants used this client to upload submissions to the service, after which it was evaluated by a scoring algorithm against the truth catalogue. Once the score had been calculated, it could be retrieved from the scoring service using the client. Teams were limited to a maximum submission rate of 30 submissions per 24 hour period.

\subsection{Scoring procedure}
\label{scoring}

The scoring algorithm \footnote{\url{https://pypi.org/project/ska-sdc/}} is written in {\sc python} and makes use of the {\sc pandas} and {\sc astropy} libraries. Scoring is performed by comparing submitted catalogues with a truth catalogue, each containing the same source properties. The first step of the scoring is to perform a positional cross-match between the true and the submitted catalogues. Matched sources from the submitted catalogue are then assigned scores according to the combined accuracy of all their measured properties. Finally, the scores of all matched sources are summed and the number of false detections subtracted, to give the overall Challenge score.

\subsubsection{Source cross-match}
\label{crossmatch}

Cross-matching is performed using the {\sc scikit} nearest neighbours classifier with the {\tt kd\_tree} algorithm, which uses a tree-based data structure for computational efficiency \citep{10.1145/361002.361007}.  The cross-match procedure considers the position of a source in the 3D cube, identified by RA, Dec and central frequency. Each coordinate set is first converted to a physical position space via the source angular diameter distance. All submitted sources with positions within which a truth catalogue source is in range are then recorded as matches. For each submitted source, this range in the spatial and frequency dimensions is determined by the beam-convolved submitted H{\sc i} size and the line width, respectively. Detections that do not have a truth source within this range are recorded as false positives. Matched detections are further filtered by considering the range of the matched truth sources. Detections which lie outside the beam-convolved H{\sc i} size and the line width of the matched truth source are at this stage also rejected and recorded as false positives.

It is possible that the cross-match returns multiple submitted sources per true source. In that case, all matches are retained and scored individually. The reasoning behind this choice is that components of H{\sc i} sources, especially in the velocity field,  could be correctly identified but interpreted as separate sources. If that were the case, classifying them as false positives would be too much of a penalty. All submitted sources matched to the same true source are inversely weighted  by the number of  multiple matches during the scoring step. It is also possible for more than one truth source to be matched with a single submitted source. In these cases, only the match between the submitted source and truth source which yields the lowest multi-parameter error (eq.~\ref{multip}) is retained. This procedure ensures that matches in crowded regions take into account the resemblance of a truth source to a submitted  source, in addition to its position. 

A final step is performed to compare the multi-dimensional error with a threshold value, above which any nominally matched submitted sources are discarded and counted as false positives. The multi-parameter error $D$ is calculated using the Euclidean distance between truth and submitted sources in normalised parameter space:

\begin{equation}
D = (D_{\rm pos}^2+D_{\rm freq}^2+D_{\rm HI\: size}^2+D_{\rm line\: width}^2+D_{\rm flux}^2)^\frac{1}{2},
\label{multip}
\end{equation}

\noindent where the errors on parameters of spatial position, central frequency, line width and integrated line flux have been normalised following the definitions in Table~\ref{errors}. The error on H{\sc i} size is at this stage normalised by the beam-convolved true H{\sc i} size in order not to lead to the preferential rejection of unresolved sources.  The multi-dimensional error threshold is set at 5, i.e. the sum in quadrature of unit normalised error values.

\subsubsection{Accuracy of sources properties}
\label{accuracy}

For all detections that have been identified as a match, properties are compared with the truth catalogue and a score is assigned per property and per source. The following properties are considered for accuracy: sky position (RA, Dec), H{\sc i} size, integrated line flux, central frequency, position angle, inclination angle and line width. Each attribute $j$ of a submitted source $i$ contributes a maximum weighted score $w_i^j$ of $1/7$, so that the maximum weighted score $w_i$ for a single matched source is 1:

\begin{equation}
w_i =\sum_{j=1}^{7}w_i^j .    
\end{equation}

\noindent The weighted score of each property of a source is determined by 

\begin{equation}
    w_i^j=\frac{1}{7}\min \left\{1, \frac{{\rm thr}_j}{{\rm err}_i^j}  \right\} ,
\end{equation}

\noindent where ${\rm err}_i^j$ is the error on the attribute and ${\rm thr}_j$ is a threshold applied to that attribute for all sources. Errors calculated in this step are detailed in Table~\ref{errors}, along with corresponding threshold values, which have been chosen using the distribution of errors obtained during tests on the Challenge data products using the SoFiA source finder.  Finally, the weighted scores of submitted sources are averaged over any duplicate matches with unique truth sources.

\begin{table}
	\centering
	
	\begin{tabular}{lcr} 

        	\hline\noalign{\smallskip}
    	  Property &Error term&Threshold\\
    	 	 \hline\noalign{\smallskip}
			   RA and Dec, $x, y$ \;\;\;\;\;&$\displaystyle D_{\rm pos}=\frac{(x-x')^2+(y-y')^2}{\hat{S'}}$&0.3{\smallskip}\\
			   H{\sc i} size, $S$ &$\displaystyle  D_{\rm HI \;size}=\frac{|S-S'|}{\hat{S'}}$&0.3{\smallskip}\\
			   Integrated line flux, $F$ &$\displaystyle  D_{\rm flux}=\frac{|F-F'|}{F'}$&0.1{\smallskip}\\
			   Central frequency, $\nu$ &$\displaystyle  D_{\rm freq}=\frac{|\nu-\nu'|}{w_{20,\,\rm Hz}'}$&0.3{\smallskip}{\smallskip}\\
			   Position angle, $\theta$ &$\displaystyle D_{\rm PA} = |\theta -\theta'|$&10{\smallskip}{\smallskip}{\smallskip}\\			    
			   Inclination angle, $i$ &$\displaystyle D_{\rm incl} = |i -i'|$&10{\smallskip}\\
			   Line width, $w_{20}$ &$\displaystyle D_{\rm line \;width}=\frac{|w_{20}-w_{20}'|}{w_{20}'}$&0.3{\smallskip}\\			    

		  	\hline \noalign{\smallskip}
	\end{tabular}
        
    	\caption{Definitions of errors and threshold values for the properties of sources. Prime denotes the attributes of the truth catalogue,  $x$, $y$ are the pixel coordinates corresponding to RA, Dec,  $\nu$ is the central frequency,  $S$ is the H{\sc i} major axis diameter and $\hat{S}$ is the beam-convolved major axis diameter,  $f$ is the source integrated line flux, $\theta$ is the position angle, $i$ is the inclination angle, and  $w_{20}$ is the H{\sc i} line width. Calculations of position angles take into account potential angle degeneracies by defining the angle difference as a point on the unit circle and taking the two-argument arctangent of the coordinates of that point: $|\theta-\theta'|={\rm atan}2[\sin(\theta-\theta'), \cos(\theta-\theta')]$}
        \label{errors}
\end{table}

\subsubsection{Final score per submission}
The final score is determined by subtracting the number of false positives $N_{\rm f}$ from the summed weighted scores $w_i$ of all  $N_{\rm m}$ unique matched sources:
\begin{equation}
    {\rm final\: score} = \sum^{N_{\rm m}}_i w_i-N_{\rm f}.
\end{equation}

\noindent False positives are linearly penalised in order to preserve equal weighting between characterisation performance and the ability to remove false detections.


\subsection{Reproducibility awards}
\label{repro}
Participating teams were encouraged to consider early on in the Challenge the overall architecture and design of their software pipelines. At the Challenge close,  teams were invited to share pipeline solutions. Reproducibility awards were then granted in acknowledgement of those teams whose pipelines demonstrated best practice in the provision of reproducible results and reusable methods. Pipelines were evaluated using a checklist developed in partnership with the Software Sustainability Institute (SSI)\footnote{ \url{https://www.software.ac.uk/} } \citep{crouch2013software}, which was provided to teams for the purposes of self-assessment during the Challenge. The checklist\footnote{ \url{https://sdc2.astronomers.skatelescope.org/sdc2-challenge/reproducibility-awards}} considered the following criteria:

\begin{description}
\item {\it Reproducibility of the solution.} Can the software pipeline be re-run easily to produce the same results? Is it:
\begin{enumerate}
\item Well-documented 
\item Easy to install 
\item Easy to use 
\end{enumerate}

\item {\it Reusability of the pipeline.} Can the code be reused easily by other people to develop new projects? Does it: 
\begin{enumerate}
\item Have an open licence
\item Have easily accessible source code  
\item Adhere to coding standards
\item Utilise tests

\end{enumerate}
\end{description}

\noindent All parts of the software pipeline developed by each team were evaluated, including packages that the teams have written and code that interacts with third party packages, but not including any third party packages themselves.


\section{Results and analysis}

In this Section we first present the overall Challenge results before reporting on  the determination of source signal-to-noise values. We then analyse the results from source finding and characterisation perspectives and present the results of the reproducibility awards.

\subsection{Challenge results}
\label{results}

The final scores of all teams who submitted a catalogue based on the full Challenge dataset are reported in Table~\ref{SDC2results}. Each team's number of detections, $N_{\rm d}$ -- composed of matches, $N_{\rm m}$, and false positives, $N_{\rm f}$ -- are also listed, along with the number of matches, the overall reliability, $R$, and completeness, $C$, calculated as follows:

\begin{equation}
    R = \frac{N_{\rm m}}{N_{\rm d}} = \frac{N_{\rm m}}{N_{\rm m}+ N_{\rm f}};
    \label{rely}
\end{equation}

\begin{equation}
    C = \frac{N_{\rm m}}{N_{\rm t}},
    \label{complete}
\end{equation}

\noindent where $N_{\rm t}$ is the number of sources in the truth catalogue. The overall characterisation accuracy of each team's method, $A$, is defined as the accuracy of source property measurement according to Section~\ref{accuracy}, averaged over all properties for all matches per team:

\begin{equation}
   A= \frac{\sum^{N_{\rm m}}_i w_i }{ N_{\rm m} }.
    \label{accuracyeq}
\end{equation}

\noindent We note that the scoring algorithm (Section~\ref{scoring_section}), designed to penalise false detections, can result in a teams' highest scoring submission containing a significantly less complete catalogue than other submissions made by the same team if the number of false positives is high. This is the case for teams Coin, HIRAXers and SHAO. With each teams' agreement, therefore, we have used the team's submission with the highest completeness for the following analysis, while leaving the leaderboard scores unchanged. This allows us more robustly to investigate the characterisation performance of these teams' methods.

\subsubsection{Conventions and units}
\label{conventions}

Several conventions and conversions are used during the characterisation of H{\sc i} spectral line data which, without clear and unambiguous specification, can lead to inconsistencies between catalogues and between physical and measured properties.  Room for error arose due to potential alternative position angle definitions and to the need to shift the rest frequency into the frame of the source.  Where teams' catalogues have followed alternative conventions or incorrect conversions, catalogue corrections have been applied after the close of the Challenge leaderboard. While teams' scores are affected slightly, leaderboard positions do not change.  The Challenge organising team used the dedicated discussion forum (Section ~\ref{thechallenge}) to resolve misunderstandings in the rules and conventions as they arose. Future SKAO Science Data Challenges will benefit from additional instructions and examples where ambiguity or unfamiliarity can be anticipated. The reporting of observed rather than derived parameters would also reduce measurement inconsistencies.

\begin{table}
	\centering
	
	\begin{tabular}{lcccccc} 

        	\hline\noalign{\smallskip}
    	  Team name & Score & $N_{\rm d}$ & $N_{\rm m}$ & $R$&$C$& $A$\\
    	 	 \hline\noalign{\smallskip}
			  MINERVA & 23254 & 32652 & 30841  &0.945 & 0.132 &0.81\\
			  FORSKA-Sweden & 22489 & 33294 & 31507& 0.946& 0.135 &0.77 \\
			  Team SoFiA & 16822 & 24923 & 23486&0.942& 0.101 &0.78 \\ 
			  NAOC-Tianlai & 14416 & 29151 &26020 &0.893&0.112 & 0.67\\	
			  HI-FRIENDS & 13903 & 21903 & 20828&0.951& 0.089 & 0.72\\
			  EPFL & 8515 &19116 & 16742 & 0.876 &0.072&0.65 \\
			  Spardha & 5615 & 18000 & 13513 &0.751&0.058&0.75 \\
			  Starmech &  2096 & 27799 & 17560&0.632&0.075& 0.70\\
			  JLRAT & 1080 & 2100 &  1918 &0.913&0.008& 0.66\\
			  Coin & -2 & 29  & 17   &0.586&0.000& 0.60\\
			  HIRAXers  & -2 & 2 & 0&0.000&0.000& - \\
			  SHAO  & -471 & 471 & 0&0.000&0.000& - \\

		  	\hline \noalign{\smallskip}
	\end{tabular}
        
    	\caption{SDC2 finalist teams' scores are reported, rounded to the nearest integer. Also reported are the number of detections $N_{\rm d}$ and matches $N_{\rm m}$ (Section~\ref{crossmatch}), and the overall reliability ($R$; eq.~\ref{rely}) and completeness ($C$; eq.~\ref{complete}) of each method. Finally, the source characterisation accuracy ($A$; equation~\ref{accuracyeq}) reports the percentage accuracy of source property measurement averaged over all properties for all sources matched per team.}
        \label{SDC2results}
\end{table}

\subsection{Signal-to-noise}
\label{snr}

The appropriate definition and calculation of source signal-to-noise values is important in order to gain an understanding of the absolute performance of teams' methods and to transfer insights gained from SDC2 to other datasets. While the value of peak signal-to-noise is easy to define, it fails to capture any information about source extent. Alternatively, the integrated signal-to-noise can be evaluated for a chosen mask across the source. The total error contribution from the mask pixels can be calculated using the usual rules of correlated error propagation.  However, due to the smoothing effect of beam sampling, the amount of true signal contained within a finite mask cannot be determined. Further, the application of smoothing kernels -- routinely used in signal processing problems to boost signal with respect to noise -- results in modification to the signal-to-noise properties of a given source. For the purpose of this analysis, therefore, we use a signal-to-noise definition based on the peak signal of a smoothed source. The definition adopted for this paper is intended to provide the most helpful insight into SDC2 results, but is not necessarily the best choice for other datasets.

\begin{figure}
	\centering
	\includegraphics[trim={0cm 0cm 1.5cm 1cm},clip,width=1\columnwidth]{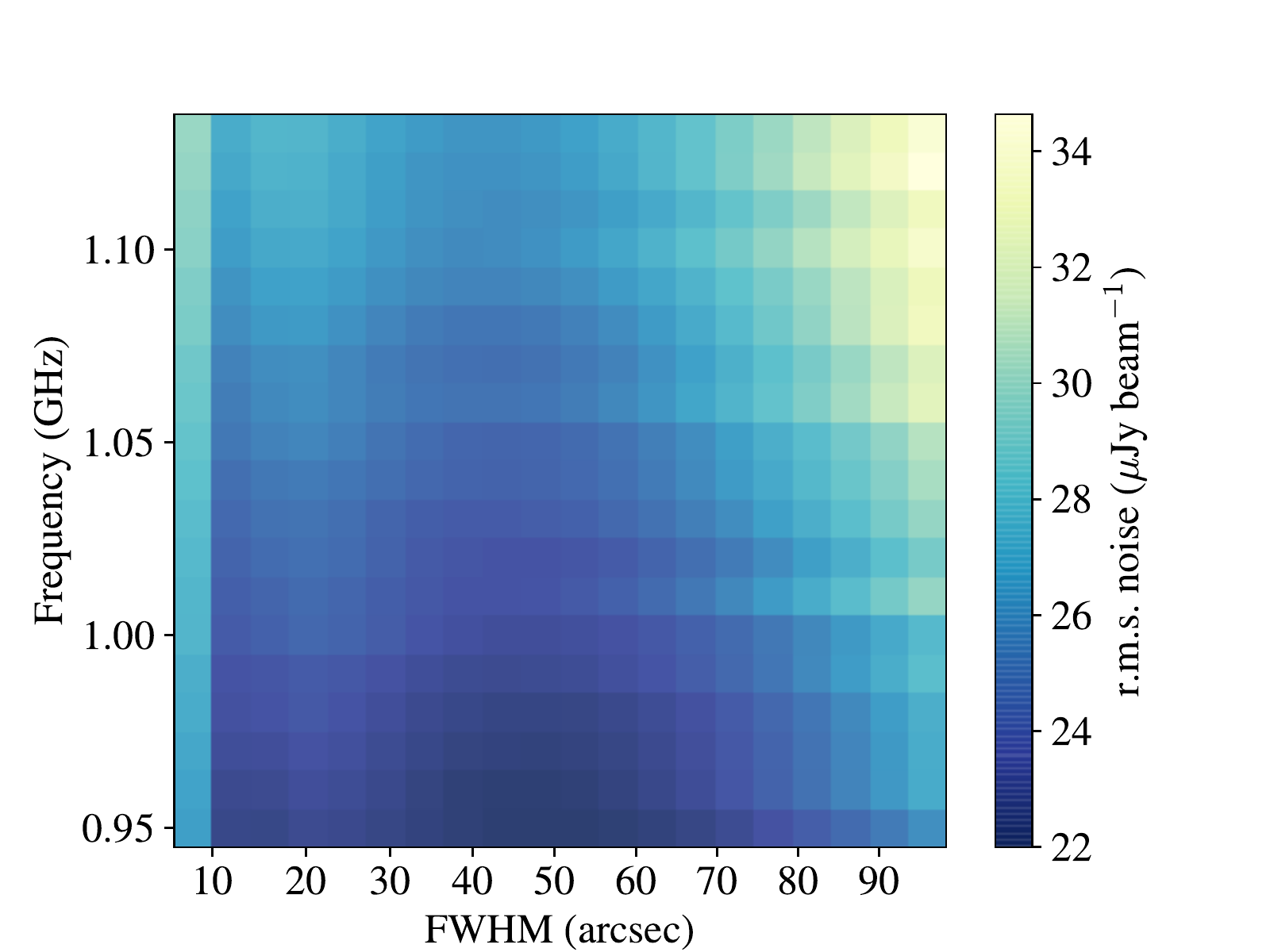}
	\caption{The r.m.s noise of the 2000h SKA-MID 20 square degree noise field is plotted as a function of frequency and of smoothing. The smoothing FWHM presented is the result of adding in quadrature the 7 arcsec beam FWHM and the FWHM of a Gaussian smoothing filter applied to the field.   }
	\label{rms}
\end{figure}

\begin{figure}
	\centering
	\includegraphics[trim={0.2cm 0cm 0.5cm 0cm},clip,width=0.9\columnwidth]{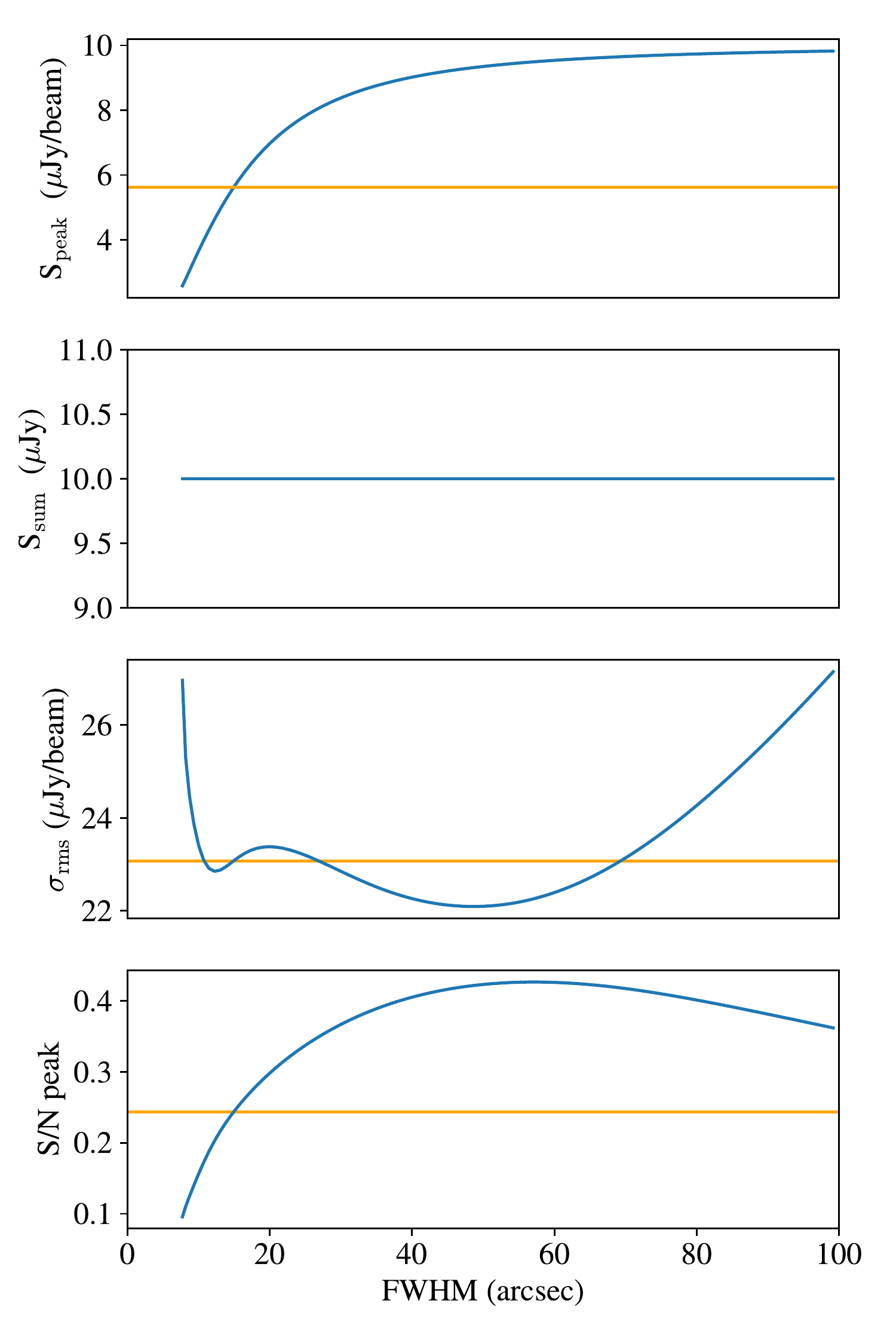}
	\caption{A simulated circular Gaussian source of FWHM 14 arcsec is convolved with a circular Gaussian `beam' of 7 arcsec FWHM and used to illustrate signal-to-noise characteristics of the SKA-MID field as a function of smoothing. A series of Gaussian smoothing filters is applied both to the beam-convolved source and to a simulated noise field representing 2000h of Band 2 SKA-MID observations of a 20 square degree field. The beam FWHM and smoothing FWHM are added in quadrature to obtain the total smoothing FWHM, which is represented by the abscissa. From top, in blue: peak smoothed source flux density; total source flux density; r.m.s. noise of the noise field; peak SNR obtained using the peak smoothed source flux density and the r.m.s noise. The orange horizontal line represents the values obtained by applying instead a filter matched to the source.}
	\label{snr_plots}
\end{figure}

A given signal in the presence of additive white Gaussian noise can be maximised with respect to the noise by applying a smoothing filter matched to the signal. In this case, the matched filter optimises the trade-off between noise-suppression and signal-suppression. In the case of an SKA-observed spatial noise field, logarithmic spacing of the array configurations results in a relatively uniform sensitivity, in units of Jy beam$^{-1}$, across a wide range of angular scales \citep{2019arXiv191212699B}. This property is evident upon Gaussian smoothing of the SDC2 simulated spatial noise field, which sees a slight reduction in beam-normalised r.m.s. noise to an approximately constant level between angular ranges $\sim$10--80 arcsec FWHM (see Fig.~\ref{rms}, which presents r.m.s. noise as a function of total spatial smoothing and frequency for a simulated 2000h SKA-MID observation of a 20 square degree field). The signal-to-noise of a source observed using the SKA can therefore be maximised in the spatial dimensions simply by applying a sufficiently large Gaussian smoothing kernel, provided that the source itself is no larger in spatial extent than the angular range of uniform sensitivity. Fig.~\ref{snr_plots} presents the effect on signal-to-noise of smoothing an SKA-observed Gaussian source using a range of Gaussian smoothing kernels.

For each SDC2 source, a signal-to-noise ratio (SNR) value was obtained by first selecting the minimum r.m.s noise value, $\sigma_{\rm rms,\nu}$, achieved by smoothing the SKA noise field at the source central frequency, $\nu$, with a Gaussian smoothing kernel. The total smoothing scale is obtained by adding in quadrature the FWHM of the corresponding smoothing kernel to the SKA beam FWHM. Making the assumption that the spatial extent of the source is smaller than the total smoothing scale, such that the integrated source flux density per channel $i$ would equal the peak value of the smoothed source per channel, the source pixel values were integrated over spatial dimensions to produce a spectral profile, $S(i)$.  A tophat filter was then applied to the source spectral profile:

\begin{equation}
S'(i) =\frac{1}{k}\sum_{u=0}^{k-1}S(i-u),
\end{equation}

\noindent and to a one-dimensional white Gaussian noise field $N(i)$ with standard deviation equal to $\sigma_{\rm rms, \nu}$:

\begin{equation}
N'(i) = \frac{1}{k}\sum_{u=0}^{k-1}N(i-u).
\end{equation}

\noindent The size of the tophat filter, $k$, was chosen to equal the number of channels in the spectral profile with values greater than 10 percent of the maximum value. The final SNR value,

\begin{equation}
    {\rm SNR} = \frac{S'_{\rm max}}{\sigma'_{\rm rms} }, 
\end{equation}

\noindent was calculated using the maximum value of the  filtered spectral profile,

\begin{equation}
S'_{\rm max} = \max\{S'(i)\},
\end{equation}

\noindent and the r.m.s. value of the filtered Gaussian noise,

\begin{equation}
    \sigma'_{\rm rms} = \sqrt{\langle N'(i)\rangle}.
\end{equation}

\begin{figure}
  
  \includegraphics[trim={0.9cm 0.cm 1.7cm 1.5cm},clip,width=8.cm]{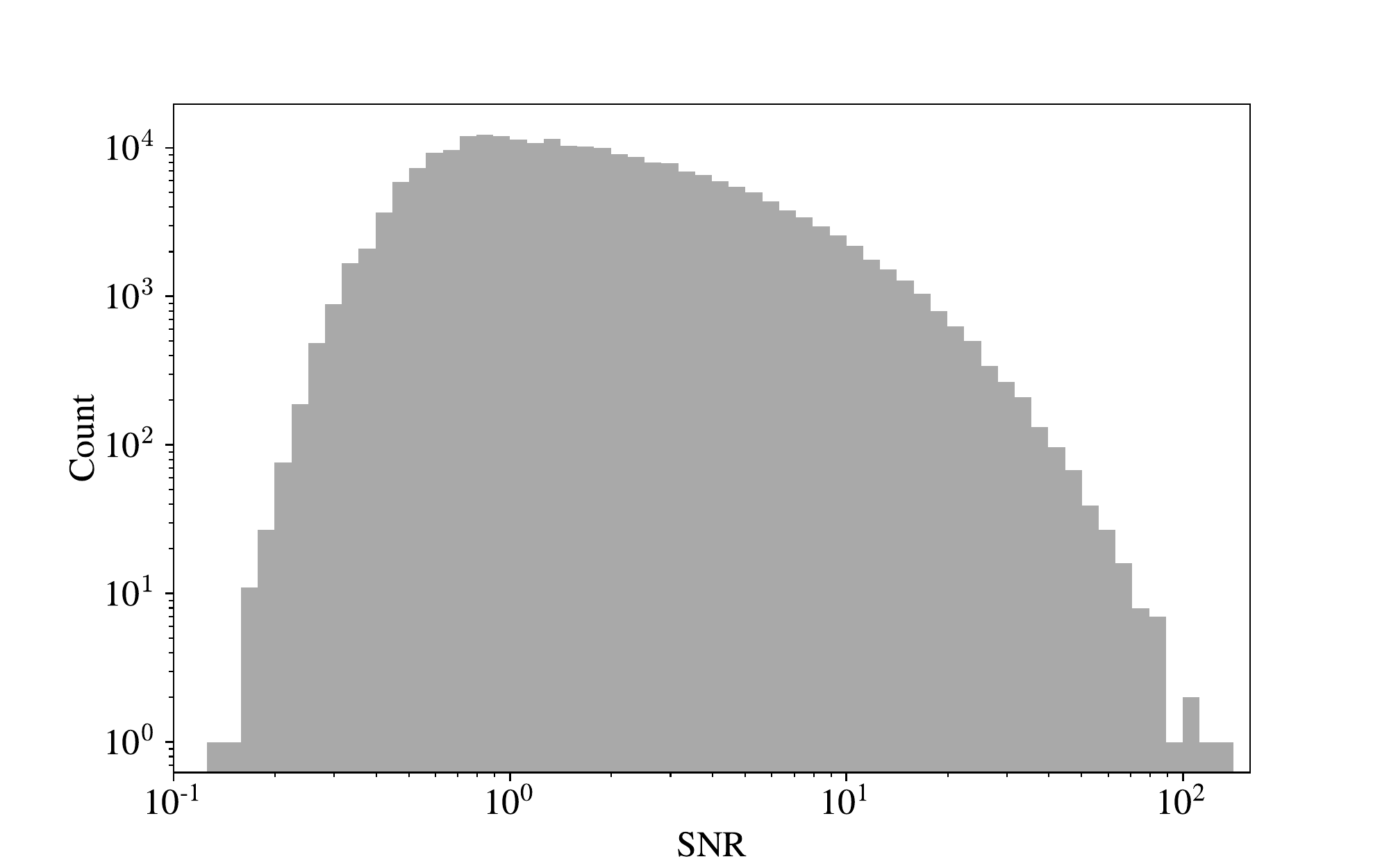}

  \caption{Truth catalogue sources are binned according to SNR values (see Section~\ref{snr} for a description of the  signal-to-noise calculation).}
  \label{snrbinned}
  \end{figure}

\noindent Fig.~\ref{snrbinned} presents binned SNR values of all sources in the full SDC2 truth catalogue.


\subsection{Source finding}

\begin{figure*}
	\centering
	\includegraphics[trim={0.cm 2.5cm 0.cm 04cm},clip,width=2\columnwidth]{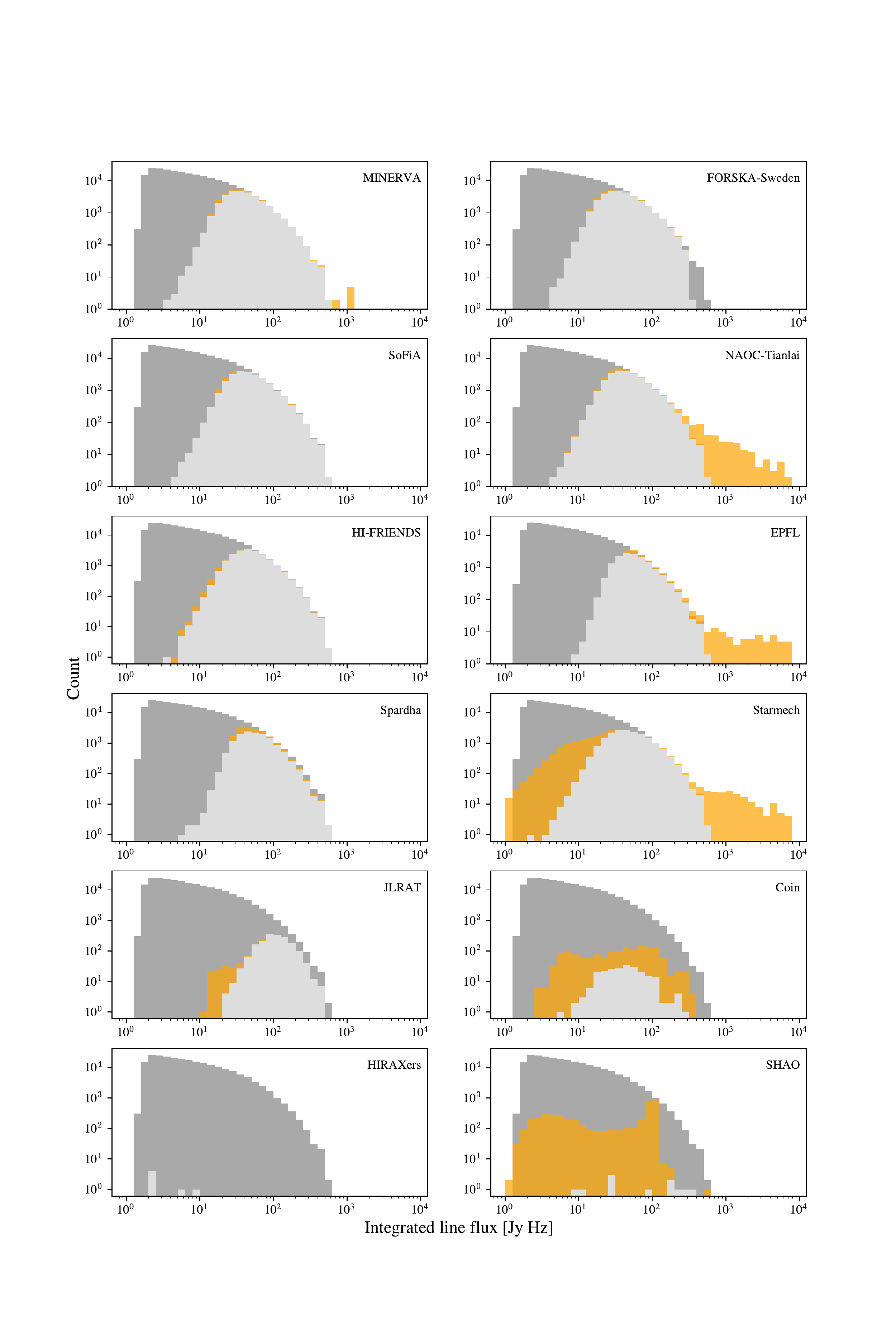}
	\caption{Sources in the full Challenge dataset binned according to integrated line flux value. For each team, all sources in the full truth catalogue (dark grey) are overplotted by the true values of matches (light grey) and by the submitted values of false detections (yellow).  }
	\label{flux}
\end{figure*}

\begin{figure}
  \centering
  \begin{tabular}{c}
  \includegraphics[trim={0.9cm 0.cm 1.7cm 1.5cm},clip,width=8.5cm]{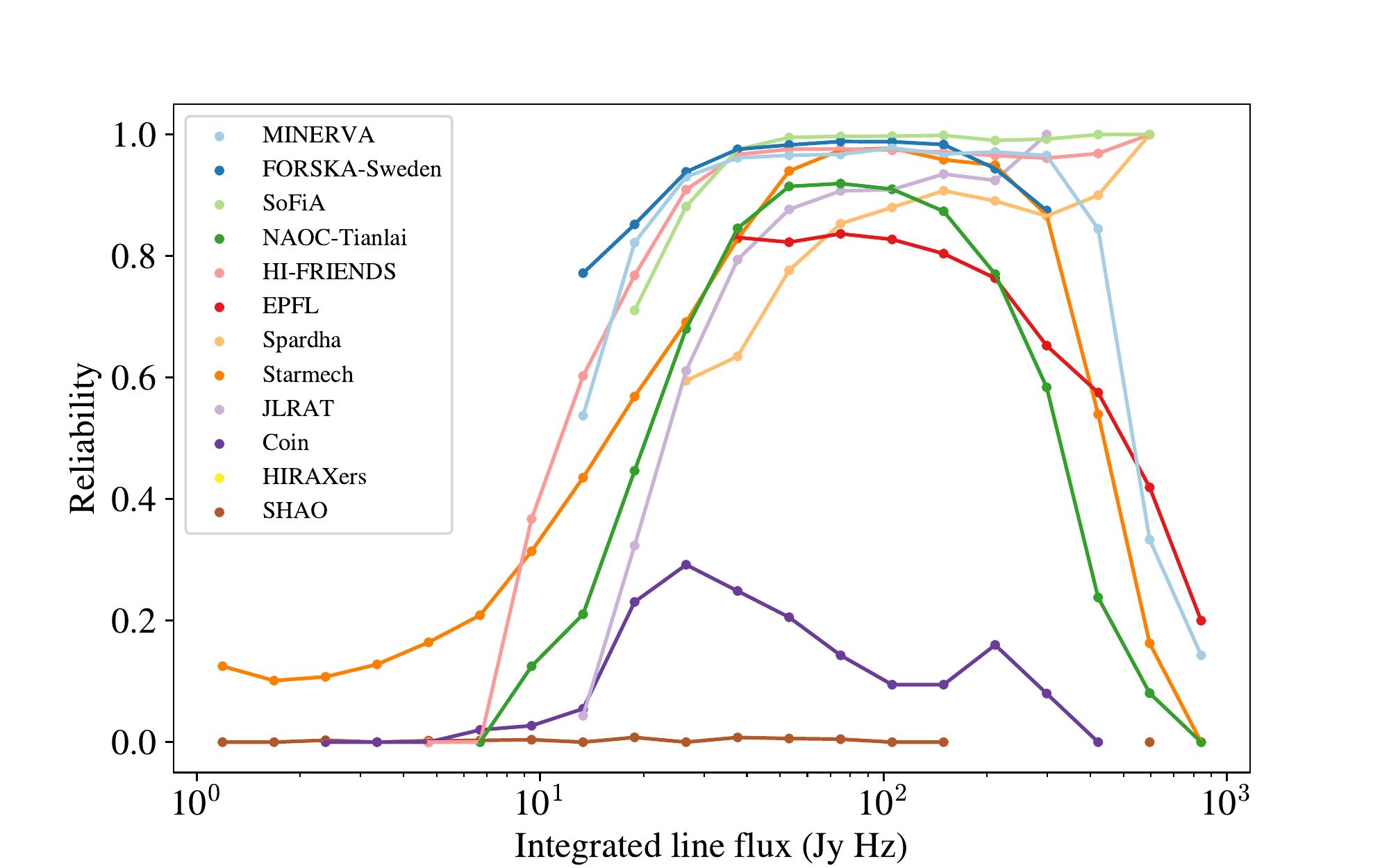}
  \\
  \includegraphics[trim={0.9cm 0.cm 1.7cm 1.2cm},clip,width=8.5cm]{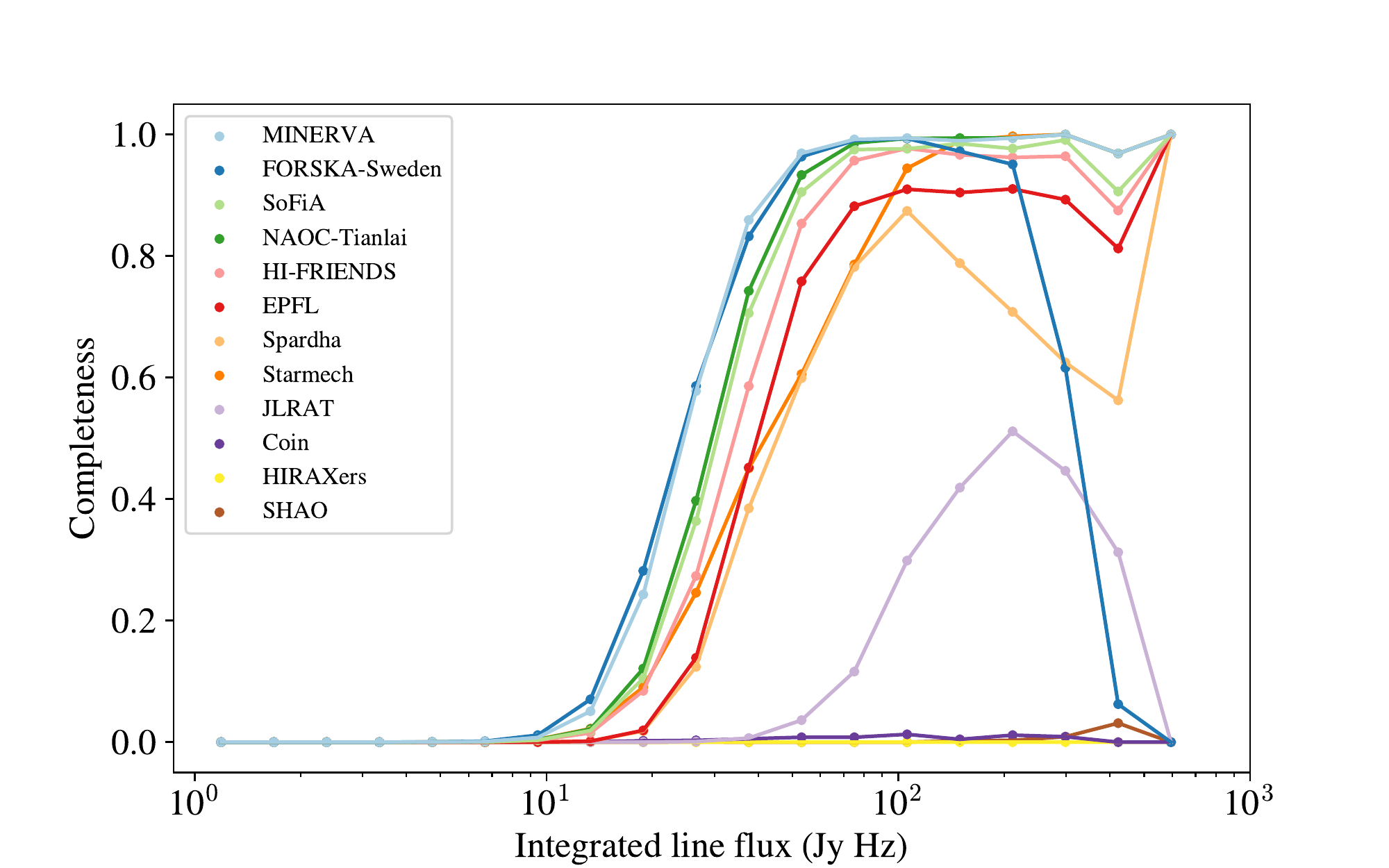}\\
    \includegraphics[trim={0.9cm 0.cm 1.7cm 1.2cm},clip,width=8.5cm]{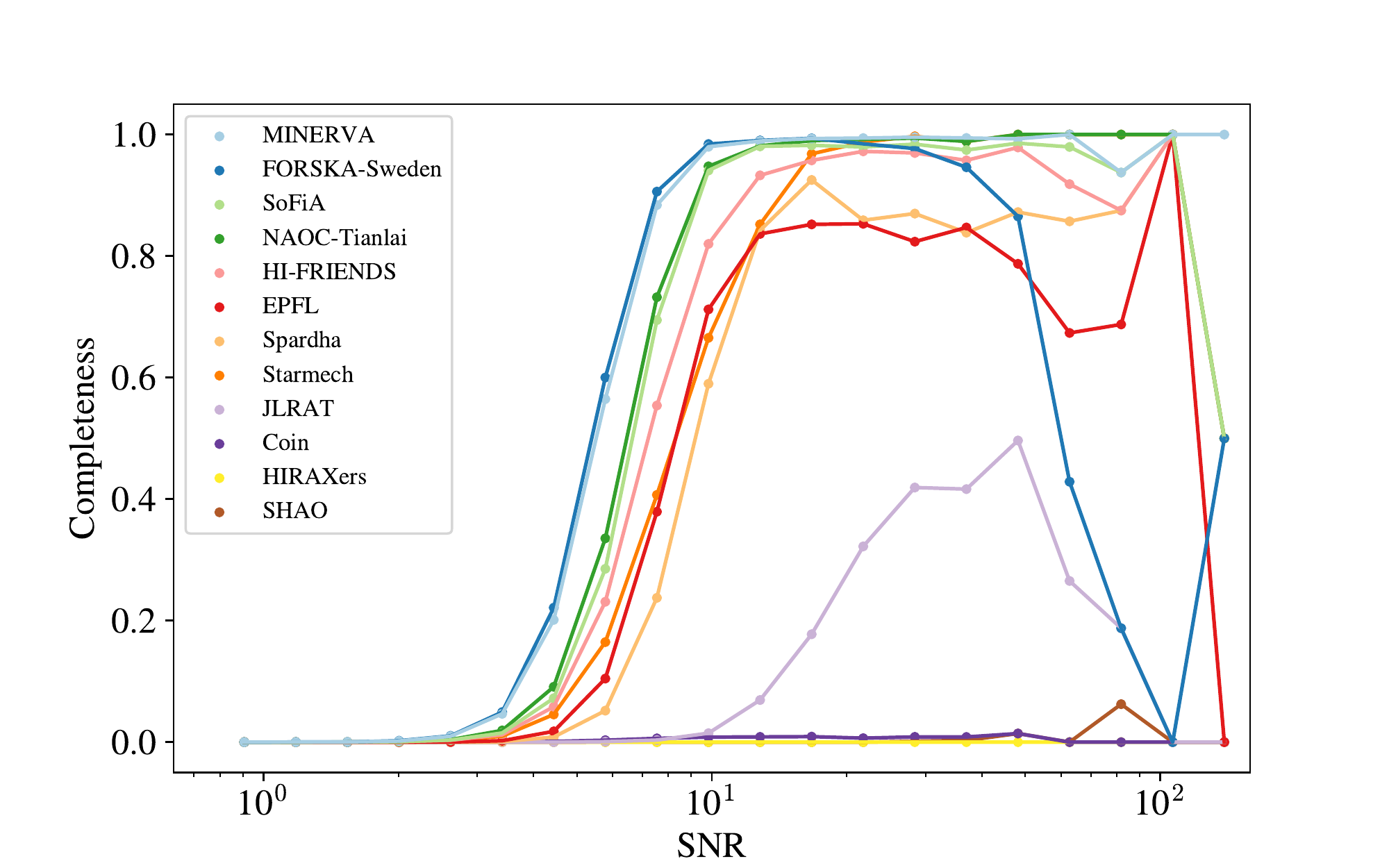}\\
  \end{tabular}
  \caption{Top: Reliability, defined as the number of matches divided by the number of detections, is plotted for each team as a function of submitted integrated line flux. Middle:  Completeness, defined as the number of matches divided by the number of truth catalogue sources, is plotted for each team as a function of true integrated line flux. Bottom: Completeness is plotted for each team as a function of true SNR (see Section~\ref{snr} for a description of the chosen signal-to-noise definition).}
  \label{relicomp}
  \end{figure}

Fig.~\ref{flux} presents for each team the number of final matches and false positives, binned according to integrated line flux along with all sources from the truth catalogue. When considering matches, truth catalogue line flux values are used; when considering false positives, the lack of corresponding truth values necessitates the use of submitted line flux values. Fig.~\ref{relicomp} presents reliability and completeness values as a function of integrated line flux, where submitted values are again used in the calculation of reliability due to the absence of corresponding truth values for false positives. Fig.~\ref{relicomp} also presents completeness as a function of SNR values.

\subsection{Source characterisation}

\begin{figure*}
  \centering
  \begin{tabular}{cc}
  \includegraphics[trim={0.5cm 0.cm 1.5cm 1.2cm},clip,width=1\columnwidth]{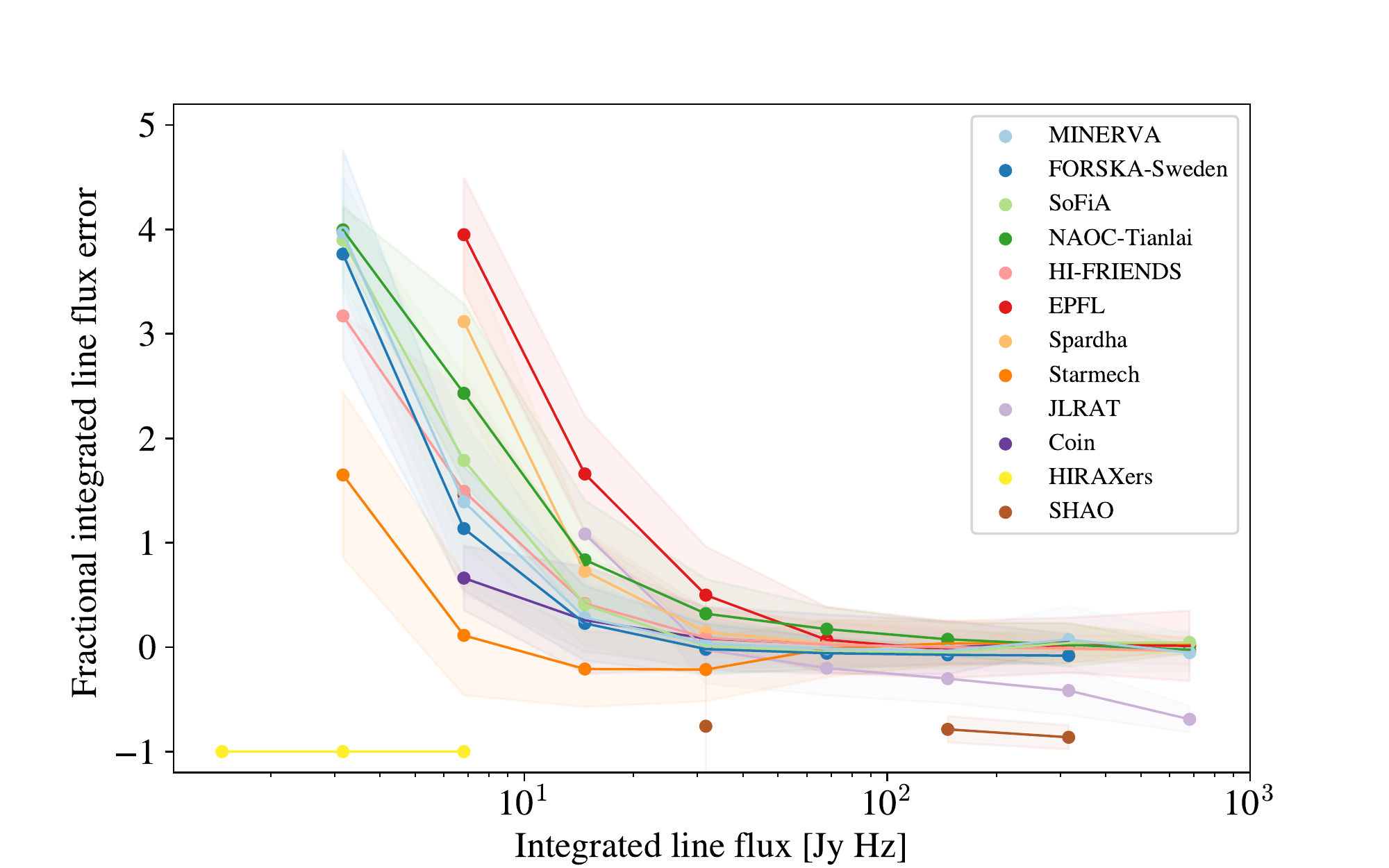}
  & \includegraphics[trim={0.5cm 0.cm 1.5cm 1.2cm},clip,width=1\columnwidth]{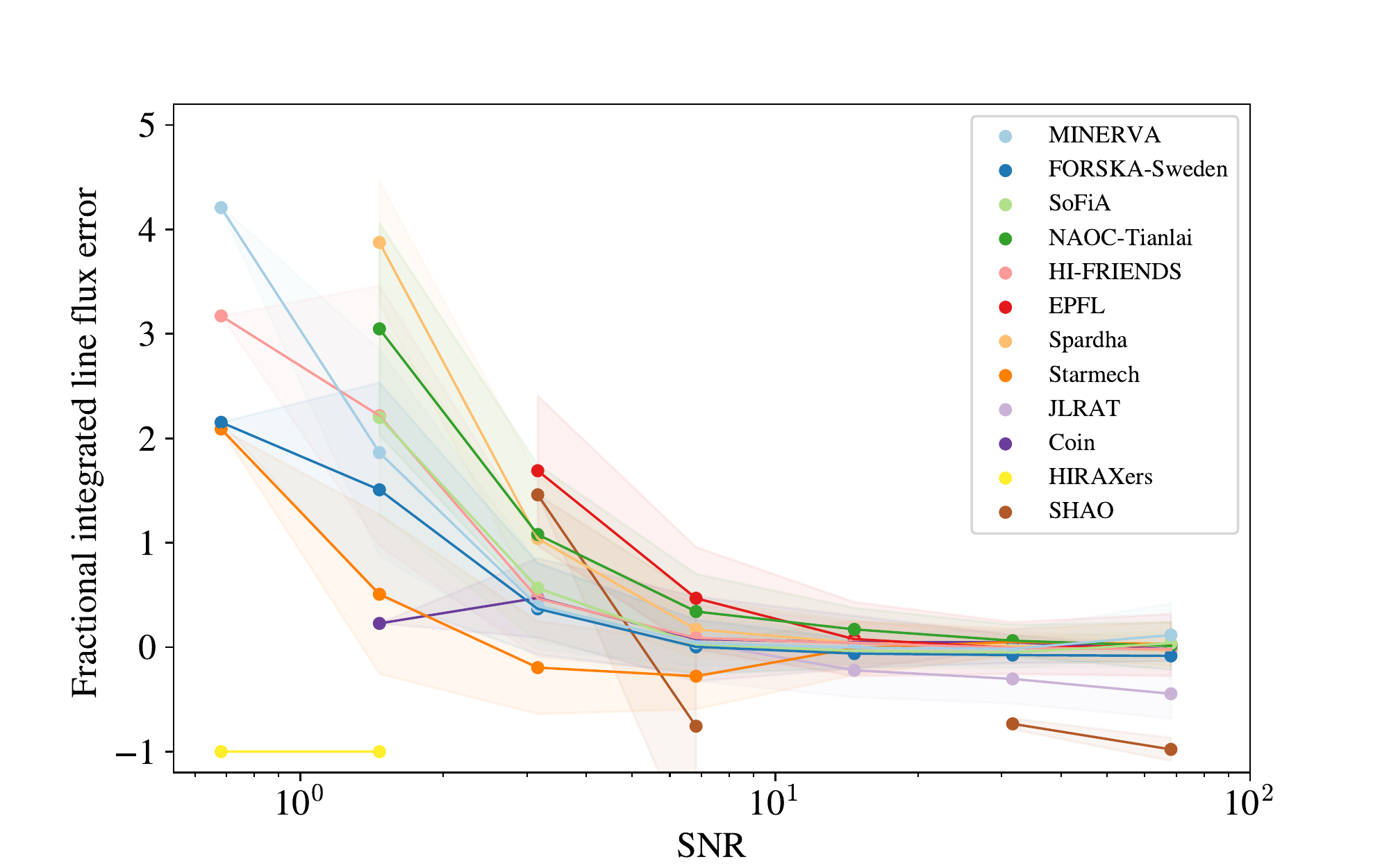}
 \\
  \includegraphics[trim={0.5cm 0.cm 1.5cm 1.2cm},clip,width=1\columnwidth]{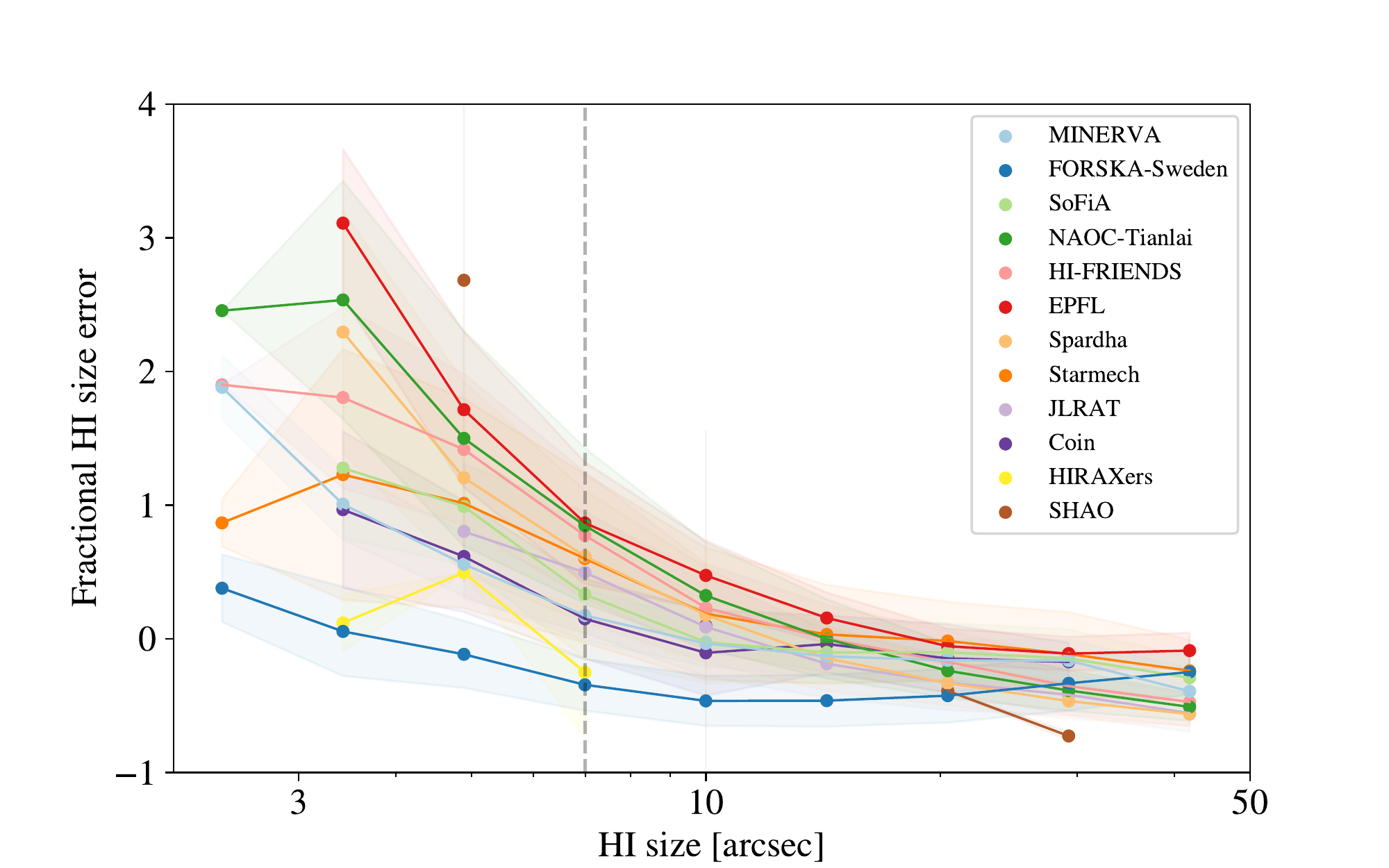}& \includegraphics[trim={0.5cm 0.cm 1.5cm 1.2cm},clip,width=1\columnwidth]{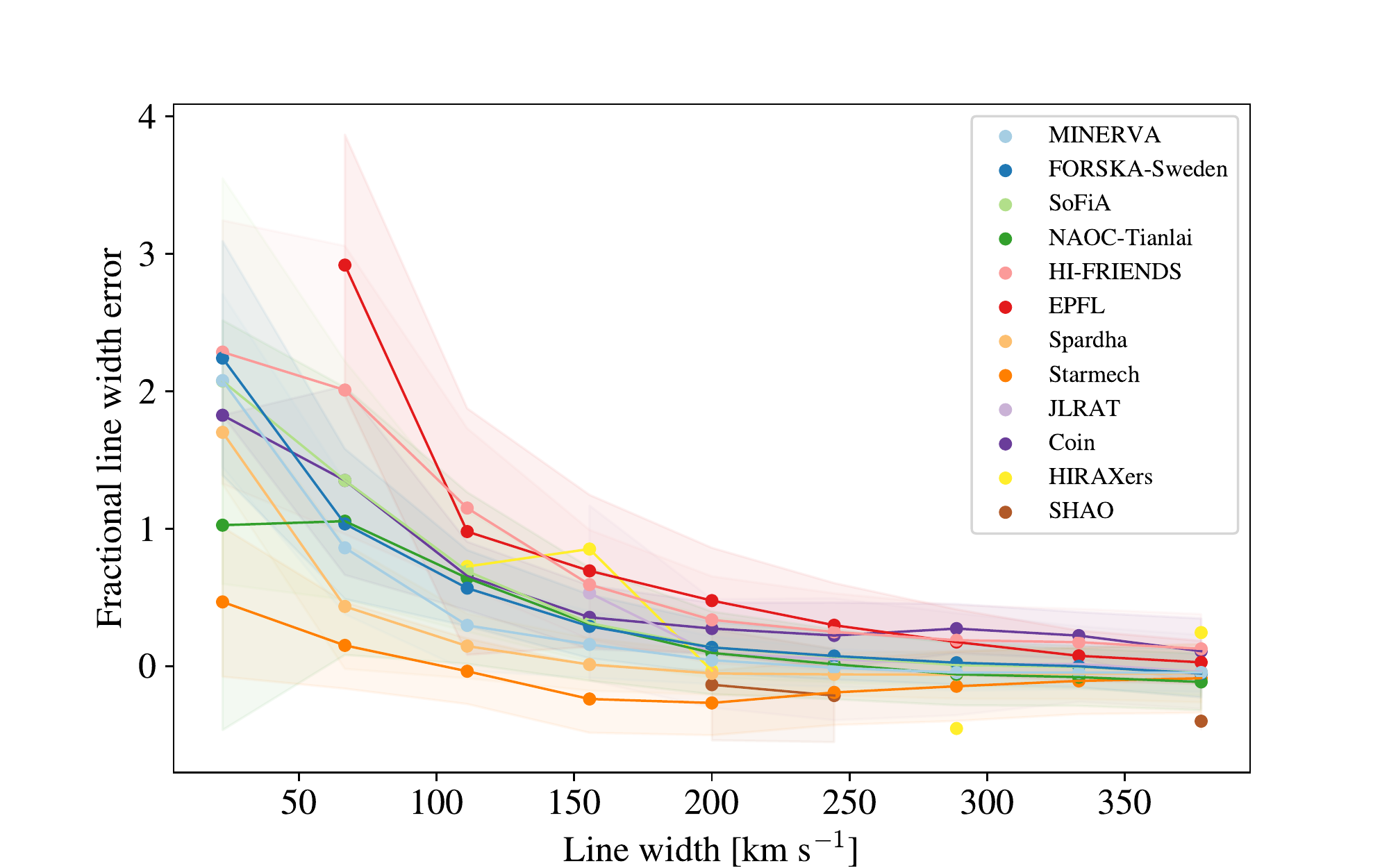}\\
    \includegraphics[trim={0.5cm 0.cm 1.5cm 1.2cm},clip,width=1\columnwidth]{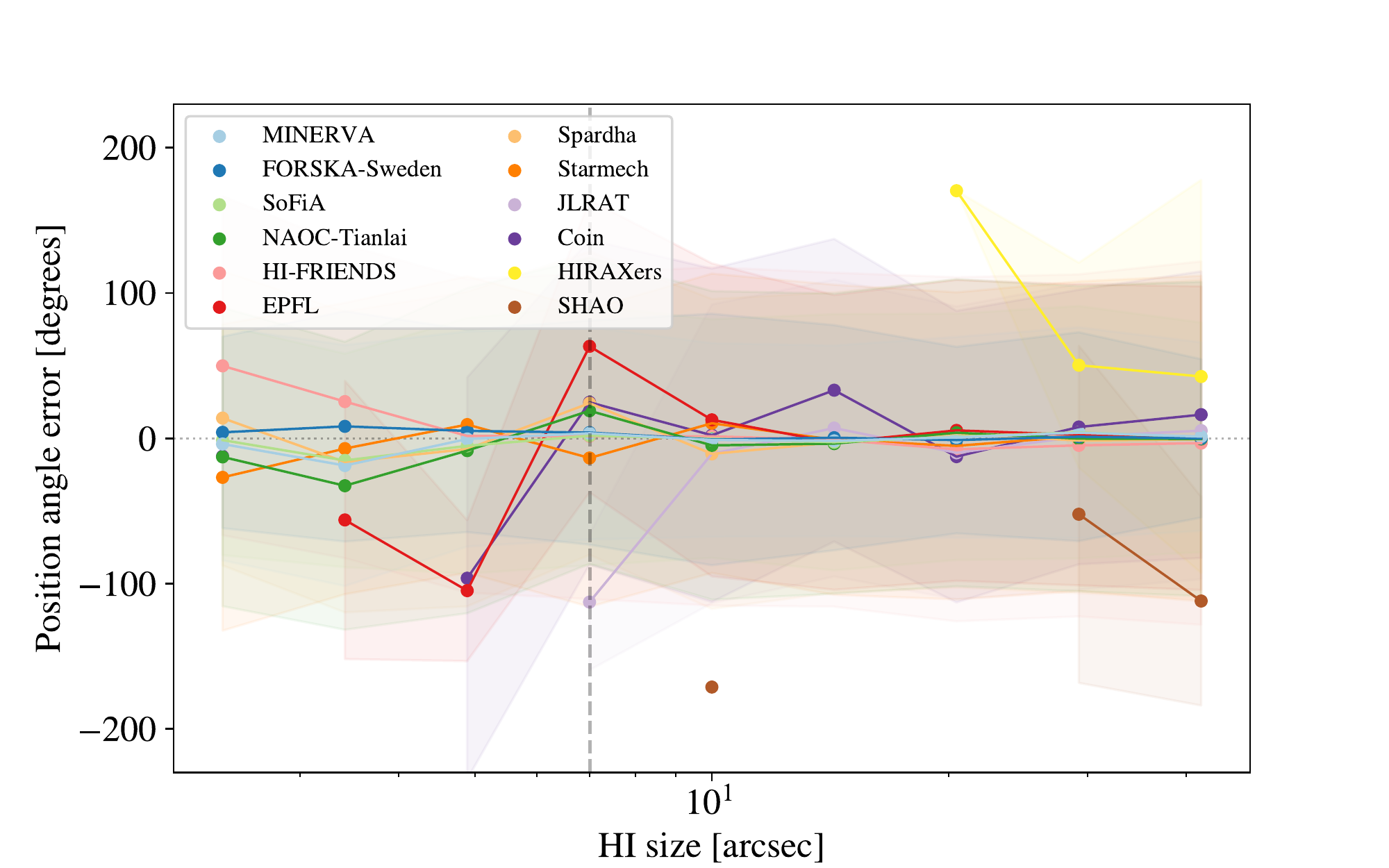}
  &
  \includegraphics[trim={0.5cm 0.cm 1.5cm 1.2cm},clip,width=1\columnwidth]{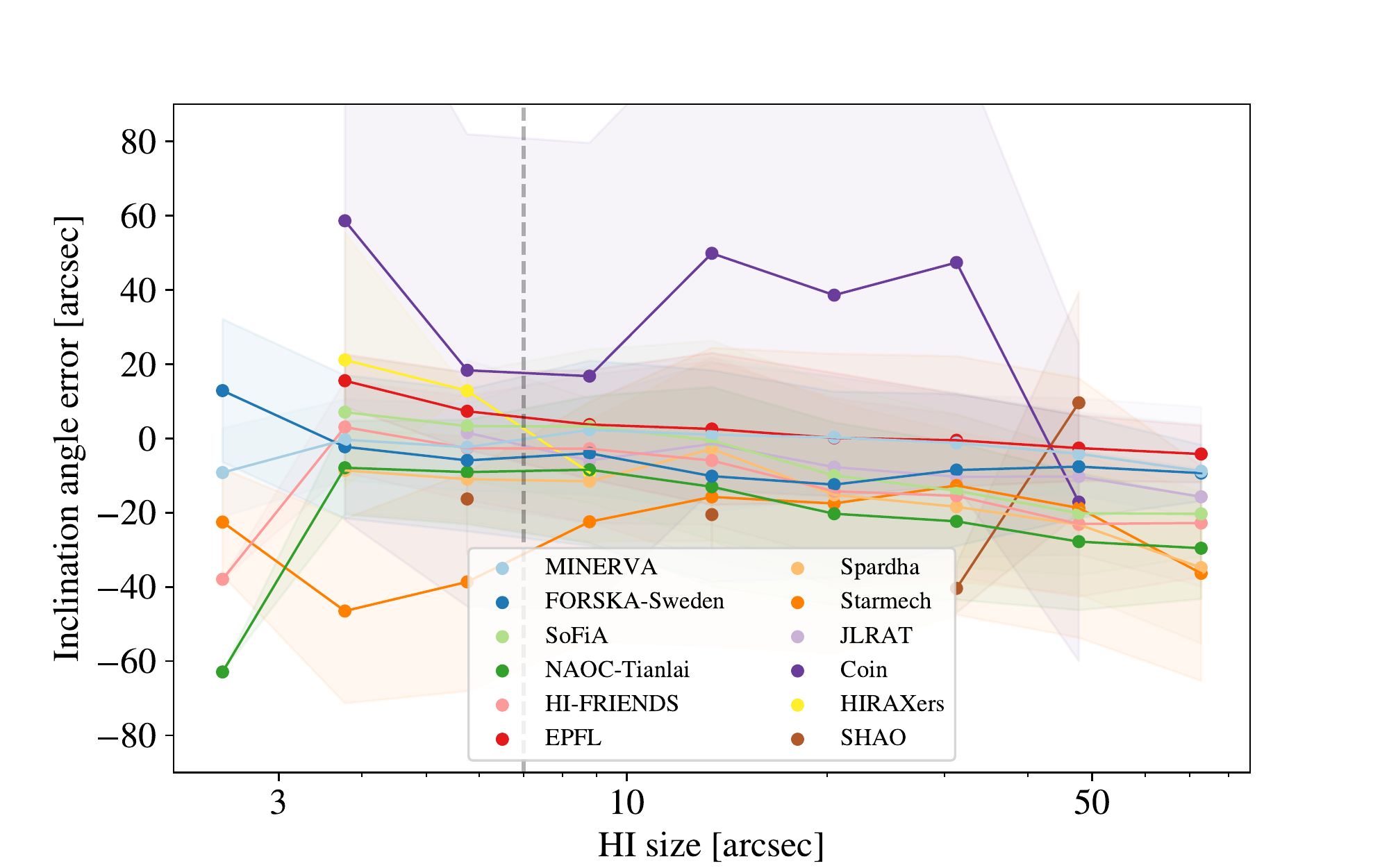}

  \end{tabular}
  \caption{Error terms (see Table~\ref{errors}), calculated without using absolute values, are plotted as a function of true property value, SNR, or spatial source size. Joined circles represent the median error per logarithmic bin, the filled regions represent the standard deviation of the error, and all plots use teams' matched submissions. A dashed line represents the beam size of the simulated observations.}
  \label{properties}
  \end{figure*}

\begin{figure}
	\centering
	\includegraphics[trim={1.cm 0.cm 1.5cm 1cm},clip,width=1\columnwidth]{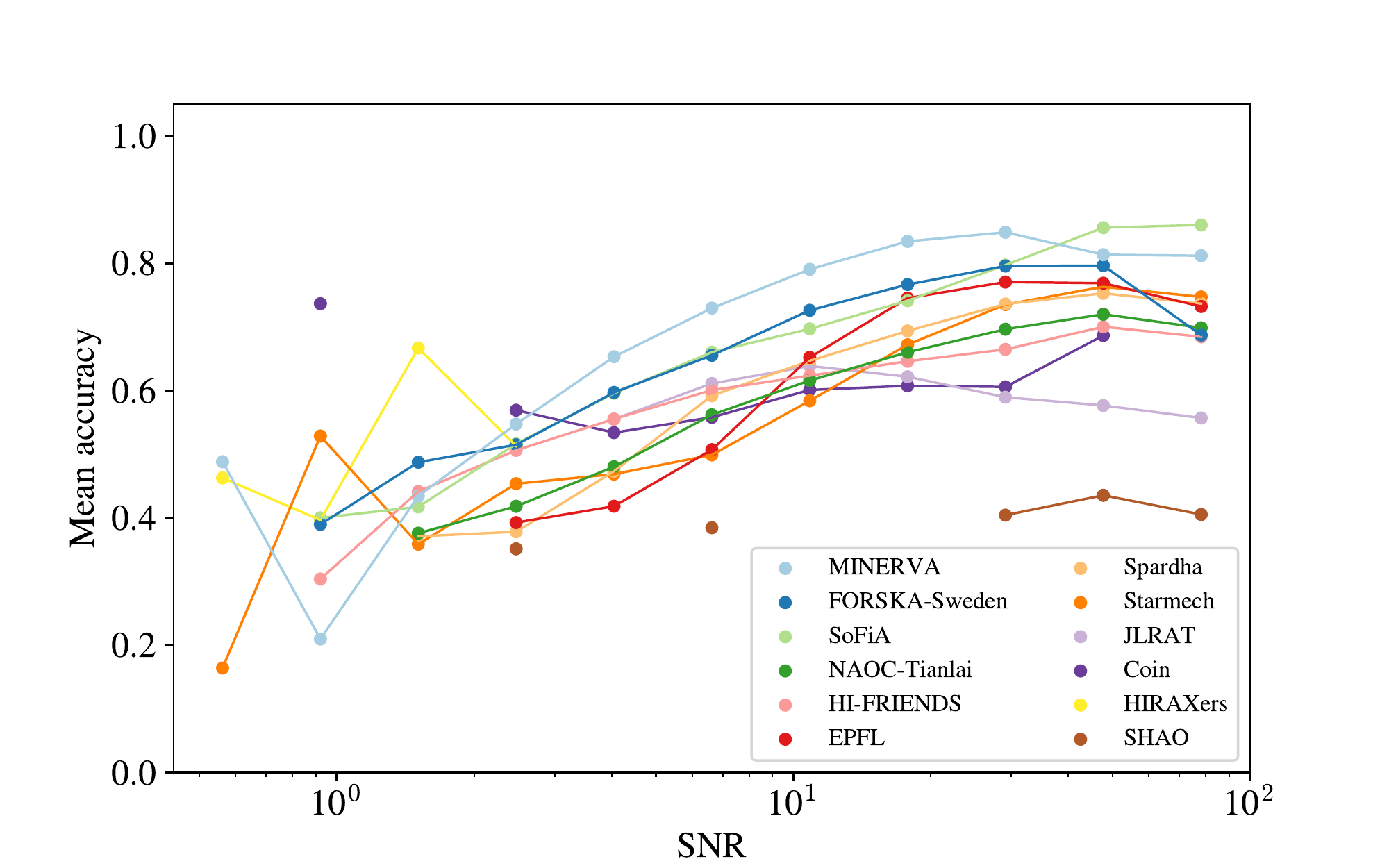}
	\caption{Source characterisation as a function of SNR. Source accuracy is determined according to Section~\ref{accuracy}, averaged over all properties except position in RA, Dec and central frequency, for all matches per team in the given SNR interval.}
	\label{mean_acc}
\end{figure}

In order to investigate the performance of teams' methods in the recovery of source properties, several relationships were investigated.   Fig.~\ref{properties} presents error terms (Table~\ref{errors}) calculated without using absolute values and plotted as a function of true property value and of SNR for flux, of true property value for size and line width measurements, and as a function of true size, for position and inclination angle measurements.  Fig.~\ref{mean_acc} presents overall source characterisation accuracy as a function of SNR. Characterisation accuracy is determined according to Section~\ref{accuracy}, averaged over all properties except position in RA, Dec and central frequency, for all matches per team in the given SNR interval.

\begin{figure*}
  \begin{tabular}{cc}
      \includegraphics[trim={0.2cm 0.cm 0.2cm 0.20cm},clip,width=1\columnwidth]{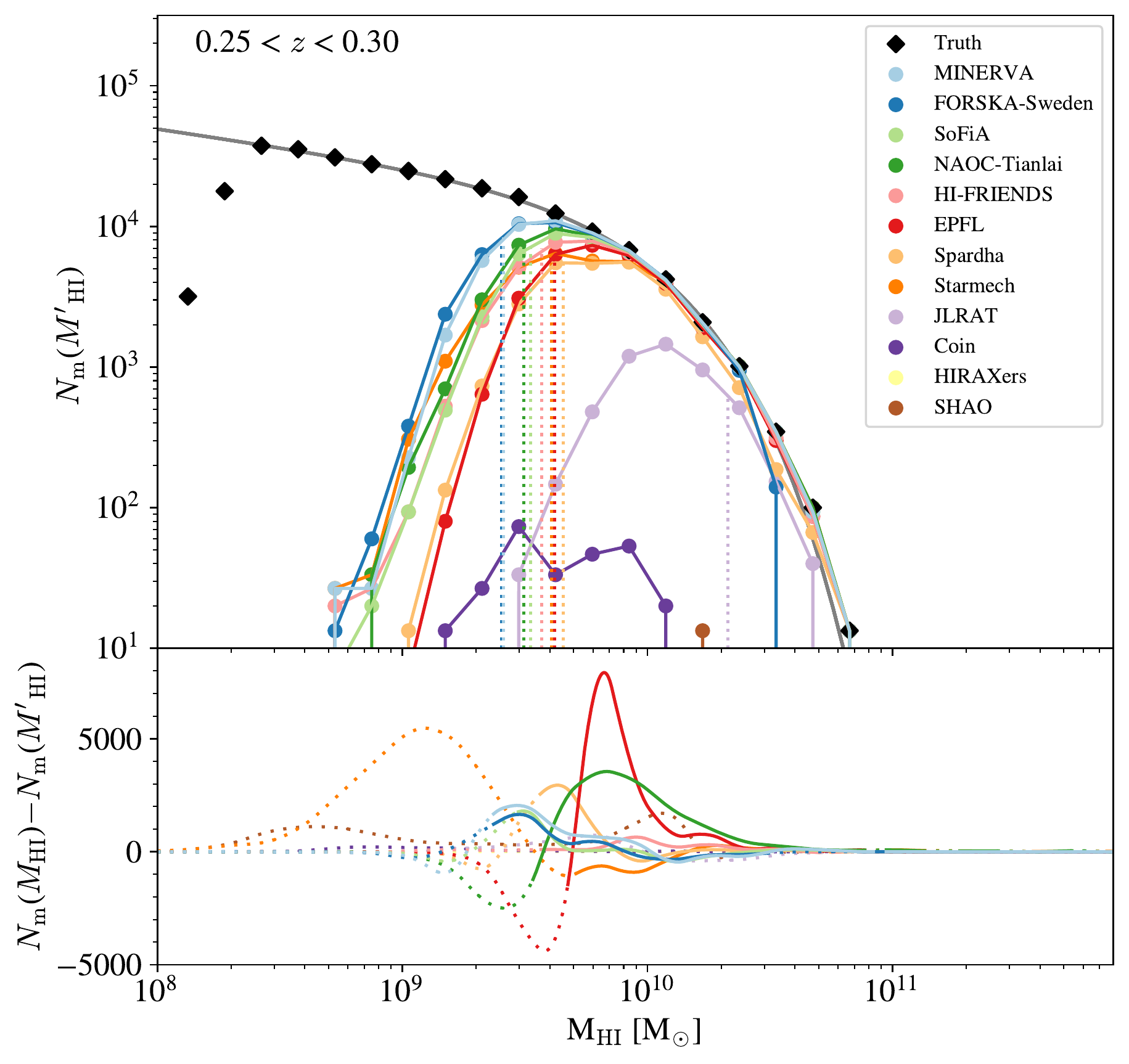} 
      &\includegraphics[trim={0.2cm 0.cm 0.2cm 0.2cm},clip,width=1\columnwidth]{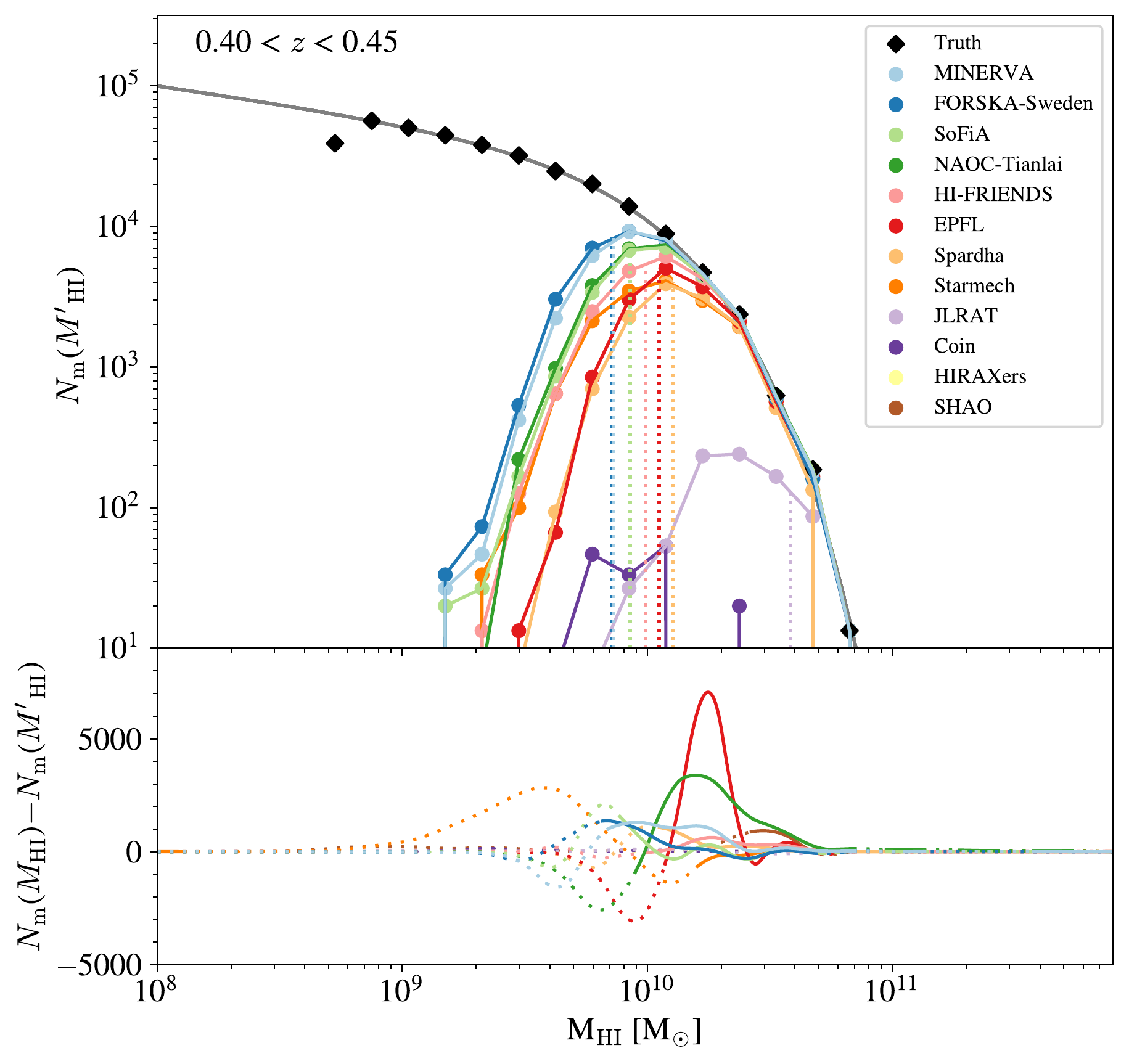} \\
      \end{tabular}
  \caption{Top panels: H{\sc i} mass distributions $N_{\rm m}(M'_{\rm HI})$ are constructed using the true values of integrated line flux and central frequency of each teams' matches (joined circles). The redshift-dependent H{\sc i} mass function (eq.~\ref{HIMF}), from which truth catalogue sources were drawn, is multiplied by the comoving volume of the given redshift interval and plotted (grey curve). Black diamonds represent the H{\sc i} mass distribution reconstructed using the full truth catalogue. Dotted lines indicate for each team the H{\sc i} mass above which completeness exceeds 50 percent. Bottom panels: the H{\sc i} mass distribution residual represents the difference between the distribution constructed from the values of teams' submissions and distribution constructed from truth values of teams' matches. Both distributions are interpolated prior to finding the residual. Completeness values are in this case calculated using teams' submitted values, and dotted and solid curves are used to delineate H{\sc i} masses where completeness falls below and above 50 percent, respectively. }

 \label{HIMF}
\end{figure*}

Fig.~\ref{HIMF} compares H{\sc i} mass distributions constructed using teams' matched sources with the function constructed by taking the input redshift-dependent H{\sc i} mass function ${\phi}(M_{\rm HI})$ (eq.~\ref{HIMF_eq}) and multiplying by the sky volume covered by a given redshift interval. True H{\sc i} masses generated during our simulation (Section~\ref{sims}) were used to obtain for each team an H{\sc i} mass distribution, $N_{\rm m}(M'_{\rm HI})$, by counting matched sources that fall within a logarithmic bin centred on true mass value $M'_{\rm HI}$.  

A second H{\sc i} mass distribution, $N_{\rm m}(M_{\rm HI})$, was constructed using submitted property values, $F$ and $\nu$, of teams' detections, which were converted to mass according to eq.~\ref{massderiv} \citep{2012MNRAS.426.3385D}. The same conversion was applied to the full truth catalogue to produce the complete H{\sc i} mass distribution, $N^C_{\rm m}(M'_{\rm HI})$, which was used to verify consistency between the input mass function and simulated observables.





Submitted and true values of teams' matches and detections, respectively, were used to plot the residual,

\begin{equation}
    \Delta N_{\rm m}(M_{\rm HI}) = N_{\rm m}(M_{\rm HI})-N_{\rm m}(M'_{\rm HI}),
\end{equation}

\noindent after applying a second order spline interpolation to both distributions.

\begin{figure}
\centering
	\includegraphics[trim={0.cm 0.cm 0.5cm .cm},clip,width=\columnwidth]{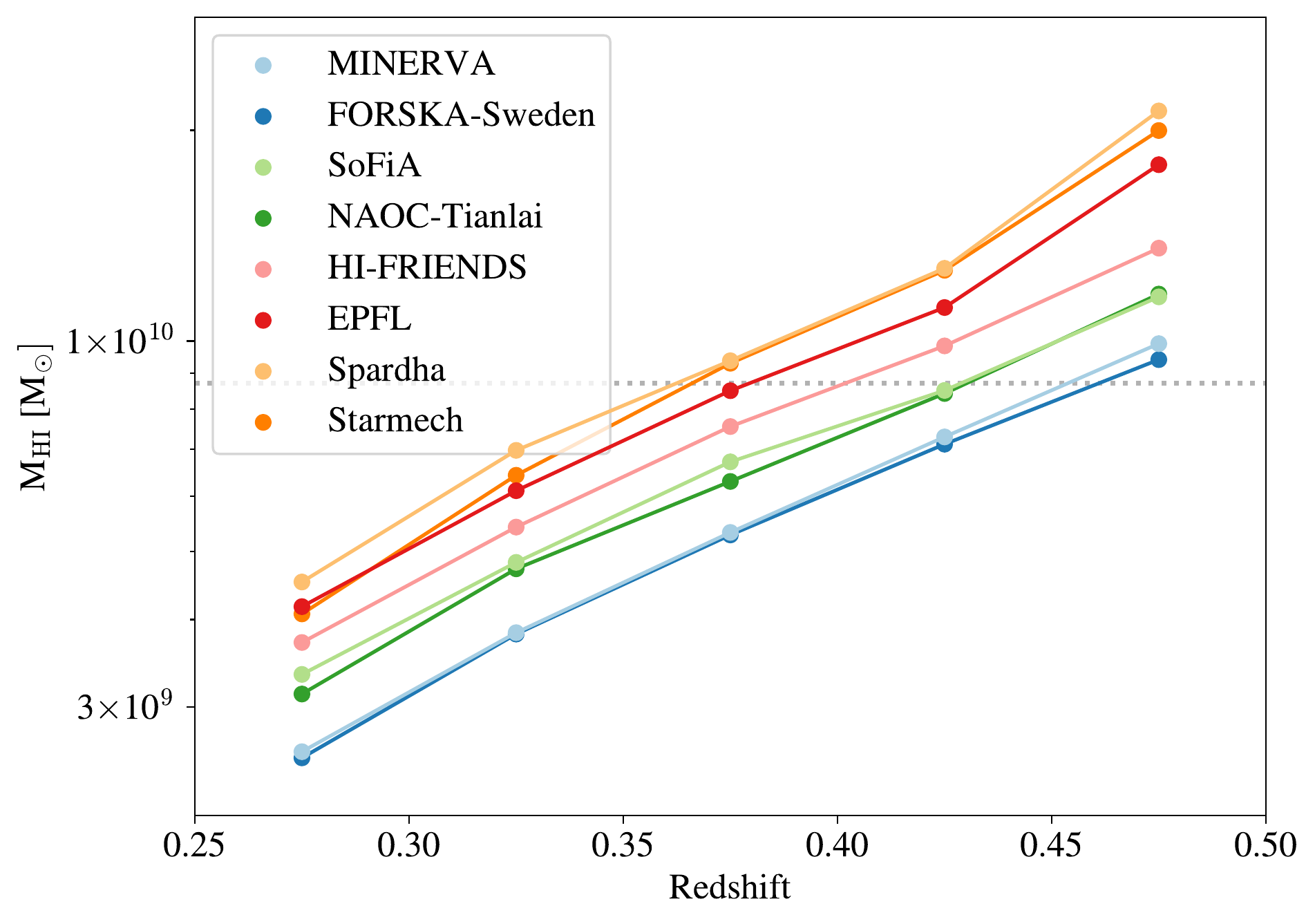}
	\caption{The H{\sc i} mass above which at least 50 percent of truth catalogue sources are recovered is plotted against redshift for the eight top scoring teams. The dotted line represents the input H{\sc i} `knee' mass, $M_{*}$ (equation~\ref{HIMF_eq}), which marks in the H{\sc i} mass function the exponential decline from a shallow power law. }
	\label{comp_50}
\end{figure}

For each team, the H{\sc i} mass distribution derived from true mass values, $N_{\rm m}(M'_{\rm HI})$, was interpolated and compared with the input H{\sc i} mass distribution, $N_{\rm m}(M_{\rm HI})$, in order to identify the H{\sc i} mass above which at least 50 percent of truth catalogue sources are recovered (Table~\ref{comp_50_tab}). Fig~\ref{comp_50} presents this mass for the top eight scoring teams as a function of redshift and compared with the H{\sc i} mass function `knee' mass  (equation~\ref{HIMF_eq}).


\subsection{Reproducibility awards}

Six teams submitted entries for the SDC2 reproducibility awards. Each pipeline was evaluated by an expert panel against the pre-defined award criteria (Section~\ref{repro}). Table~\ref{repawards} reports the awards granted to each team.

\begin{table*}
	\centering
	\begin{tabular}{lll} 
        	\hline\noalign{\smallskip}
        	\smallskip
        	 Team name & Reproducibility award & Pipeline\\

    	 	 \hline\noalign{\smallskip}
			  EPFL  & Bronze&\url{https://github.com/epfl-radio-astro/LiSA}\\
			  FORSKA-Sweden & Silver&\url{https://github.com/FraunhoferChalmersCentre/ska-sdc-2}\\
			  HI-FRIENDS & Gold&\url{https://github.com/HI-FRIENDS-SDC2/hi-friends}\\
			  NAOC-Tianlai &Bronze&\url{https://github.com/kfyu/SDC2-tianlai}\\	
			  SHAO  & Bronze& \url{https://github.com/astrosumit/SDC2-SHAO} \\
			  Team SoFiA & Silver&\url{https://github.com/SoFiA-Admin/SKA-SDC2-SoFiA}\\ 
		

		  	\hline \noalign{\smallskip}
	\end{tabular}
        
    	\caption{Reproducibility awards were made to six teams who submitted pipelines demonstrating best practice in the provision of reproducible results and reusable methods. Entries were evaluated by an expert panel using a pre-defined set of criteria (Section~\ref{repro}). }
        \label{repawards}
\end{table*}

\section{Discussion}

Challenge teams employed a variety of methods to tackle the simulated SKA MID H{\sc i} dataset.  In this section we discuss the findings in terms of individual and collective method capabilities.

\subsection{Source finding and characterisation}

The overall results (Table~\ref{SDC2results}) show a wide range of performance both within and between methods. Reference to Table~\ref{SDC2methods} indicates that strategies for false positive rejection are important. Further, the use of a refined training dataset, as employed by team MINERVA, may be crucial. 

While reliability and completeness (Fig.~\ref{relicomp}) generally show an increase with increasing flux and SNR, several teams show a drop-off at the brighter flux end. This is partly explained by a low number of sources, resulting in statistical noise. Reliability, in addition, will be particularly affected by the presence of brighter artefacts arising from imperfect continuum subtraction. Unreliability could in turn lead to a lower level of completeness in the corresponding flux bin, if source-finding methods themselves become correspondingly uncertain. For the top two scoring teams, a completeness of at least 50 percent is achieved down to a limit of SNR$\sim$5 and an integrated flux limit of $\sim$20 Jy Hz.

The analysis of individual source property recovery (Fig.~\ref{properties}) finds that of all properties, position angle is the most difficult to recover, with a standard deviation on the errors often covering most of the position angle range. This is understandable considering the large fraction of partially unresolved sources, and some teams are able to recover position angle well for resolved source sizes.  Inclination angle, which gives rise to the radial velocity for a given rotational velocity  (equation~\ref{vrot}), and can therefore be approximated by making use of line width, flux and size measurements, does not suffer the same problem.  Source characterisation could be improved by choosing a suitably high detection threshold. For example, analysis of characterisation accuracy as as function of SNR (Fig.~\ref{mean_acc}) finds a clear trend. The winning team, MINERVA, dominates across most of the SNR range, maintaining an average accuracy above 0.8 from SNR$\sim$10--60, and remaining around 10 percent higher than the next team from SNR$\sim$3--60. At the very highest SNR, however, Team SoFiA achieved the greatest averaged accuracy, while the MINERVA performance falls slightly.

\subsubsection{Noise biases}
The analysis of integrated line flux measurements finds in general a positive excess at lower values. This demonstrates the problem of so-called `flux boosting' as a result of increasing number counts in the presence of local noise fluctuations \citep{1998PASP..110..727H}.  In terms of SNR, flux boosting becomes apparent at SNR$\sim$7 but remains minimal for the top three scoring teams, which see a flux boosting effect of $\sim$40 percent at SNR$=3$. Similar noise biases may be apparent in the measurement of H{\sc i} size and line width, where there is a general tendency to overestimate smaller sizes and underestimate larger sizes. Some teams used the SDC2 development dataset to calibrate pipeline output against the available truth catalogue. For example, team SoFiA used polynomial fits to affected parameters as a function of flux, in order to derive corrections for flux, H{\sc i} size and line width. While corrections can remove the bias, intrinsic scatter, which is likely to be considerable at low SNR, will remain (see e.g. \cite{1998PASP..110..727H}).  The overestimation of H{\sc i} size is compounded by the finite resolution of the simulated observation: the fractional error on H{\sc i} size understandably rises steeply as the true size decreases below the 7 arcsec beam size. Despite this limitation, some teams are significantly more accurate in constraining the source size limit.

\subsubsection{H{\sc i} mass recovery}


The H{\sc i} mass distributions presented in Fig.~\ref{HIMF} are constructed without making corrections for survey sensitivity, which is a non-trivial task that falls outside the scope of the Challenge. Our analysis is therefore intended to demonstrate the depth of H{\sc i} mass that can be probed by respective methods, and the discrepancy that may arise between number counts of observed and intrinsic masses of detected sources.

A 50 percent completeness threshold was chosen to characterise H{\sc i} mass recovery depths following \cite{2002ApJ...567..247R}, who, using an H{\sc i}-selected galaxy sample from the Arecibo Dual-Beam Survey \citep{2000ApJS..130..177R}, found a negligible difference between the mass function derived using only sources above the 50 percent `sensitivity limit' and the function derived using all sources.  Fig.~\ref{comp_50} demonstrates that the two top scoring teams' methods are able to probe the H{\sc i} knee mass with a 50 percent completeness out to a redshift of approximately 0.45, or 1740 Mpc of comoving distance. For comparison, the ALFALFA survey -- with a footprint of $\sim$6900 deg$^2$ -- has probed the knee mass out to distances of approximately 200 Mpc. 


With the caveat that line width completeness corrections have not been performed on the mass distributions constructed using teams' submitted values, we use Fig.~\ref{HIMF} also to demonstrate the relative error between distributions constructed using the true and submitted values of teams' detections. The top three scoring teams attain a relatively high degree of accuracy for detected sources, each seeing an overestimation in the mass distribution of less than 0.1 dex at the point where completeness falls below 50 percent. 


\subsection{Machine learning vs non-machine learning}

Supervised machine learning (ML) methods, particularly convolutional neural networks (CNN), proved a popular technique during the Challenge, and featured in the pipelines of the two top scoring teams. Of particular note is the significant success by winning team MINERVA in both the finding and characterisation parts of the Challenge. The winning technique, which used ML both to find and characterise sources, achieved a ten percent improvement over the next team in characterisation accuracy across a SNR range $\sim$4--30. Methods involving traditional signal processing techniques also achieved high scores, including the SoFiA package, which was used not only by the third-placed team of its developers, but also in the source characterisation of the second placed team and by several others.

\subsubsection{Generalisation}

The results demonstrate the promise of ML in the analysis of very large and complex datasets. As seen in similar community challenges (e.g.~\citealt{metcalf2019}), ML methods are often able to outperform traditional methods.  This success is not without its caveats. In order for supervised ML models to transfer successfully to real data, they must be able to generalise beyond the parameter distribution that has been sampled by the training data \citep{burges1998tutorial}. Overfitting by models with large numbers of parameters can be avoided using a sufficiently large set of training data. A more difficult problem is that of covariate shift: when the distributions of training and real datasets are intrinsically different. This is a common issue for astronomy (see e.g. \citealt{freeman2017unified,10.1093/mnras/staa166,2021arXiv210611211A}), where techniques are often being developed in preparation for data that is yet to be recorded. Models are instead trained using simulated data, which cannot capture unknown characteristics of the future observations. Limitations to the realism of the SDC2 data products (Section~\ref{limitations}) are likely in turn to introduce limitations in the ability of SDC2 ML models to transfer to real data. An increased number of real H{\sc i} observations used to generate the H{\sc i} emission cube will reduce the risk of model overfitting. Further characterisation of RFI and other instrumental effects during the commissioning phase of the SKAO telescopes will enable the simulation of ever more realistic datasets for training purposes, and transfer learning \citep{pan2009survey,10.1093/mnras/stz1883} could close the gap further still. In future SDCs, the inclusion of a data product produced using a different distribution could provide a test for model robustness to covariate shift. 

Non-machine learning methods, generally making use of far fewer parameters than ML models and less reliant on the availability of training data, may transfer more successfully from simulated to real data. This advantage appears to be evidenced by the comparative successes of team-SoFiA and HI FRIENDS -- both of which used the {\sc sofia} software package -- at the brighter end of the integrated flux and SNR ranges across reliability, completeness, and characterisation accuracy  (see Figs.~\ref{flux}, \ref{relicomp} and \ref{mean_acc}). By contrast, the ML-based pipelines used by teams MINERVA and FORSKA-Sweden have produced a number of false positives and false negatives, respectively, in the detection of the very brightest sources. The ML-based pipelines  also appear to show a fall in characterisation accuracy at the very highest SNR values. It is possible that the paucity of very bright samples in the training datasets has prevented ML methods from modelling very well the features of the brightest sources. On the other hand, it is likely that the small number of bright samples in the Challenge dataset has led to the prioritisation during pipeline optimisation of greater accuracy for fainter populations, since the large number of fainter sources produce a much greater impact on the score.


\subsection{Method complementarity}

The strategy employed by winning team MINERVA underscores one of the most important outcomes of the Challenge: that of method complementarity. By combining the outputs of two independent pipelines the teams were able to recover sources from a larger amount of the flux--line width parameter space than by using a single pipeline alone (Fig.~\ref{fig_minerva}), and could further exploit the independence of the pipelines to reduce bias and variance in source measurements. The success of this strategy demonstrates that, given a selection of sufficiently independent and well-performing methods, stacking -- where the predictions made by a group of independent machine learning methods are used as inputs into a subsequent learning model --  could improve generalisation from training data to new data (see also \citealt{WOLPERT1992241,2017JInst..12.5005A,10.1093/mnras/stw1454}).

The promise of a multi-method approach is further demonstrated by the performance of different methods in different aspects of the Challenge.  Teams Starmech and Coin, for example, though occupying the lower half of the leaderboard, performed particularly well in the recovery of line flux and H{\sc i} size, respectively (Fig.~\ref{properties}). Teams NAOC-Tianlai, HI-FRIENDS, EPFL, though missing out on the top three positions of the leaderboard, all demonstrated a high accuracy in the recovery, variously, of flux, source size and inclination angle. HI-FRIENDS also achieved highest overall reliability, while Team ForSKA, a very close second on the leaderboard, achieved the highest level overall completeness (Table.~\ref{SDC2results}).  If the measurement of source properties is considered a separate problem from source finding, and the measurement of different source properties considered a many-problem task in itself, then a so-called bucket-of-models approach \citep{10.1093/mnras/stv1608} could harness the capabilities of different methods to further improve performance beyond any individual method.

\subsection{Scoring metrics}
\label{metrics}

In the case of SDC2, the scoring algorithm has been designed to evaluate source finding and characterisation performance together. We note that the choice of any scoring metric will necessarily have an impact on the analysis that teams will perform. Strategies designed to maximise such a score might not be the best ones for other scientific goals: a search for fewer, highly resolved sources will take a very different approach from one aiming to produce a complete catalogue.  The Challenge leaderboard score, if looked at in isolation, can obscure strong performance by teams on source characterisation. This is a consequence of the strong penalty for false positives. Given the strong degree of method complementarity, a challenge scoring system that can reflect specialised solutions to a problem may further exploit complementarity as a quality of a collection of independent methods.

\begin{table}
	\centering
	\begin{tabular}{lccccc} 
        	\hline\noalign{\smallskip}
        	\smallskip
        	 Team name & \multicolumn{5}{c}{ Redshift interval}\\
        	 
    	   & 0.25 & 0.30 & 0.35 & 0.40& 0.45\\
    	  & --0.30 & --0.35 & --0.40 & --0.45& --0.50\\
    	 	 \hline\noalign{\smallskip}
			  MINERVA &2.60 &3.82& 5.27& 7.12&10.04\\
			  FORSKA-Sweden & 2.52 &3.80 &5.15& 6.91&  9.57\\
			  Team SoFiA & 3.32&4.77 &6.68 &8.52  &11.59 \\ 
			  NAOC-Tianlai &3.12&4.67& 6.33& 8.40&  11.69 \\	
			  HI-FRIENDS &3.67& 5.37& 7.51 & 9.94&  13.55\\
			  EPFL &4.14 &6.10& 8.45& 11.21& 17.60 \\
			  Spardha & 4.78& 6.98 & 9.47 & 12.55 &20.91\\
			  Starmech & 3.97& 6.52 &9.41 &12.44 & 20.22\\
			  JLRAT &- &46.03 &- & 46.77& 72.57  \\
			  Coin & -& 69.44& - & 70.11&  72.52\\
			  HIRAXers  & - & - & -& -&- \\
			  SHAO  & - & - & -& -&- \\
		

		  	\hline \noalign{\smallskip}
	\end{tabular}
        
    	\caption{The H{\sc i} mass (in units of $10^{9}$ M$_{\odot}$) above which at least 50 percent of truth catalogue sources are recovered is reported per redshift interval for the SDC2 finalist teams. }
        \label{comp_50_tab}
\end{table}

\subsection{Open Science}

The SDC2 reproducibility awards were designed to recognise Open Science best practice in the preparation and dissemination of analysis software pipelines. By providing public access to codes written to address SDC2, six teams were able to enhance the reproducibility and reusability of their methods. Noteable examples of best practice included the use of clear and comprehensive documentation, quick-start examples, command line interface excerpts,  open-source licensing and descriptive variable names. Practices employed by the Gold-standard HI-FRIENDS pipeline included the use of the workflow management system {\sc snakemake} (see Section~\ref{hifriends}) to design the overall workflow and suggest well-structured code directories, to manage the installation of software dependencies, and to generate a workflow graph image, all of which support the reusability and portability of the code. The advantages of well-documented and easily accessible codes are underscored by the popularity during the Challenge of the publicly available and regularly maintained SoFia package, which was used by six of the participating teams. 

Reproducible and reusable analysis pipelines help to address some of the challenges of conducting research under a deluge of data while leveraging the many new technologies available to deal with the data. However, preparing software for public access can require a significant time investment. As we look ahead to the exascale era of data \citep{2020RSPTA.37890060S}, adequate funding to allow for software package maintenance and development will be essential.


\subsection{Data handling}

Teams were able to handle the large Challenge dataset with minimal difficulty thanks to the generous provision of computational resources by the SDC2 partner facilities (Section~\ref{hpcs}). By dividing the dataset into smaller portions and running parallelised codes, teams could comfortably process the full Challenge dataset in under 24 hours of wall clock time.  Efficiency savings will become ever more important as volumes of observational data grow and analysis pipelines proliferate; the use of fewer resources to analyse data will not only allow future SKA Regional Centres to support a greater number of researchers, but will also reduce energy consumption during processing.

\subsection{Lessons learned}

We summarise here the opportunities for improvement in Challenge delivery that would further support the achievement of the overall goals of the SDC series:

\begin{enumerate}
    \item Additional guidance for the use of radio astronomy convention and conversions (see Section~\ref{conventions}).  
    \item Consideration of the use of multiple scoring metrics to reflect different aspects of a challenge (see Section~\ref{metrics}). 
    \item A smaller set of criteria for a reproducibility component of a challenge could prove more accessible for teams to achieve (see Section~\ref{repawards}).
\end{enumerate}

\section{Conclusions}

The second SKAO Science Data Challenge has brought together scientists and software experts from around the world to tackle the problem of finding and characterising H{\sc i} sources in very large SKAO datasets. The high level of engagement coupled with multidisciplinary collaboration has enabled the goals of the Challenge to be met, with over 100 finalists gaining familiarity with future SKAO spectral line data in order to drive forward new data processing methods and improve on existing techniques.

Interpretation of the results from SDC2 is limited by three main factors:

\begin{enumerate}
    \item The Challenge dataset is a simulation and cannot fully represent real future SKA observations. Dataset realism is limited most significantly by oversimplication of the noise (see Section~\ref{limitations}).
    \item The Challenge did not aim to provide a standardised cross comparison of methods; only a single dataset was used and no attempt was made to control for team effort or domain expertise. 
    \item Team methods were developed as a means to maximise a score calculated according to the Challenge definition. Depending on the scientific goal, alternative metrics may be measured, for which other strategies may be explored.
    
\end{enumerate}

\noindent With these caveats in mind, the main outcomes from the Challenge are summarised below:

\begin{enumerate}
    \item Twelve international teams, using a variety of methods (Section~\ref{methods}) were able to complete the full Challenge. 
    \item Simulated data products representing a 2000\,h spectral line observation by SKA MID telescopes were produced for the Challenge (Section~\ref{sims}), and are now publicly available together with  accompanying truth catalogues\footnote{\url{https://sdc2.astronomers.skatelescope.org/sdc2-challenge/data}}. We encourage the use of these data products by the science community in order to support the preparation and planning for future SKAO observations. 
    \item The generous contribution from supercomputing partner facilities (Section~\ref{hpcs}) has been integral to the success of the Challenge. Thanks to the provision of resources for hosting, processing and access to Challenge data, it has been possible to provide a realistically large H{\sc i} data product in an accessible way. The support has also provided the opportunity to test several aspects of the future SRC model of collaboratively networked computing centres, from web technologies involved in the SDC2 scoring service (Section~\ref{scoring_section}), to the access processes in place for resource users. 
    \item The provision of a realistically large H{\sc i} data product has allowed participants to explore approaches for dealing with very large datasets. By interacting with the full Challenge dataset, finalist teams were able to investigate optimisation and efficiency savings in readiness for future SKAO observational data products. 
    \item Analysis of teams' submissions (Section~\ref{results}) has shown that sources are recovered with over 50 percent completeness down to a SNR limit of $\sim$5 and an integrated flux limit of $\sim$20 Jy Hz by the top scoring teams. Keeping in mind the caveats above, this translates to the ability to probe the H{\sc i} mass function down to $\sim3\times 10^9$ M$_{\odot}$ at $0.25< z <0.30$ and to $\sim1\times 10^{10}$ M$_{\odot}$ at $0.45< z <0.50$. The `knee' mass of the H{\sc i} mass function can be probed out to $z\sim 0.45$ by the same methods for the chosen redshift evolution. 
    \item The analysis of submitted catalogues also provides a qualitative and quantitative understanding of the biases inherent to sensitivity-limited survey results. Biases arising from the presence of local noise fluctuations resulted in overestimation of flux at SNR$\lesssim$7. Source size and line width also showed a positive bias with fainter objects and smaller sizes. 
    \item Six teams took part in the SDC2 reproducibility awards, which ran alongside the main Challenge and were designed to recognise best practice in the preparation of reproducible and reusable pipelines. All six teams received an award, with team HI-FRIENDS receiving a Gold award for an exemplary software pipeline.
    \item  New applications of machine learning-based techniques -- used by the two top scoring teams -- have shown particular promise in the recovery and characterisation of H{\sc i} sources. The results suggest a dependency on sufficient training data, evidenced by a drop in performance at the bright flux end, where a paucity of very bright training sources exists. A more uniformly distributed training sample may address this problem.  Further work using real observations from SKAO commissioning activities and from precursor instruments will examine how well machine learning models can transfer from simulated training data to real observational data.
    \item The existing SoFiA software package also performed very well, achieving third place in the Challenge and also being used by several other teams, including by the second placed team for source characterisation. That the package proved so popular further demonstrates the value of clearly documented and easily accessible codes, in addition to its accuracy and efficiency. This challenge highlights the need for such software packages,  built and designed by astronomers to tackle specific problems, to receive the funding to be well maintained.
    \item Perhaps the most important finding of the Challenge is that of method complementarity. Also seen in the first SKAO Science Data Challenge \citep{Bonaldi_2020}, the relative performance of individual teams varied across aspects of the Challenge. It is likely that a combination of methods will produce the most accurate results.  This finding is underscored by the strategy employed by the winning team, MINERVA. By optimising the combined predictions from two independent machine learning methods, the team was able to record an improvement in score 20 percent above either method alone (see Fig.~\ref{fig_minerva}).  The result demonstrates the promise of ensemble learning in exploiting very large astronomical datasets.

\end{enumerate}



\section*{Acknowledgements}
We would like to thank members of the SKAO HI Science Working Group for useful feedback. We are grateful for helpful discussions with the Software Sustainability Institute. The simulations make use of data from WSRT HALOGAS-DR1. The Westerbork Synthesis Radio Telescope is operated by ASTRON (Netherlands Institute for Radio Astronomy) with support from the Netherlands Foundation for Scientific Research NWO. The work also made use of ‘THINGS’, the HI Nearby Galaxy Survey (Walter et al. 2008), data products from which were kindly provided to us by Erwin de Blok after multi-scale beam deconvolution performed by Elias Brinks. We would like to thank INAF for the hosting of SDC2 data products. LA is grateful for the support from UK STFC via the CDT studentship grant ST/P006809/1. 
This project has received funding from the European Research Council (ERC) under the European Union’s Horizon 2020 research and innovation programme (grant agreement no. 679627; project name FORNAX). JMvdH and KMH acknowledge support from the European Research Council under the European Union’s Seventh Frame- work Programme (FP/2007–2013)/ERC Grant Agreement no. 291531 (HIStoryNU). SSI.
The works of the NAOC-Tianlai team members have been supported by the National Key R\&D Program grants 2018YFE0120800,2017YFA0402603, 2018YFA0404504, 2018YFA9494691, The National Natural Science Foundation of China (NSFC) grants 11633004, 11975072, 11835009, 11890691, 12033008, the Chinese Academy of Science (CAS) QYZDJ-SSW-SLH017, JCTD-2019-05, and the China Manned Space Projects CMS-CSST-2021-A03, CMS-CSST-2021-B01.
Team FORSKA-Sweden acknowledges support from Onsala Space Observatory for the provisioning of its facilities support. The Onsala Space Observatory national research infrastructure is funded through Swedish Research Council (grant No 2017-00648). Team FORSKA-Sweden also acknowledges support from the Fraunhofer Cluster of Excellence \emph{Cognitive Internet Technologies}.
CH, MB acknowledge support by the Deutsche Forschungsgemeinschaft (DFG, German Research Foundation) under Germany's Excellence Strategy -- EXC 2121 „Quantum Universe“ -- 390833306.
MP acknowledges the support of the CEFIPRA foundation under project 6504-3.
We acknowledge financial support from SEV-2017-0709, CEX2021-001131-S, AEI/ 10.13039/501100011033. LD, JG, KMH, JM, MP, SSE, LVM, AS from RTI2018-096228-B-C31, PID2021-123930OB-C21 AEI/ 10.13039/501100011033 FEDER, UE. LVM, JG, SSE acknowledge The European Science Cluster of Astronomy and Particle Physics ESFRI Research Infrastructures project that has received funding from the European Union’s Horizon 2020 research and innovation program under Grant Agreement No. 824064. LVM, JG, and JM RED2018-102587-T AEI/ 10.13039/501100011033. LVM, JG, SSE, JM acknowledge financial support from the grant IAA4SKA (Ref. R18-RT-3082) from the Economic Transformation, Industry, Knowledge and Universities Council of the Regional Government of Andalusia and the ERDF from the EU, TED2021-130231B-I00 AEI/ 10.13039/501100011033 EU NextGenerationEU/PRTR. LVM, JG, KMH acknowledges financial support from the coordination of the participation in SKA-SPAIN, funded by the Ministry of Science and Innovation (MCIN). LD from PTA2018-015980-I AEI/ 10.13039/501100011033. MP from the grant DOC01497 funded by the Economic Transformation, Industry, Knowledge and Universities Council of the Regional Government of Andalusia and by the Operational Program ESF Andalucía 2014-2020.
MTS acknowledges support from a Scientific Exchanges visitor fellowship (IZSEZO\_202357) from the Swiss National Science Foundation. 
AVS thanks Martin Kunz and Bruce Bassett for the valuable discussions. 
Team Spardha would like to acknowledge SKA India Consortium, IUCAA and Raman Research Institute for providing the support with the computing facilities. Team Spardha would also acknowledge National Supercomputing Mission (NSM) for providing computing resources of ‘PARAM Shakti’ at IIT Kharagpur, which is implemented by C-DAC and supported by the Ministry of Electronics and Information Technology (MeitY) and Department of Science and Technology (DST), Government of India.

\section*{Supercomputing partner facilities}

\noindent We would like to make a special acknowledgment of the very generous support from the SDC2 computing partner facilities (Section~\ref{hpcs}), without which a realistic and accessible Challenge would not have been possible. We acknowledge support from the Australian SKA Regional Centre (AusSRC) and the Pawsey Supercomputing Centre. This work was granted access to the HPC/AI resources of IDRIS under the allocations AP010412412, AP010412365 and AP010412404 made by GENCI. The authors acknowledge use of IRIS (https://www.iris.ac.uk) resources delivered by the SCD Cloud at STFC’s Rutherford Appleton Laboratory (https://www.scd.stfc.ac.uk/Pages/STFC-Cloud-Operations.aspx). This work was supported by a grant from the Swiss National Supercomputing Centre (CSCS). This work used resources of China SKA Regional Centre prototype (An, Wu, Hong, Nat Astron, 2019, 3, 1030) funded by the National Key R\&D Programme of China (2018YFA0404603) and Chinese Academy of Sciences (114231KYSB20170003). The Enabling Green E-science for the Square Kilometre Array Research Infrastructure (ENGAGE-SKA) team acknowledges financial support from grant POCI-01-0145- FEDER022217, funded by Programa Operacional Competitividade e Internacionalização (COMPETE 2020) and the Fundação para a Ciência e a Tecnologia (FCT), Portugal. This work was also funded by FCT and Ministério da Ciência, Tecnologia e Ensino Superior(MCTES) through national funds and when applicable co-funded EU funds under the project UIDB/50008/2020-UIDP/50008/2020 and UID/EEA/50008/2019. The authors acknowledge the Laboratory for Advanced Computing at University of Coimbra for providing HPC, computing, consulting resources that have contributed to the research results reported within this paper or work. 
This work used the Spanish Prototype of an SRC (SPSRC) at IAA-CSIC, which is funded by SEV-2017- 0709, CEX2021-001131-S, RTI2018-096228-B-C31 AEI/ 10.13039/501100011033, EQC2019- 005707-P AEI/ 10.13039/501100011033 ERDF, EU, TED2021-130231B-I00 AEI/ 10.13039/501100011033 EU NextGenerationEU, PRTR SOMM17\_5208\_IAA funded by the Regional Government of Andalusia. We acknowledge the computing infrastructures of INAF, under the coordination of the ICT office of Scientific Directorate, for the availability of computing resources and support.

\section*{Data availability}

The SDC2 simulated datasets are publicly available from the SDC2 website: \url{https://sdc2.astronomers.skatelescope.org/}




\bibliographystyle{mnras}
\bibliography{sdc2,misc_refs, references} 

\hrulefill

\footnotesize
\noindent $^{1}$SKA Observatory, Jodrell Bank, Lower Withington, Macclesfield, SK11 9FT, UK\\
$^{2}$Jodrell Bank Centre for Astrophysics, Department of Physics \&
Astronomy, The University of Manchester, Manchester M13 9PL,
UK\\
$^{3}$Shanghai Astronomical Observatory, Key Laboratory of Radio Astronomy, CAS, 80 Nandan Road, Shanghai 200030, China\\
$^{4}$DIO, Observatoire de Paris, CNRS, PSL, 75104, Paris, France\\
$^{5}$Institute for Astronomy, University of Edinburgh, Royal Observatory, Blackford Hill, Edinburgh, EH9 SHJL, UK\\
$^{6}$Department of Astronomy, Astrophysics and Space Engineering, Indian Institute of Technology Indore, Indore 453552, India\\
$^{7}$National Astronomical Observatory, Chinese Academy of Sciences, 20A Datun Road, Beijing 100101, P. R. China\\
$^{8}$School of Physics and Astronomy, Queen Mary University of London, London E1 4NS, UK\\
$^{9}$Centre for Strings, Gravitation and Cosmology, Department of Physics, Indian Institute of Technology Madras, Chennai 600036, India\\
$^{10}$Department of Physics \& Institute of Astronomy, University of Cambridge, Cambridge, United Kingdom\\
$^{11}$LERMA, Observatoire de Paris, PSL research Université, CNRS, Sorbonne Université, 75104, Paris, France\\
$^{12}$Instituto de Astrofísica de Andalucía (CSIC), Glorieta de la Astronomía s/n, 18008 Granada, Spain\\
$^{13}$Department of Information Technology and Electrical Engineering, University of Naples Federico II, 21 Via 
Claudio, I-80125, Napoli, Italy \\
$^{14}$Centre for Astrophysics Research, University of Hertfordshire, Hatfield, Hertfordshire, United Kingdom\\
$^{15}$Centro Brasileiro de Pesquisas Físicas (CBPF), 22290-180 URCA, Rio de Janeiro (RJ), Brazil\\
$^{16}$Institute of Physics, Laboratory of Astrophysics, École Polytechnique Fédérale de Lausanne (EPFL), Observatoire de Sauverny, 1290 Versoix, Switzerland\\
$^{17}$Faculty of Computational Mathematics and Cybernetics of Lomonosov, Moscow State University, Moscow, Russia\\
$^{18}$Fraunhofer-Chalmers Centre \& Fraunhofer Center for Machine Learning, SE-412 88, Gothenburg, Sweden\\
$^{19}$University of Hamburg, Hamburg Observatory, Gojenbergsweg 112, 21029 Hamburg, Germany\\
$^{20}$Instituto de Física de Cantabria, CSIC-UC, Av. de Los Castros s/n, E-39005 Santander, Spain\\
$^{21}$ASTRON, the Netherlands Institute for Radio Astronomy, Postbus 2, 7990 AA, Dwingeloo, The Netherlands\\
$^{22}$Kapteyn Astronomical Institute, University of Groningen, P.O. Box 800, 9700 AV Groningen, The Netherlands\\
$^{23}$Department of Electrical and Electronics Engineering, PES University, Bangalore 560085, India\\
$^{24}$Department of Physics, School of Mathematics and Physics, The University of Queensland, Brisbane QLD 4072, Australia\\
$^{25}$ICRAR M468, The University of Western Australia, 35 Stirling Highway, Crawley, WA 6009, Australia\\
$^{26}$INAF – Osservatorio Astronomico di Cagliari, Via della Scienza 5, 09047 Selargius, CA, Italy\\
$^{27}$Department of Astrophysics, School of Physics and Astronomy, Tel Aviv University, Tel Aviv 69978, Israel\\
$^{28}$Department of Physics, College of Sciences, Northeastern University, Shenyang 110819, China\\
$^{29}$Raman Research Institute, C. V. Raman Avenue, Sadashivanagar, Bengaluru 560080, India\\
$^{30}$Astronomy Centre, Department of Physics and Astronomy, University of Sussex, Brighton, BN1 9QH, UK\\
$^{31}$International Space Science Institute (ISSI), Hallerstrasse 6, CH-3012 Bern, Switzerland \\
$^{32}$Department of Physics, Indian Institute of Technology Kharagpur, Kharagpur  721302, India\\
$^{33}$CSIRO Space and Astronomy, PO Box 1130, Bentley WA 6102, Australia\\
$^{34}$Australian SKA Regional Centre (AusSRC)\\
$^{35}$Special Astrophysical Observatory of RAS, Nizhny Arkhyz, 369167, Russia\\
$^{36}$Department of Space, Earth and Environment, Chalmers University of Technology, Onsala Space Observatory, SE-439 92 Onsala, Sweden\\
$^{37}$Département de Physique Théorique and Center for Astroparticle Physics, University of Geneva\\
$^{38}$CAS Key Laboratory of FAST, National Astronomical Observatories, Chinese Academy of Sciences, Beijing 100101, China\\
$^{39}$School of Physics and Astronomy, Sun Yat-Sen University, 2 Daxue Road, Tangjia, Zhuhai, U1YP8R, China\\
$^{40}$Peng Cheng Laboratory, No.2, Xingke 1st Street, Shenzhen 518000, People’s
Republic of China\\
$^{41}$Department of Physics \& Astronomy, Macalester College, 1600 Grand Avenue, Saint Paul, MN 55105, USA\\
$^{42}$Collège de France, 11 Place Marcelin Berthelot, 75005, Paris, France\\
$^{43}$Université de Strasbourg, CNRS UMR 7550, Observatoire astronomique de Strasbourg, 67000 Strasbourg, France\\
$^{44}$Steward Observatory, University of Arizona, 933 North Cherry Avenue, Tucson, AZ 85721, USA\\
$^{45}$South African Radio Astronomy Observatory (SARAO), 2 Fir Street, Black River Park, Observatory 7925, South Africa\\
$^{46}$Ruhr University Bochum, Faculty of Physics and Astronomy, Astronomical Institute, 44780 Bochum, Germany\\
$^{47}$Department of Astronomy, Tsinghua University, Beijing 100084, P. R. China\\
$^{48}$Canadian Institute for Theoretical Astrophysics, University of Toronto, 60 St. George Street, Toronto, ON M5S 3H8, Canada \\
$^{49}$Space Research Institute of Russian Academy of Sciences, Profsoyuznaya 84/32, 117997 Moscow, Russia\\
$^{50}$Laboratoire Univers et Particules de Montpellier (LUPM)-CNRS, UNIVERSITÉ DE MONTPELLIER LUPM CC 072 - Place Eugène Bataillon 34095 Montpellier Cedex 5, France\\
$^{51}$GEPI, Observatoire de Paris, CNRS, Université Paris Diderot, 5 Place Jules Janssen, 92190, Meudon, France\\
$^{52}$Department of Physics \& Electronics, Rhodes University, PO Box 94, Grahamstown, 6140, South Africa\\






\bsp	
\label{lastpage}
\end{document}